\documentclass[twocolumn]{aa}

\usepackage[T1]{fontenc}
\usepackage[utf8]{inputenc}

\usepackage{graphicx}
\usepackage{natbib}
\usepackage{ulem}
\usepackage{textcomp}
\usepackage{gensymb}
\usepackage{longtable}
\usepackage{threeparttable}
\usepackage{multicol}
\usepackage{multirow}
\usepackage{textgreek}
\usepackage{float}
\usepackage{todonotes}
\usepackage[Symbol]{upgreek}
\usepackage{amsmath}
\usepackage{etoolbox}
\usepackage{txfonts}
\usepackage{url}
\usepackage{xcolor}

\definecolor{ferngreen}{rgb}{0.31, 0.47, 0.26}

\usepackage[most]{tcolorbox}
\usepackage{hyperref}
\usepackage{xcolor}
\hypersetup{
  colorlinks   = true, 
  urlcolor     = blue, 
  linkcolor    = blue, 
  citecolor   = blue 
}
\usepackage[all]{hypcap} 

\begin{document} 
	\title{Turbulent magnetic fields in the merging galaxy cluster MACS\,J0717.5$+$3745: polarization analysis}
	\titlerunning{MACS\,J0717.5$+$3745: polarization analysis}
	\authorrunning{Rajpurohit et al.}

\author{K. Rajpurohit\inst{1,2,3}, M. Hoeft\inst{3}, D. Wittor\inst{4,1},  R. J. van Weeren\inst{5}, F. Vazza\inst{1,2,4},  L. Rudnick\inst{6}, S. Rajpurohit\inst{7}, \\ W. R. Forman\inst{8}, C. J. Riseley\inst{1,2,9},  M. Brienza\inst{1,2}, A. Bonafede\inst{1,2,4}, A. S. Rajpurohit\inst{10}, P. Dom\'{i}nguez-Fern\'{a}ndez\inst{11}, \\J. Eilek\inst{12,13}, E. Bonnassieux\inst{1,2}, M. Br\"uggen\inst{4}, F. Loi\inst{14}, H. J. A. R\"ottgering\inst{5}, A. Drabent\inst{3}, N. Locatelli\inst{1,2}, A. Botteon\inst{5}, G. Brunetti\inst{2}, and T. E. Clarke\inst{15}}

\institute{Dipartimento di Fisica e Astronomia, Universit\'a di Bologna, Via P. Gobetti 93/2, 40129, Bologna, Italy\\
 {\email{kamlesh.rajpurohit@unibo.it}}
\and
INAF-Istituto di Radio Astronomia, Via Gobetti 101, 40129, Bologna, Italy
\and
Th\"uringer Landessternwarte (TLS), Sternwarte 5, 07778 Tautenburg, Germany
\and
Hamburger Sternwarte, Universit\"at Hamburg, Gojenbergsweg 112, 21029, Hamburg, Germany
\and
Leiden Observatory, Leiden University, P.O. Box 9513, 2300 RA Leiden, The Netherlands
\and
Minnesota Institute for Astrophysics, University of Minnesota, 116 Church St. S.E., Minneapolis, MN 55455, USA
\and
Molecular Foundry, Lawrence Berkeley National Laboratory, CA 94720, USA
\and
Harvard-Smithsonian Center for Astrophysics, 60 Garden Street, Cambridge, MA 02138, USA
\and
CSIRO Astronomy and Space Science, PO Box 1130, Bentley, WA 6102, Australia
\and
Astronomy \& Astrophysics Division, Physical Research Laboratory, Ahmedabad 380009, India
\and
Department of Physics, School of Natural Sciences UNIST, Ulsan 44919, Korea
\and
INAF-Osservatorio Astronomico di Cagliari, Via della Scienza 5, 09047 Selargius, CA, Italy
\and
Department of Physics, New Mexico Tech, Socorro, NM 87801, USA 
\and
 National Radio Astronomy Observatory, Socorro, NM 87801, USA
\and
U.S. Naval Research Laboratory, 4555 Overlook Avenue SW, Washington, D.C. 20375, USA
}

    \abstract
  {We present wideband ($1{-}6.5\,\rm GHz$) polarimetric observations, obtained with the Karl G. Jansky Very Large Array (VLA), of the merging galaxy cluster MACS\,J0717.5$+$3745, which hosts one of the most complex known radio relic and halo systems. We use both Rotation Measure Synthesis and QU-fitting, and find a reasonable agreement of the results obtained with these methods, in particular, when the Faraday distribution is simple and the depolarization is mild. The relic is highly polarized over its entire length (850\,kpc), reaching a fractional polarization ${>}30\%$ in some regions. We also observe a strong wavelength-dependent depolarization for some regions of the relic. The northern part of the relic shows a complex Faraday distribution suggesting that this region is located in or behind the intracluster medium (ICM). Conversely, the southern part of the relic shows a Rotation Measure very close to the Galactic foreground, with a rather low Faraday dispersion, indicating very little magnetoionic material intervening the line-of-sight. From spatially resolved polarization analysis, we find that the scatter of Faraday depths correlates with the depolarization, indicating that the tangled magnetic field in the ICM causes the depolarization. We conclude that the ICM magnetic field could be highly turbulent. At the position of a well known narrow-angle-tailed galaxy (NAT), we find evidence of two  components clearly separated in Faraday space. The high Faraday dispersion component seems to be associated with the NAT, suggesting the NAT is embedded in the ICM while the southern part of the relic lies in front of it. If true, this implies that the relic and this radio galaxy are not necessarily physically connected  and thus, the relic may be not powered by the shock re-acceleration of fossil electrons from the NAT. The magnetic field orientation follows the relic structure indicating a well-ordered magnetic field. We also detect polarized emission in the halo region; however the absence of significant Faraday rotation and a low value of Faraday dispersion suggests the polarized emission, previously considered as the part of the halo, has a shock(s) origin. }


   \maketitle

\section{Introduction}
\label{sec:intro}
Magnetic fields are pervasive throughout the Universe and play a vital role in numerous astrophysical processes: from our solar system up to filaments and voids in the large-scale structure \citep[e.g.,][]{Klein2015,Beck2015}. Even the largest virialized structures in the Universe, galaxy clusters, are permeated by magnetic fields \citep[see][for reviews]{Carilli2002,Govoni2004,Donnert2018}. However, the actual strength, topology, and evolution of these fields is poorly constrained. 

The strongest evidence of cluster-wide magnetic fields comes from radio observations that have revealed large megaparsec-size, diffuse synchrotron emitting sources known as radio relics and halos \citep[see][for a recent review]{vanWeeren2019}. The presence of large-scale magnetic fields may have important implications for the different processes observed in galaxy clusters. A detailed analysis of the diffuse radio sources in clusters may help to shed light on the origin of the magnetic fields, for example to determine if -as widely believed- very weak seed fields are amplified by a dynamo process in the intracluster medium (ICM) and to determine how the magnetic fields impact the physics of the ICM. 
 
Radio relics are found in the periphery of merging galaxy clusters and they often show irregular and filamentary morphologies \citep[e.g.,][]{Owen2014,vanWeeren2017b,Rajpurohit2018,Gennaro2018,Rajpurohit2020a}. Relics trace shock waves occurring in the ICM during cluster merger events \citep[e.g.,][]{Sarazin2013,Ogrean2013,vanWeeren2016a,Botteon2016b,Botteon2018}. 

It is  believed that the cosmic ray electrons (CRe), which form the radio relics via synchrotron emission, originate from a first-order Fermi process, namely, Diffusive Shock Acceleration \citep[DSA, e.g.,][]{Blandford1987,Drury1983,Ensslin1998,Hoeft2007}. The DSA at the shocks causing relics may also re-accelerate a pool of mildly relativistic fossil electrons, previously injected by active galactic nuclei \citep[AGN: e.g.,][]{Bonafede2014,vanWeeren2017a}. The existence of mildly relativistic electrons in front of the shock may help to reconcile the low acceleration efficiency with the high radio luminosity for relics with weak shocks \citep[Mach number $\leq2$, e.g.,][]{Kang2011,Botteon2020a}. However, the exact spectral energy distribution and spatial distribution of such fossil electrons in the ICM are mostly unconstrained.

Radio relics are strongly polarized at frequencies above 1\,GHz, some with a polarization fraction as high as $65\%$ \citep[e.g.,][]{vanWeeren2010,vanWeeren2012a,Owen2014,Kierdorf2016,Loi2020,Rajpurohit2020b,DiGennaro2021}. The inferred magnetic field directions are often found to be well aligned with the shock surface. However, the exact mechanism causing the high degree of polarization and the aligned polarization angle is still unclear. The alignment could be caused by a preferentially tangential magnetic field orientation or by the compression of a small-scale tangled magnetic field at shock \citep{Laing1980,Ensslin1998}.

\setlength{\tabcolsep}{15.0pt}   
\begin{table*}[!htbp]
\caption{Image properties }
\centering
\label{Table 2}
 \begin{threeparttable} 
\begin{tabular}{c c c c c | c  c }
\hline\hline
\multirow{1}{*}{Band} & \multirow{1}{*}{Configuration}  & \multirow{1}{*}{Name} &\multirow{1}{*}{Restoring Beam}  &  \multicolumn{2}{c}{RMS noise}\\ 
 \cline{5-6} 
 &&&&Stokes $I$ ($\sigma_{\rm rms}$) & Stokes $QU$ ($\sigma_{QU}$)\\
&&&&$\upmu\,\rm Jy\,beam^{-1}$&$\upmu\,\rm Jy\,beam^{-1}$\\

\hline
   & BCD&IM1 &$2.0\arcsec \times 2.0\arcsec$&1.4&1.2\\
VLA C-band $^{\dagger}$& BCD&IM2 &$4.0\arcsec \times 4.0\arcsec$&1.8&1.4\\
(4.5-6.5\,GHz) &  BCD&IM3&$5.0\arcsec \times 5.0\arcsec$&1.9&1.5 \\
& BCD&IM4&$12.5\arcsec \times 12.5\arcsec$&7.2&5.9\\
\hline   
&  ABCD&IM5 &$2.0\arcsec \times 2.0\arcsec$&1.1&0.9 \\
VLA S-band$^{\dagger}$&  ABCD&IM6 &$4.0\arcsec \times 4.0\arcsec$& 1.7&1.0 \\
(2-4\,GHz)&ABCD&IM7&$5.0\arcsec \times 5.0\arcsec$&1.8&1.1\\
 &  ABCD&IM8&$12\farcs5 \times 12\farcs5$& 8.2&6.1\\
\hline
&  ABCD&IM9 &$2.0\arcsec \times 2.0\arcsec$&3.2&2.7 \\
VLA L-band$^{\dagger}$& ABCD&IM10&$4.0\arcsec\times 4.0\arcsec$&6.0&3.2\\
(1-2\,GHz)&  ABCD&IM11&$5.0\arcsec \times 5.0\arcsec$&6.8&3.4\\
 &  ABCD&IM12&$12.5\arcsec \times 12.5\arcsec$&12.4&10.1\\
 \hline
\\
\end{tabular}
\begin{tablenotes}[flushleft]
  \footnotesize
   \vspace{-0.2cm}
   \item\textbf{Notes.} For full wideband Stokes $IQU$ maps, imaging was performed using multi-scale clean, $\tt{nterms}$=2 and $\tt{wprojplanes}$=500.  All Stokes $IQU$ images are made with {\tt Briggs} weighting with ${\tt robust}=0$ and different uv-tapering. For making all Stokes $IQU$ images at $4\arcsec$, $5\arcsec$, and $12.5\arcsec$ resolutions, we used  $\tt{nterms}$=1 and ${\tt robust}=0.0$. The single channel images were re-gridded to the same pixel size. Due to slight difference in the beam size from 1-6.5\,GHz, all Stokes \textit{IQU} cubes (used for RM-Synthesis and QU-fitting) were convolved to the same beam size using {\tt CASA} task {\tt imsmooth}; $^{_\dagger}$For data reduction steps, we refer to \cite{vanWeeren2016b,vanWeeren2017b}.  
       \end{tablenotes}
    \end{threeparttable} 
\label{Tabel:imaging}
\end{table*}

Unlike relics, radio halos are typically unpolarized sources located at the center of a cluster. The radio emission from halos roughly follows the X-ray emission \citep[e.g.,][]{Pearce2017,Rajpurohit2018,vanWeeren2017a}. The currently favored scenario for the formation of radio halos involves the reacceleration of CRe to higher energies by turbulence induced during mergers \citep[e.g.,][]{Brunetti2001, Petrosian2001,Brunetti2014}.

Polarized emission at the cluster outskirts is crucial to understand the magnetic field properties of the ICM \citep[see,][for a review]{Carilli2002,Govoni2004}. The orientation and topology of magnetic fields at merger shocks is important to better understand the physics of shock acceleration,  because the efficiency of particle acceleration might be a strong function of the magnetic field topology upstream of the shock  \citep[e.g.,][]{2020MNRAS.495L.112W}. However, magnetic fields in the ICM are notoriously difficult to measure, and the low-density regions in the cluster outskirts are even more challenging to probe \citep{Johnson2020}.

In this work, we describe the results obtained from polarimetric analysis of Karl G. Jansky Very Large Array (VLA) L-, S-, and C-band observations covering the 1-6.5\,GHz frequency range. The enormous wideband information and remarkable  resolution allow us to carry out spatially resolved polarimetric studies, providing crucial insights into the ICM magnetic fields.

The outline of this paper is as follows: in Sect.\,\ref{obs}, we describe the observations and data reduction. The polarization images are presented in Sect.\,\ref{results}. This is followed by a detailed analysis and discussion from Sects.\,\ref{sec:Faraday} to \ref{halopol}. We summarize our main findings in Sect.\,\ref{summary}. 

Throughout this paper, we assume a $\Lambda$CDM cosmology with $H_{\rm{ 0}}=70$ km s$^{-1}$\,Mpc$^{-1}$, $\Omega_{\rm{ m}}=0.3$, and $\Omega_{\Lambda}=0.7$. At the cluster's redshift, $1\arcsec$ corresponds to a physical scale of 6.4\,kpc. All output images are in the J2000 coordinate system and are corrected for primary beam attenuation.

\begin{figure*}[!thbp]
\centering
\includegraphics[width=1.0\textwidth]{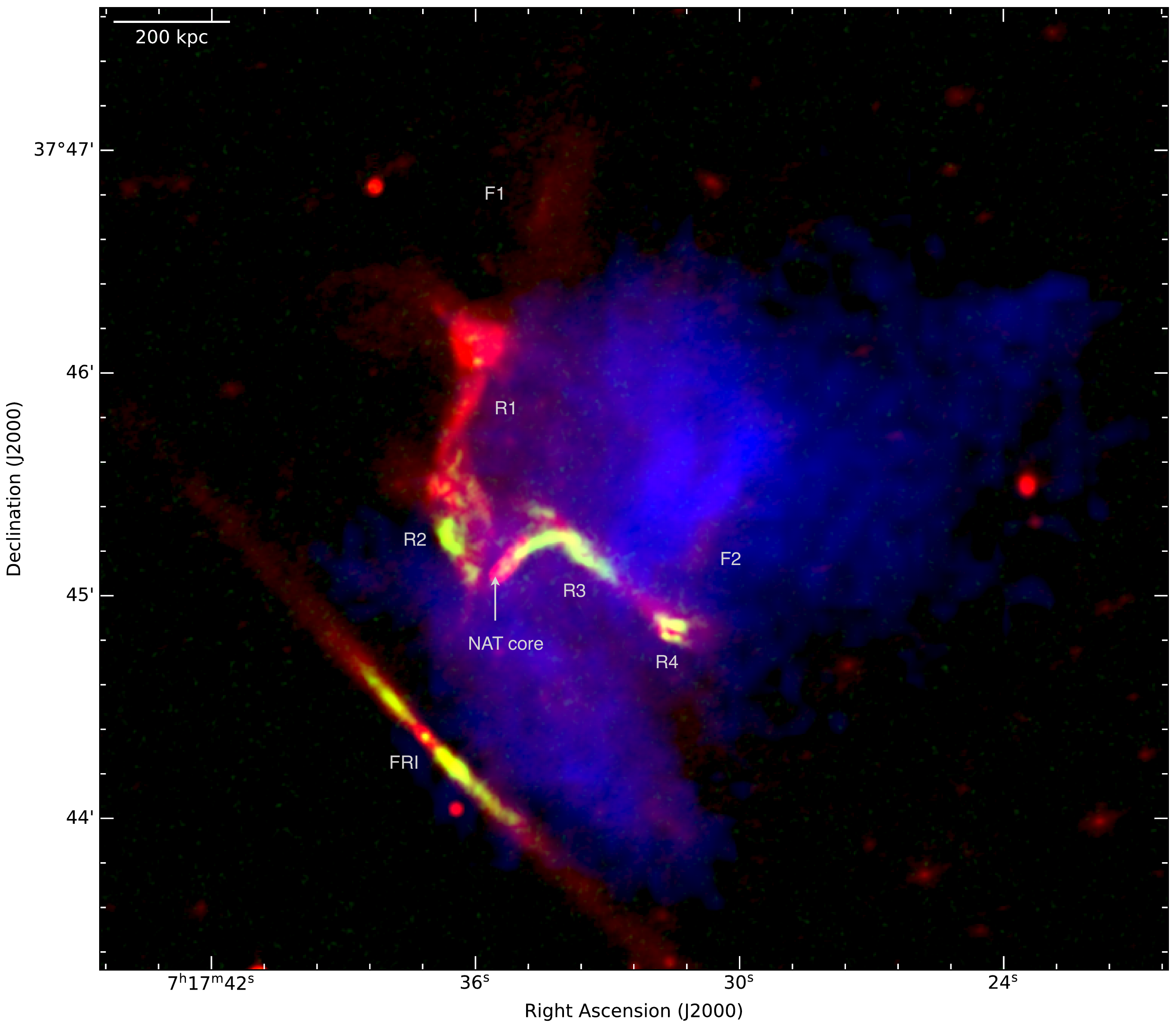}
\caption{Composite total power, polarization intensity, and X-ray image of the relic in MACS\,J0717+3745 at $2\arcsec$ resolution. The total power and L-band polarization emission are shown in red and green-yellow, respectively. The intensity in blue shows the X-ray emission. The lack of polarized emission in the north region (R1) indicates depolarization. The image properties are given in Table\,\ref{Tabel:imaging}, IM5 and IM9.}
\label{IPX}
\end{figure*} 

\section{MACS J0717.5+3745}

The galaxy cluster MACS\,J0717.5$+$3745 ($l=180.25^{\degree}$ and $b=+21.05^{\degree}$) located at $z=0.5458$, hosts one of the most complex and powerful known relic-halo systems. \citep[e.g.,][]{Bonafede2009a,vanWeeren2009,PandeyPommier2013,vanWeeren2017b,Rajpurohit2021a,Rajpurohit2021b}. The relic consists of four subregions, which have historically been referred to as R1, R2, R3, and R4; see Fig\,\ref{IPX}. The relic is known to be polarized above 1.4\,GHz and the polarization fraction varies along the relic \citep{Bonafede2009a}. 

High-resolution images from the VLA reveal that the relic has several filaments on scales down to $30\,\rm kpc$. Some of these filaments originate from the relic itself, while a few of them, F1 and F2 (see Fig.\,\ref{IPX}) appear more isolated. Recently, it has been reported that the relic consists of several fine overlapping structures with different spectral indices  \citep{Rajpurohit2021a}. The curvature distribution suggests that the relic is very likely seen less edge-on with a viewing angle close to about $45\degree$.

At the center of the relic, there is an embedded Narrow Angle Tail (NAT) galaxy (see Fig.\,\ref{IPX}). At high frequencies (above 1\,GHz), the tails of the NAT seem to fade into the R3 region of the relic \citep{vanWeeren2017b}. However, at low frequencies (below 750 MHz), the tails are apparently bent to the south of the R3 region \citep{Rajpurohit2021a}. It is not clear if this morphological connection between the NAT and the relic is physical or they are simply two different structures projected along the same line-of-sight \citep{vanWeeren2017b,Rajpurohit2021a}. If this connection is physical this may suggest that the relic is powered by the shock re-acceleration of mildly relativistic fossil electrons from the NAT. 

The cluster also hosts a powerful steep spectrum radio halo with a largest linear size of about 2.6\,Mpc \citep{Bonafede2009a,vanWeeren2009,PandeyPommier2013,vanWeeren2017b,Bonafede2018,Rajpurohit2021b}. High resolution total power images taken with the VLA have shown the presence of several radio filaments of 100-300\,kpc in extent within the halo region \citep{vanWeeren2017b}. Uncommonly, for radio halos, the halo in MACS\,J0717.5$+$3745 has previously been reported to be polarized at the level of 2-7\% at 1.4\,GHz \citep{Bonafede2009a}. However, it is unclear if the polarized emission reported in the central part of the halo is associated with the halo or not.

 \begin{figure*}[!thbp]
    \centering
    \includegraphics[width=1.0\textwidth]{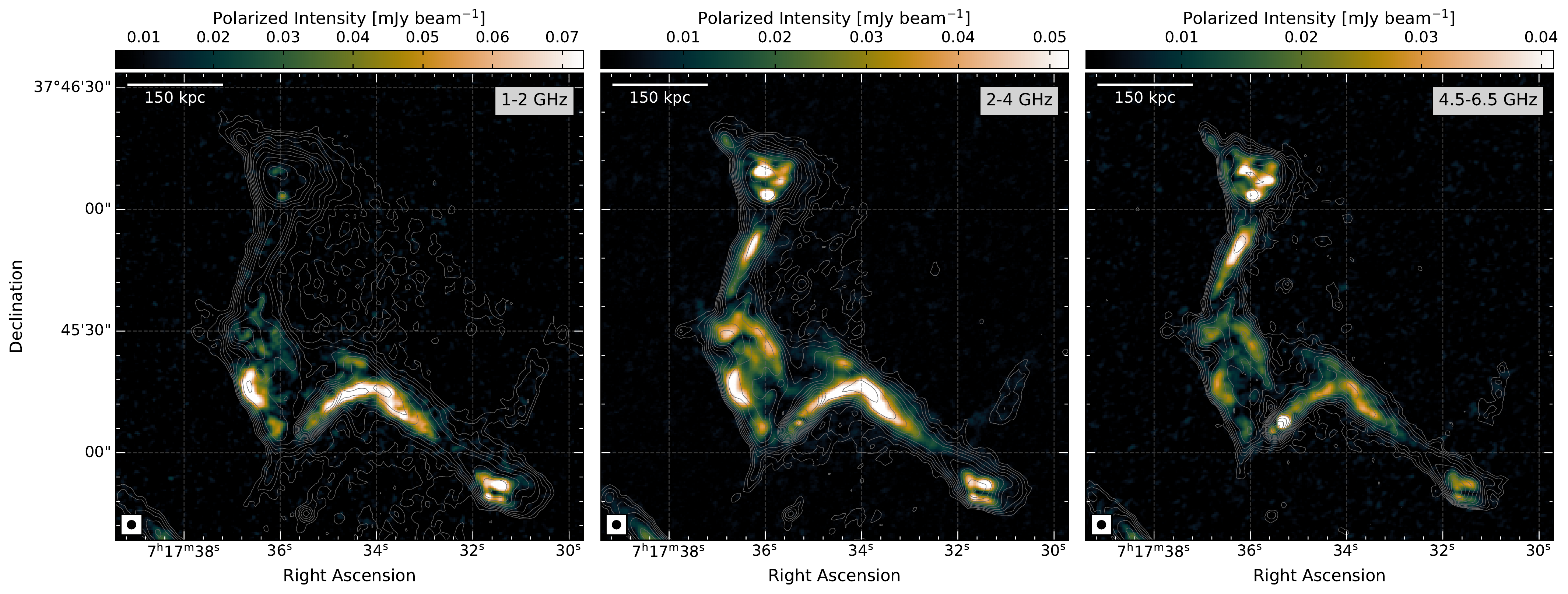}
    \vspace{-0.7cm}
    \caption{ Polarization intensity images of the relic in MACS\,J0717.5$+$3745 at $2\arcsec$ resolution, showing that the polarization emission is distributed in a clumpy manner. The image also revels fine, small-scale filaments visible in the total power emission. Contour levels are drawn at $\sqrt{[1,2,4,8,\dots]}\,\times\,5\sigma_{{\text{ rms}}}$  and are from the VLA L-, S-, and C-band Stokes $I$ images. The beam sizes are indicated in the bottom left corner of the each image. The image properties are given in Table\,\ref{Tabel:imaging}, IM1, IM5, and IM9.}
      \label{fig3}
  \end{figure*}


\section{Observations and Data Reduction}
\label{obs}

VLA wideband observations of MACS\,J0717.5$+$3745 were obtained in L-band (ABCD-array configuration), S-band (ABCD-array configuration), and C-band (BCD-array configuration), covering the frequency range from 1 to 6.5\,GHz. For observation details and description of the data reduction procedure, we refer the reader to \cite{vanWeeren2016b,vanWeeren2017b}. While previous analysis of the data only considered the total power, the VLA observations were taken in full-polarization mode, allowing us to investigate the polarization properties of the cluster.

The data reduction and imaging were performed with {\tt CASA}. The data were calibrated for antenna position offsets, elevation-dependent gains, parallel-hand delay, bandpass, and gain variations using 3C147. For polarization calibration, the leakage response was determined using the unpolarized calibrator 3C147. The cross-hand delays and the absolute polarization position angle were corrected using 3C138. Finally, the calibration solutions were applied to the target field and the resulting calibrated data were averaged by a factor of 4 in frequency per spectral window. Several rounds of self-calibration were performed to refine the gain solutions. After the individual data sets were calibrated, the observations from the different configurations (for the same frequency band) were combined and imaged together. 

We produced Stokes $I$, $Q$, and $U$ images of the target field from the data at L-band, S-band, and C-band, including data from all array configurations. Deconvolution was done with {\tt CLEAN} masks generated in the PyBDSF \citep{Mohan2015}. Imaging was always performed with Briggs weighting \citep{Briggs1995} using \verb|robust|~$=0.0$ and all images were corrected for primary beam attenuation, see Table~\ref{Tabel:imaging} for the properties of the images obtained. For Faraday analysis, the full 1-6.5\,GHz channel with Stokes $IQU$ cubes were imaged to a common resolution of $4\arcsec$, $5\arcsec$ and $12.5\arcsec$ resolutions. These $IQU$ cubes were inspected and any spectral channels which showed large artifacts or a large increase in noise compared to the average were excluded. We note that the highest common resolution possible with our L-, S- and C-bands VLA data was $4\arcsec$, therefore $2\arcsec$ resolution Stokes $IQU$ cubes were not used for Faraday analysis.

The polarized flux density was computed from the Stokes $Q$ and $U$ flux densities according to the definition of polarization as a complex property 
\begin{equation}
 P = Q + iU,
\end{equation}
where the absolute of $P$ results in the polarized flux density. Since both, $Q$ and $U$, are affected by noise in the measurement, we correct the polarized flux density, $|P|$, for the Rician bias {\citep{Wardle1974,George2012} as:
\begin{equation}
 |P|=\sqrt{Q_{\rm meas}^{2}+U_{\rm meas}^2-2.3\,\sigma_{QU}^2},
\end{equation}
where $\sigma_{QU}$ is the average rms of Stokes $Q$ and $U$ images and the index `meas' indicates the measured property, unavoidably including a noise, for clarity. The uncertainties in the flux density measurements were estimated as: 
\begin{equation}
 \Delta S_\nu
 =  
 \sqrt {(f \cdot S_\nu)^{2}+{N}_{{\rm{ beams}}}\cdot\sigma_{\rm{rms}}^{2}}
 \: ,
\end{equation}
where $f$ is an absolute flux density calibration uncertainty, $S_\nu$ is the flux density, $\sigma_{{\rm{ rms}}}$ is the RMS noise, and $N_{{\rm{beams}}}$ is the number of beams covered by the source. We assume absolute flux density uncertainties of 4\% for the VLA L-band and 2.5\% for the VLA S- and C-bands \citep{Perley2013}.


\section{Polarized emission}
\label{results}

In Fig.\,\ref{fig3}, we show the high-resolution, that is $2\arcsec$, polarized intensity maps of the relic for the VLA L-, S-, and C-band. Polarized emission from the relic subregions (R1, R2, R3, and R4) is detected in all three frequency bands. 

The polarized emission more or less follows the structure seen in the total intensity images; however, the polarized emission seems to be more `clumpy' compared to the total power emission.  Moreover, we find that there are  fluctuations in the polarization intensity, in particular for the northern part of the relic (R1 and R2), on scale as small as 10\,kpc. The polarized intensity map in Fig.\,\ref{fig3} also indicates that in L-band the polarized flux density in the northern part of the relic is low compared to the southern part (R3 and R4). To investigate this further, we create maps for the fractional polarization $p=|P|/I$. 

The L-, S-, and C-band high resolution ($2\arcsec$) fractional polarization maps of the relic are shown in Fig.\,\ref{fig3a}. At such a high resolution, the relic is polarized over its entire length in C-band. We find that in all three bands the polarization fraction across the relic varies significantly, from unpolarized to a maximum fractional polarization of about 50\,\% in C-band. 

An overview of the polarization properties of the diffuse radio sources in the cluster is given in Table\,\ref{Tabel:Tabel2}. We measure the average fractional polarization in the four subregions of the relic. These regions are shown in the left panel of Fig.\,\ref{regions}. The spatially averaged polarization fraction over R1 is $(21 \pm 2)\,\%$ in C-band. The fractional polarization drops quickly towards lower frequencies, reaching as low as $(3 \pm 1)\,\%$ in L-band. Similar trends are noticed for the R2 region of the relic. We observe large local fluctuations in the polarization fraction in particular for R1 and R2.

From Table\,\ref{Tabel:Tabel2}, the average fractional polarization of the R3 region (($28 \pm 3$)\,\% at C-band) is highest compared to the rest of the relic. In the region where the NAT is located we find a very low fractional polarization in all three bands. For the relic, the average polarization fraction at R4 is the lowest ($15 \pm 1$\,\% at C-band). 

In the right panel of Fig.\,\ref{regions}, we show the polarization fraction profiles for the relic extracted from the $2\arcsec$ map at 3\,GHz. The corresponding regions are depicted in the inset. Overall, there is a hint that the fractional polarization decreases in the downstream regions. Recently, \cite{DiGennaro2021} found that the polarization fraction decreases toward the downstream of the entire Sausage relic. Simulations show that such trends are expected in a turbulent medium while the opposite trends (i.e., polarization fraction increasing toward the downstream region of a shock front) is expected if the medium is uniform \citep{Paola2021a}. In the R2 region of the relic, the degree of polarization first decreases for about 60\,kpc and then increases (from 60\,kpc to 100\,kpc). This could be due to projection effects because there are filamentary structures at that location \citep{vanWeeren2017b}.

Looking at the high-resolution fractional polarization images from C-band to L-band (Fig.\,\ref{fig3a}), the average fractional polarization of the relic in MACS\,J0717.5$+$3745 also increases in L-band and S-band from R1 to R3. We find that the degree of polarization decreases generally with increasing wavelength. 

We also create polarization maps at $5\arcsec$ resolution. The resulting maps at 1.5\,GHz, 3\,GHz, and 5.5\,GHz are shown in Fig. \,\ref{fig3b}. We note that that these polarization intensity maps are obtained by applying the Rotation Measure Synthesis  \citep[RM-synthesis:][discussed in Sect.\,\ref{sec:RMsynthesis}]{Brentjens2005}. In these maps, we find additional low surface brightness polarized emission, in particular in filamentary features F1 and F2 (see Fig.\,\ref{IPX} for labeling). The two filaments are highly polarized at all the observed frequencies, reaching values as high as ($30 \pm 2$)\,\% in L-band, see Table\,\ref{Tabel:Tabel2}.

At $5\arcsec$, the polarization fraction across the relic varies from about 2\,\%, that is the lowest value for a detection of polarized emission in our maps, up to about 40\,\% between 1 and 5.5\,GHz. In Table\,\ref{Tabel:Tabel2}, we also report the average fractional polarization measured from $5\arcsec$ maps in the relic subregions. These values are consistent with those reported by \cite{Bonafede2009a} but are lower than measured from our $2\arcsec$ resolution maps. This implies that the degree of polarization increases with increasing resolution, mainly by a factor of about 1.4 (for a resolution improving by a factor of 2.5). At low resolution, regions with different polarization characteristics become blurred within a single resolution element, leading to a loss of the observed polarized signal. This effect is known as beam depolarization, and will be less if the source is imaged at a higher resolution. Fig. \,\ref{fig3b} also shows a clear and sharp distinction between the main relic and the filaments (F1 and F2), apparently protruding from the relic. 

As shown in Figs.\,\ref{fig3a} and \,\ref{fig3b}, the average fractional polarization across the relic increases from R1 to R3 also in L-band and S-band. Moreover, the southern part of the relic is still significantly polarized in L-band. In contrast, the northern part of the relic seems to be depolarized from C-band to L-band.

In Fig.\,\ref{DP}, we show depolarization maps of the relic at $2\arcsec$ and $5\arcsec$ resolutions. The depolarization fraction is defined as ${\rm DP}=p_{\rm 1.5\,GHz}/p_{\rm 5.5\,GHz}$, where $p_{\rm 1.5\,GHz}$ and $p_{\rm 5.5\,GHz}$ are the fractional polarization values at 1.5\,GHz and 5.5\,GHz, respectively. As evident, for the northern part of the relic, we find strong depolarization (${\rm DP}{<}0.4$) between 1.5\,GHz and 5.5\,GHz. In particular, the R1 region of the relic is almost fully depolarized at 1.5\,GHz. For the southern part, the depolarization fraction (${\rm DP}{>}0.6$) is less significant, compared to the northern part. 
There is also clear beam depolarization when comparing the $2\arcsec$ and $5\arcsec$ resolution maps, see Table\,\ref{Tabel:Tabel2} for the average depolarization fraction values across the subregions of the relic.

 \begin{figure*}[!thbp]
    \centering
    \includegraphics[width=1.0\textwidth]{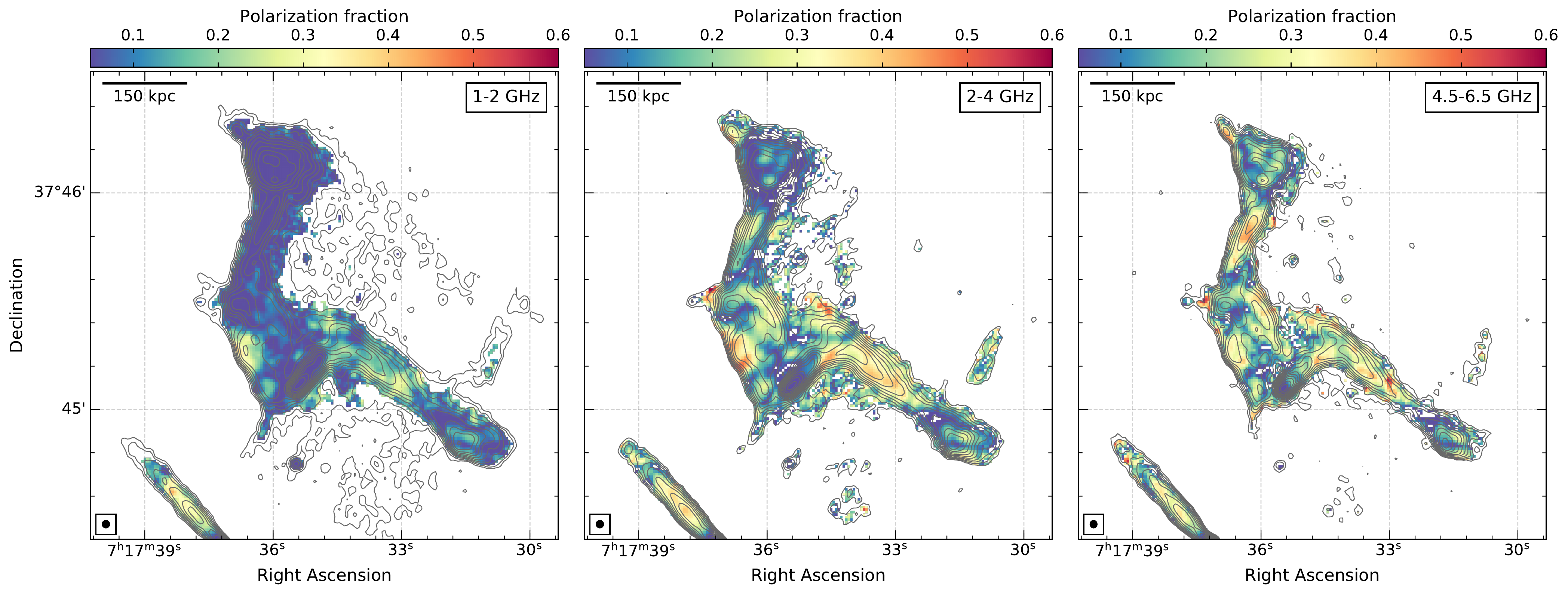}
   \vspace{-0.7cm}
    \caption{
    High resolution ($2\arcsec$) fractional polarization maps at VLA L-, S-, and C-band. The relic is polarized at all of the observed frequencies, reaching values up to 50\,\% in some regions. The northern part of the relic strongly depolarizes at 1.5\,GHz. Contour levels are drawn at $\sqrt{[1,2,4,8,\dots]}\,\times\,5\sigma_{{\text{rms}}}$ and are from the S-band Stokes $I$ image. These maps are corrected for the Rician bias. The beam sizes are indicated in the bottom left corner of the each image. The image properties are given in Table\,\ref{Tabel:imaging}, IM1, IM5, and IM9. }
    \label{fig3a}
  \end{figure*}

 \setlength{\tabcolsep}{15pt}
\begin{table*}[!htbp]
\caption{Polarization properties of the diffuse radio emission in the cluster MACS\,J0717.5$+$3745.}
\centering
\begin{threeparttable} 
\begin{tabular}{ c | c | c | c | c | c | c | c  c }
\hline \hline 
\multirow{1}{*}{Source} & \multicolumn{6}{c|}{VLA} & \multicolumn{2}{c}{Depolarization fraction} \\ 
 \cline{2-7} 
&\multicolumn{2}{c|}{C-band} & \multicolumn{2}{c|}{S-band} & \multicolumn{2}{c|}{L-band} & \multicolumn{2}{c}{(DP)}\\
  \cline{2-7} \cline{8-9}
  &\multicolumn{2}{c|}{$\langle p_{{\rm{5.5\,GHz}}} \rangle$} & \multicolumn{2}{c|}{$\langle p_{{\rm{3.0\,GHz}}} \rangle$} & \multicolumn{2}{c|}{$\langle p_{{\rm{1.5\,GHz}}} \rangle$}&\\

 &2\arcsec & 5\arcsec & 2\arcsec & 5\arcsec & 2\arcsec & 5\arcsec & 2\arcsec & 5\arcsec\\
     \cline{2-7} 
  &\% & \% & \% & \% & \% & \% &  & \\
  \cline{2-3} \cline{3-4}\cline{5-6}\cline{7-8}
  \hline  
R1 &$21\pm2$& $15\pm2$&$12\pm2$ & $9\pm1$&$3\pm1$&$2\pm1$& 0.23&0.12\\ 
R2 &$21\pm2$&$17\pm2$& $20\pm2$ &$16\pm2$& $9\pm1$&$7\pm2$&0.54&0.45\\ 
R3 &$28\pm3$&$20\pm2$& $28\pm2$ &$20\pm2$& $16\pm2$&$11\pm1$&0.70&0.60\\ 
R4 &$15\pm1$&$10\pm1$& $14\pm1$ &$9\pm1$& $9\pm1$&$6\pm1$&0.78&0.76\\ 
F1 &$-$&$30\pm2$&$-$&$26\pm2$&$-$&$13\pm1$&-&-\\ 
F2 &$-$&$24\pm2$&$-$&$22\pm2$&$-$&$14\pm1$&-&-\\ 
\hline 
\end{tabular}
\begin{tablenotes}[flushleft]
\footnotesize
\item{\textbf{Notes.}} {The fractional polarizations and ${\rm DP}$ are average values measured from $2\arcsec$ resolution L-, S-, and C-band fractional polarization and depolarization maps. The regions where the fractional polarization were extracted are indicated in the left panel of Fig.\,\ref{regions}.}
\end{tablenotes}
\end{threeparttable} 
\label{Tabel:Tabel2}   
\end{table*} 


\section{Faraday Rotation analysis}
\label{sec:Faraday}

The polarization properties of diffuse sources in clusters are vital to understand their origin and formation scenario. Moreover, they may serve as a powerful tool to disentangle the contribution from different emission regions which are otherwise blended along the line-of-sight in continuum emission.

One of the most important physical effects to consider when discussing radio polarimetric observations is Faraday rotation. It occurs when a radio wave on its way to the observer passes through a magnetoionic medium which causes the polarization angle, $\psi_{\rm obs}$, to vary as a function the wavelength ($\lambda$). The strength of the Faraday effect is measured by the Rotation Measure (RM):
\begin{equation}
  {\rm RM} 
  = 
  \frac{ {\rm d} \psi_{\rm obs} }{ {\rm d} \lambda^2 }
  \: .
  \label{RM}
\end{equation}
We note that we use RM to exclusively indicate how rapidly the polarization angle changes with $\lambda^2$. In observations, the polarization angle is obtained from the Stokes parameters $Q$ and $U$ of linearly polarized emission via
\begin{equation}
  \psi_{\rm obs}
  =
  {\frac{1}{2}}\arctan \left( \frac{U}{Q} \right)
  \: .
  \label{RM1}
\end{equation}\

 A magnetoionic medium causes a rotation of the polarization angle according to 
 \begin{equation}
     \Delta \psi 
     =
     \Delta \phi \,
     \lambda^2 
     \, .
 \end{equation}
 The difference in Faraday depth ($\Delta \phi$) is given by integrating a section of the light traveling path ($\Delta l$) 
 \begin{equation}
     \Delta \phi
     =
     0.81 \,{\rm rad \, m^{-2} \: 
    \int_{\Delta l} n_{{\rm{e}}} \, \mathit{B}_{\parallel} \mathop{{\rm{d}}\mathit{l}}}
    \, ,
 \label{RM1a}
 \end{equation}
 where the thermal electron density ($n_{\rm e}$), the magnetic field component along the line-of-sight ($B_{\parallel}$), and the path length ($l$) are in units of $\rm cm^{-3}$, $\upmu{\rm{G}}$, and pc, respectively. The Faraday depth denotes the integral given in Eq.\,\ref{RM1a} when the path length is taken from the observer to an arbitrary point along the line-of-sight. If there is only one source of emission along the line-of-sight the Faraday depth of the source position equals the RM obtained from the polarization analysis. If the emission has a more complex distribution in Faraday depth the derivative in Eq.\,\ref{RM} does not allow to conclude on the Faraday depth distribution in a simple, straight forward way.

 Polarization studies of extragalactic sources have shown that a significant number of extended radio sources cannot be described by a single component in Faraday depth \citep[e.g.,][]{OSullivan2012,Anderson2015}. Hence, the angle of Faraday rotation for multiple rotating or emitting screens along the line-of-sight is characterized by a distribution of Faraday depths ($\phi$) instead of a single component. For a mixed Faraday rotating and synchrotron emitting medium, the observed polarized intensity may originate from a large range of Faraday depths.

 The RM, as given in Eq.~\ref{RM}, determined in the observers frame, would differ from similar measurements carried out closer to the source, for example in the rest-frame of the source,  since the photons get redshifted from the source to the observer due to the cosmological expansion \citep{Kim2016,2018MNRAS.477.2528B}. Specifically, if in the rest-frame of the source, located at redshift $z_{\rm RM}$, a RM of ${\rm RM}_{\rm int}$ is determined, the cosmological expansion would cause that in the observers frame an RM of
\begin{equation}
  \rm RM_{\rm obs} 
  = 
  \frac{ {\rm RM}_{\rm int} } { (1+z_{\rm RM})^2 },
  \label{eq::RMrest} 
\end{equation} 
is found, if no magnetoionic medium is present along the line of sight.

Analyzing the Faraday rotation is a powerful method by which to investigate extragalactic magnetic fields. Observations of the polarization angle as a function of frequency may provide crucial information about the magnetization of the source and of the medium intervening along the line-of-sight. 

Fundamental for this analysis is measuring the polarization angle over a wide range of wavelength. As discussed above, the polarization angle may depend in a complex way on wavelength if there is polarized emission with a wide spread of Faraday depths along the line-of-sight \citep{Burn1966,1991MNRAS.250..726T,Sokoloff1998}. Faraday rotation may originate inside the radio emitting region, if enough thermal gas is mixed with the synchrotron radiating plasma (internal Faraday dispersion). Alternatively, it could be of external origin if magnetized thermal gas is present along the line-of-sight (external Faraday dispersion). 

The VLA L-, S-, and C-band data allow us to carry out a detailed wideband polarization study of the compact and diffuse radio sources in MACS\,J0717.5$+$3745. We used two methods to infer the Faraday distribution: Rotation Measure synthesis (RM-synthesis) and QU-fitting.

 \begin{figure*}[!thbp]
    \centering
    \includegraphics[width=0.42\textwidth]{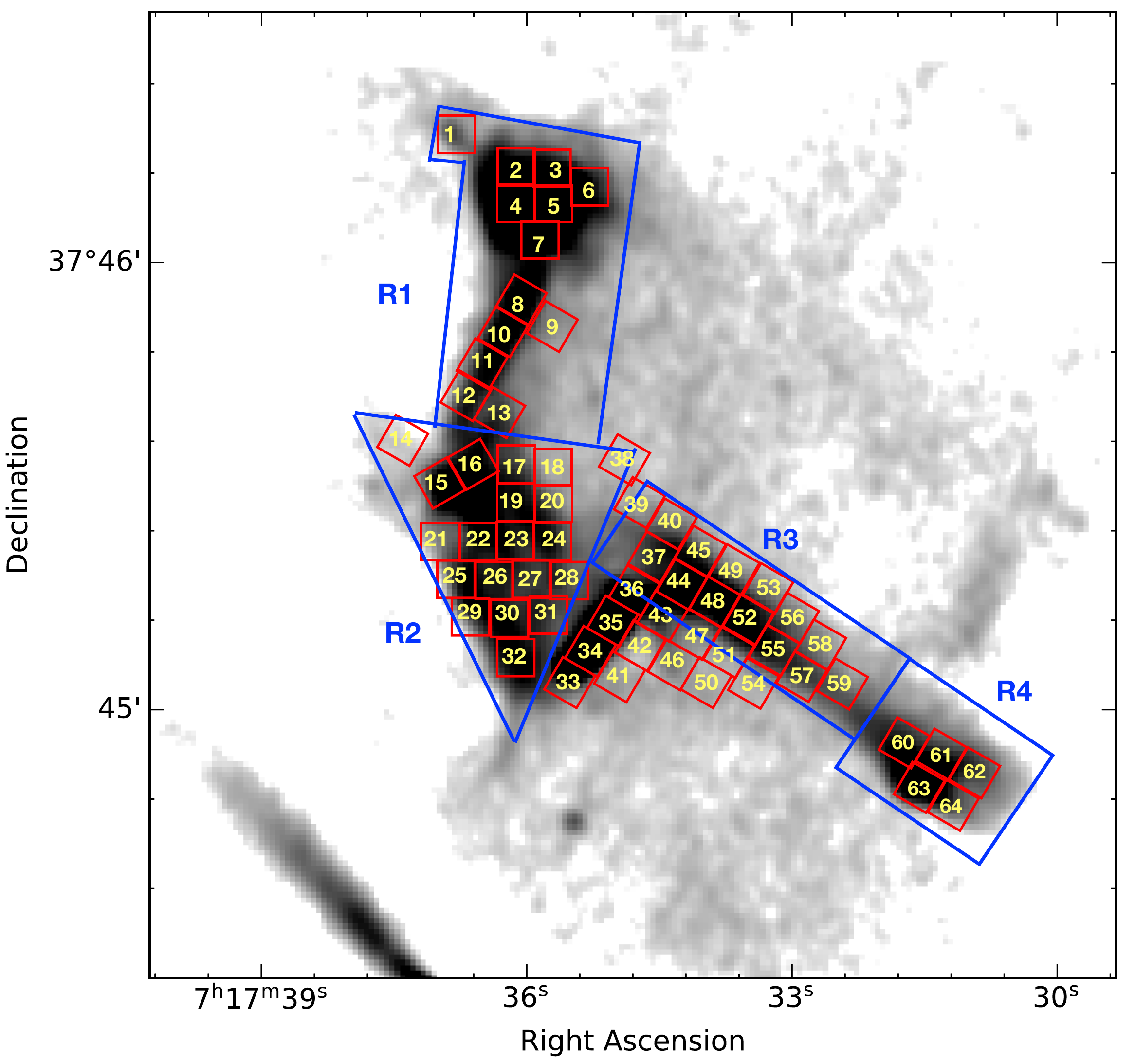}
    \includegraphics[width=0.54\textwidth]{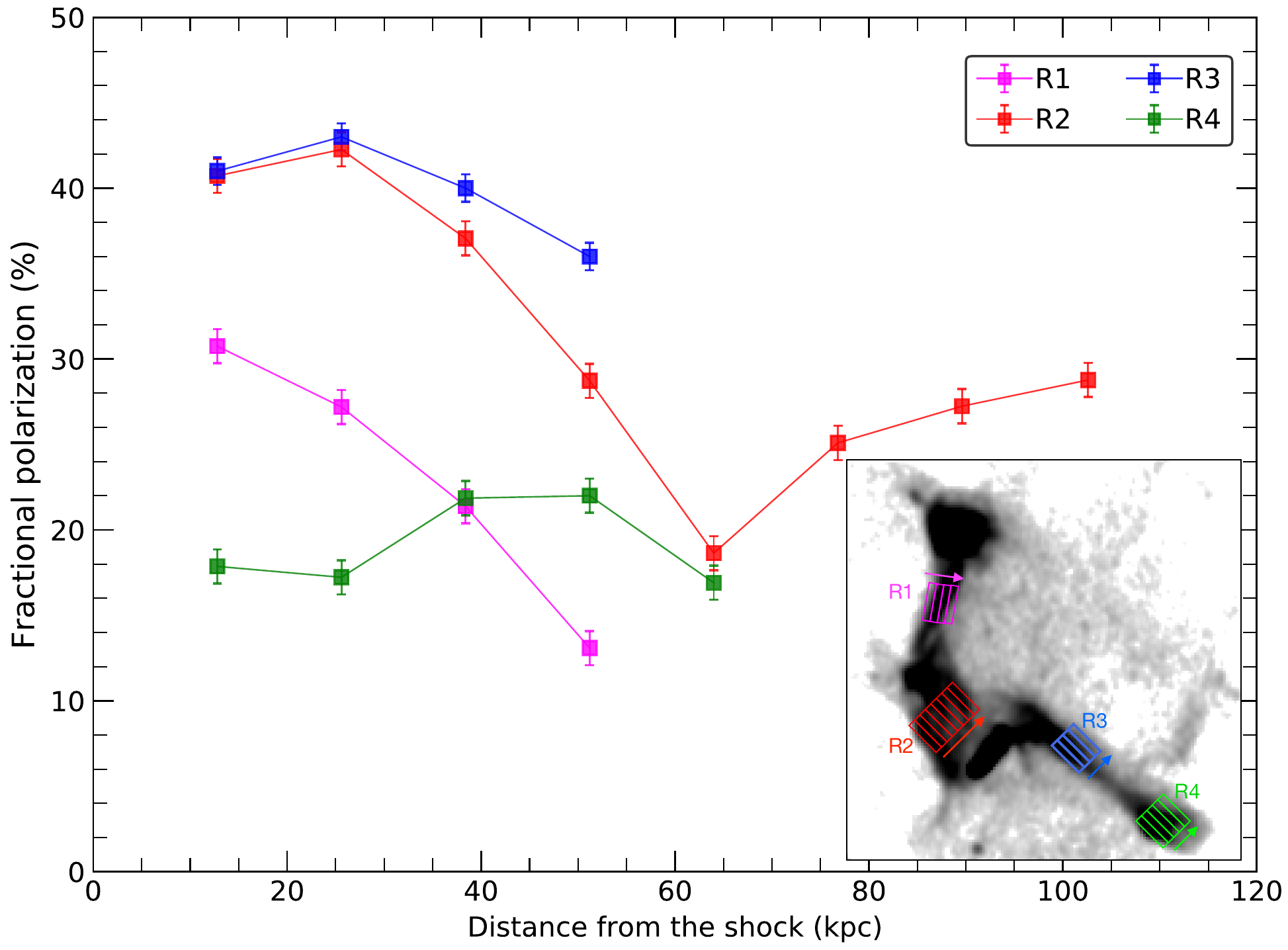}
   \vspace{-0.2cm}
    \caption{\textit{Left}: VLA L-band image, blue regions show the relic subregions where the average fractional polarization was extracted. The fractional polarization at R3 in Table\,\ref{Tabel:Tabel2} is obtained by excluding the contribution from the NAT. Red boxes where flux densities and Faraday dispersion functions were extracted for QU-fitting and RM-synthesis, respectively. Each red box has a width of $5\arcsec$, corresponding to a physical size of 32\,kpc. \textit{Right}: The fractional polarization profiles across the relic extracted from the rectangular regions, depicted in the inset, of width $2\arcsec$. There is a hint that the fractional polarization decreases in the downstream regions (i.e., shown with arrows).}
      \label{regions}
  \end{figure*}


\subsection{RM-synthesis}
\label{sec:RMsynthesis}

 %
 The RM-synthesis technique, developed by \citet{Brentjens2005}, is based on the theoretical description of \cite{Burn1966}. 

 %
 The intensity of linearly polarized emission and its polarization angle $\psi$ can be expressed as a complex number
 \begin{equation}
   P = I \, p_{0} \, e^{2i \psi},
 \end{equation}
 where $I$ is the total intensity of the source and $p_{0}$ is the fraction of polarized emission. Following \citet{Burn1966}, the wavelength dependent polarization, $P(\lambda^2)$, can be written as a Fourier transform 
 \begin{equation}
   P(\lambda^{2})
   =
   \int_{-\infty}^{\infty} 
   F(\phi) \,
   e^{2i\phi\,\lambda^{2}} \,
   {\rm d}\phi,
   \label{eq:P2}
 \end{equation} 
 where $\phi$ is the Faraday depth which here became an independent variable, forming the Faraday space.  $F(\phi)$ is known as the Faraday dispersion function (FDF) and describes the amount of polarized emission originating from a certain Faraday depth. $F(\phi)$ can be measured as
 \begin{equation}
   F(\phi)
   =
   \frac{1}{\pi}
   \int_{-\infty}^{\infty} P(\lambda^{2}) \,\rm e^{-2i\phi\,\lambda^{2}}\,d\lambda^{2}.
   \label{eq:P3}
 \end{equation} 
 RM-synthesis calculates $F(\phi)$ by the Fourier transformation of the observed polarization as a function of wavelength-squared. The Rotation Measure Spread Function (RMSF), analogous to the synthesized image beam, describes the instrumental response to the polarized signal in Faraday space. We refer to \citet{Brentjens2005} for details of this technique. 
 
 The RMSF is determined by the total coverage in $\lambda^{2}$-space of the observations. Since the finite frequency band produces a broad RMSF with sidelobes, deconvolution is advantageous. We used the deconvolution algorithm ${\tt RM\,CLEAN}$ \citep{Heald2009} for this purpose.

 %
 RM-synthesis was carried out using the {\tt pyrmsynth}\footnote{\url{https://github.com/mrbell/pyrmsynth}} code. We performed RM-Synthesis on the Stokes $Q$ and $U$ cubes at two different resolutions, namely, $4\arcsec$ and $12.5\arcsec$. The $4\arcsec$ resolution cube was used to study the relic region while the low resolution $12.5\arcsec$ to study the low surface brightness polarized emission features that are not detected at high resolution. The RM-synthesis cube synthesized a range of Faraday depths from $\rm -800\,rad\,m^{-2}$ to $\rm +800\,rad\,m^{-2}$, with a bin size of $\rm 2\,rad\,m^{-2}$. We used the entire L-, S-, and C-band data. These data give a sensitivity to the polarized emission up to a resolution in Faraday depth ($\delta{\phi}$) equal to
 \begin{equation}
  \delta{\phi}\,  
  \approx 
  \frac{2 \sqrt{3}}{\Delta{\lambda}^{2}} \, \, = 39 \,\rm rad \, \rm m^{-2},
 \end{equation}
 where, $\Delta{\lambda}^{2}$ = $\lambda^{2}_{\rm max} - \lambda^{2}_{\rm min}$. The high Faraday-space resolution may allow us to separate multiple, narrowly-spaced, Faraday-space components. We also run {\tt pyrmsynth} individually only on the L-, S-, and C-band data. The resulting polarization intensity images for L-, S-, and C-band, overlaid with the total intensity contours are shown in Fig.\,\ref{fig3b}. 

   \begin{figure*}[!thbp]
   \centering
   \includegraphics[width=1.0\textwidth]{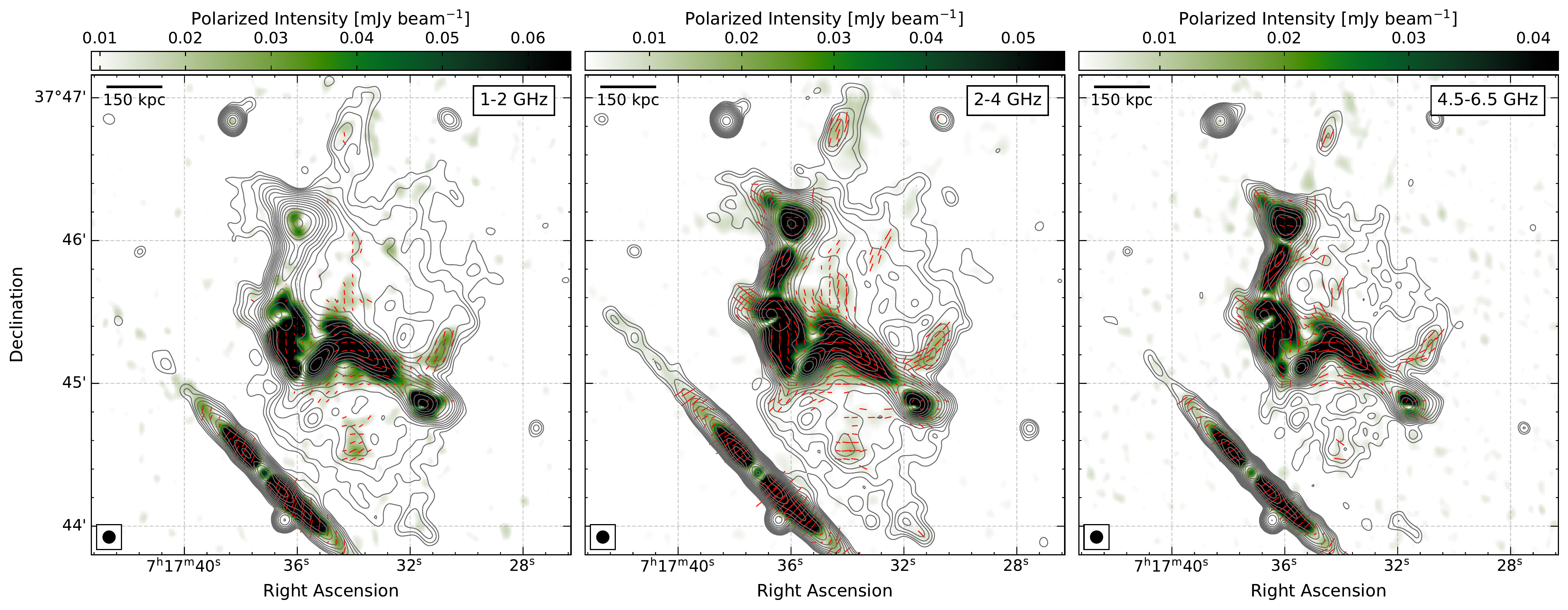}
   \vspace{-0.6cm}
   \caption{ 
     Polarization intensity maps ($5\arcsec$ resolution) at VLA L-, S-, and C-band after performing RM-synthesis. Red lines represent the magnetic field vectors. Their orientation represents the projected B-field corrected for Faraday rotation and contribution from the Galactic foreground. The vector lengths are proportional to the polarization percentage and their lengths are corrected for Ricean bias. No vectors were drawn for pixels below $5\sigma$ in the polarized intensity image. The distinct filaments, namely F1 and F2 (for labeling see Fig.\,\ref{IPX}), and some regions embedded  in the halo emission are polarized between 10-38\% between 1 and 6.5 GHz. At all the observed frequencies, the B-field across the relic and other features is highly-ordered. Contour levels are drawn at $\sqrt{[1,2,4,8,\dots]}\,\times\,5\,\sigma_{{\rm{ rms}}}$ and are from the VLA L-band, S-band and C-band Stokes $I$ images at 1.5\,GHz, 3\,GHz and 5.5\,GHz, respectively. The image properties are given in Table\,\ref{Tabel:imaging}, IM2, IM6, and IM10. The beam sizes are indicated in the bottom left corner of the each image.
     }
   \label{fig3b}
 \end{figure*} 

 %
 The Faraday distribution of the relic in  MACS\,J0717.5$+$3745 has not been studied in literature in detail. \cite{Bonafede2009a} performed a simple linear fit to the polarization angle as a function of $\lambda^{2}$ and found poor agreement between the data and the linear ansatz. In the left panel of Fig.\,\ref{rm_map}, we show the high-resolution Faraday depth map of the relic. The map represents, for each pixel (sky coordinate), the Faraday depth $\phi_{\rm max}$ at which the FDF has its maximum. At the position of MACS\,J0717.5$+$3745, the average Galactic RM contribution is $+16\,\rm rad\,m^{-2}$ \citep{Oppermann2012}. This value is also consistent with the RM we observe for the southern foreground galaxy, ${\rm{\phi}_{\rm FRI}} = +16\pm0.1\,\rm rad\,m^{-2}$. 

 For the relic, the peak Faraday depth ($\phi_{\rm max}$) values vary across the relic between $-30$ to $+40$\,rad\,m$^{-2}$. The peak Faraday depth distribution tends to be patchy with patch sizes of about 10-50\,kpc. For the southern part of the relic, the observed peak Faraday depth ranges mainly from $+7$ to $+25$\,rad\,m$^{-2}$. Stronger variations in the peak Faraday depth are visible for the northern part of the relic, in particular the R1 region. 

 To further investigate the Faraday distribution in the relic, we use 64 square-shaped regions with an edge length of $5\arcsec$ covering the entire relic. These regions are shown in the left panel of Fig.\,\ref{regions}. Each region defines a `box' in the Faraday cube, when taking the Faraday depth axis into account. For each box, we obtain a FDF, see Fig.\,\ref{QURM} for examples. We find that the FDF of most of the boxes is dominated by a pronounced single component, except for a few boxes, for example boxes 4, 5, and 34.

 %
 For the southern part of the relic (boxes 33-64), the peak Faraday depth in the boxes is well defined and relatively uniform; for example see panel (a) of Fig.\,\ref{QURM}. The analysis confirms that the southern part of the relic shows a peak Faraday depth very close to the Galactic foreground, implying very little Faraday rotating material intervening the line-of-sight to the emission region in the cluster. 

 For the northern part of the relic (boxes 1-32), in particular for the R1 region, the analysis reveals strong peak Faraday depth variations, with no particular coherent structure. As seen in the left panel of Fig.\,\ref{QURM}, the Faraday dispersion functions extracted across the northern part of the relic tend to be broader and less symmetric than those extracted from the southern part. The broader FDFs hint to the presence of emission at different Faraday depths. There are two basic scenarios leading to a complex FDF: either the emission is extended along the line-of-sight embedded in a magnetoionic medium or there is a magnetoionic medium with a complex Faraday depth distribution in front of the emission. Of course, there can be mixture of both.  
 
\begin{figure*}[!thbp]
    \centering
    \includegraphics[width=1.0\textwidth]{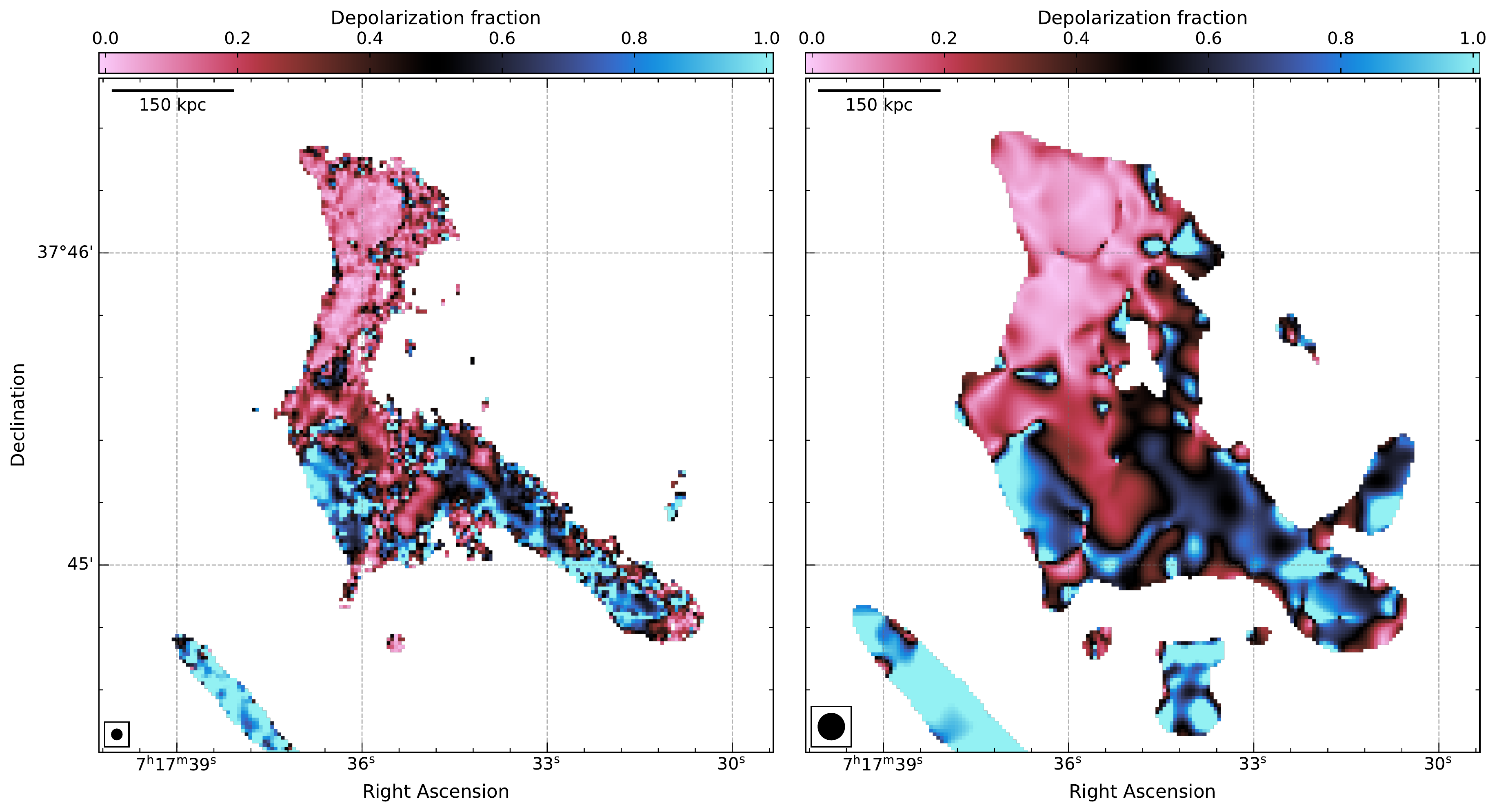}
   \vspace{-0.2cm}
    \caption{Depolarization maps of the relic  between 1.5 and 5.5 GHz at $2\arcsec$ (left) and $5\arcsec$ (right) angular resolution. The depolarization fraction is defined as ${\rm DP}=p_{\rm 1.5\,GHz}/p_{\rm 5.5\,GHz}$. ${\rm DP}=0$ implies full depolarization while ${\rm DP}=1$ means no depolarization. These maps demonstrate that the northern part of the relic is strongly depolarized, in particular the R1 region of the relic.}
      \label{DP}
  \end{figure*}    
 
\begin{figure*}[!thbp]
   \centering
   \includegraphics[width=1.0\textwidth]{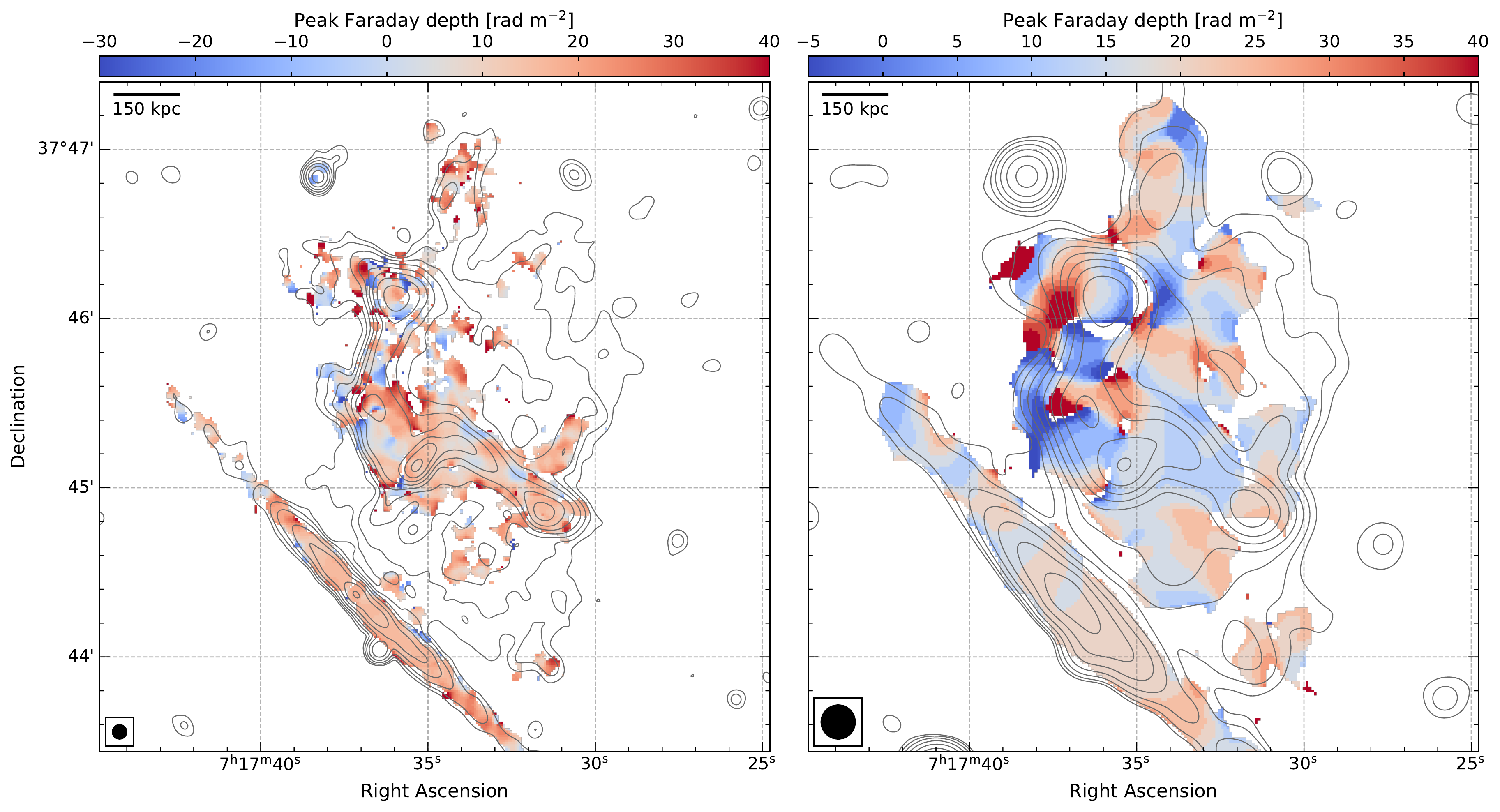}
      \vspace{-0.2cm}
   \caption{Faraday depth ($\phi_{\rm max}$) maps of the relic in MACS\,J0717.5$+$3745 measured over 1.0-6.5\,GHz using RM-synthesis technique.\textit{Left}: High-resolution (4\arcsec) Faraday map. The $\phi_{\rm max}$ distribution across the relic, in particular in the northern part is patchy with coherence lengths of 10-50\,kpc.  \textit{Right}: Low-resolution (12.5\arcsec) Faraday depth map. The measured $\phi_{\rm max}$ values in the polarized halo region are similar to the R3 region of the relic, indicating very little Faraday Rotation intervening material. This suggests that these regions are located on the near side of the cluster.}
   \label{rm_map}
\end{figure*}  


\subsection{QU-fitting}
\label{qufit} 

 %
 
 An alternative approach to interpret broadband polarimetric data is to approximate the observed quantities $Q(\lambda^2)$ and $U(\lambda^2)$, in the following dubbed as `QU-spectra', over the broad wavelength range using an analytic model with a small number of free parameters. We refer to this method as `QU-fitting'. This technique is particularly powerful when the FDF is rather simple and can be described by an analytic model which can be guessed, for example, from the geometry of the source, the most likely morphology of the magnetic field, or knowledge about the medium intervening along the line-of-sight to the source \citep[e.g.,][]{Farnsworth2011,Ozawa2015,Anderson2015,Anderson2016,Pasetto2018,Anderson2018}. 
 
  \begin{figure*}[!thbp]
   \centering
    \includegraphics[width=0.45\textwidth]{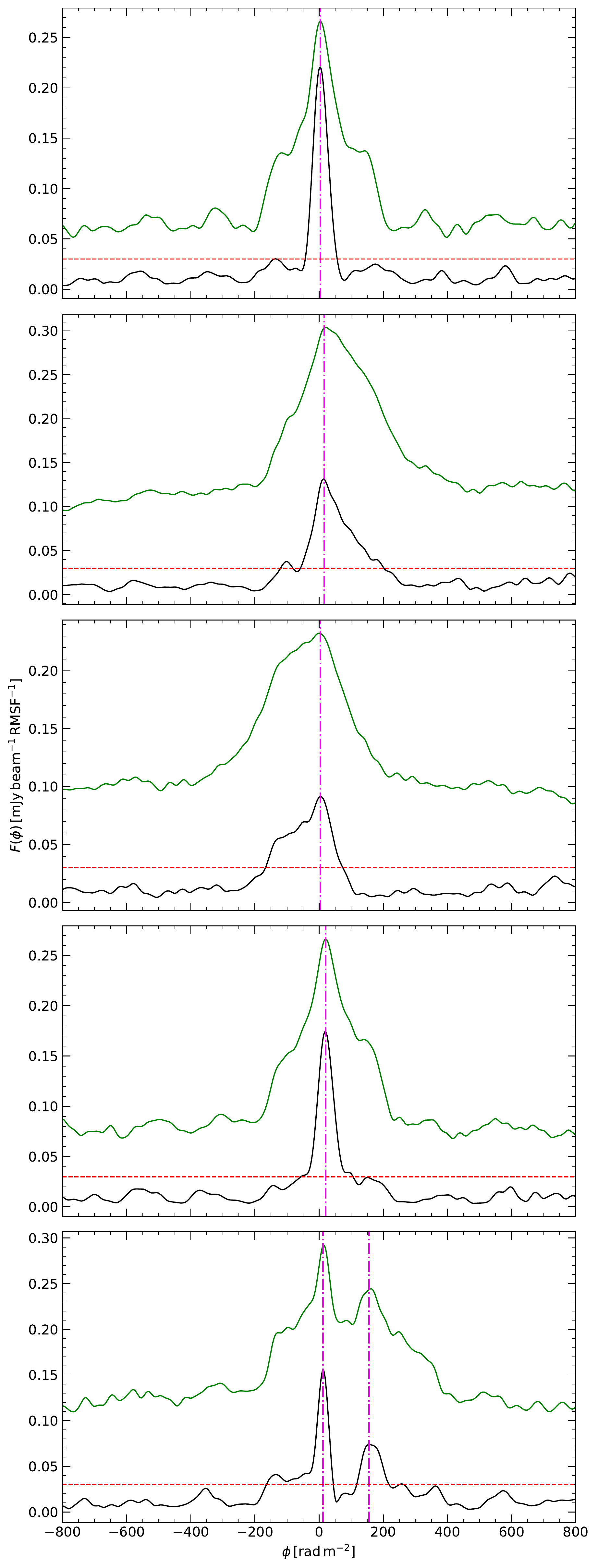}
   \includegraphics[width=0.438\textwidth]{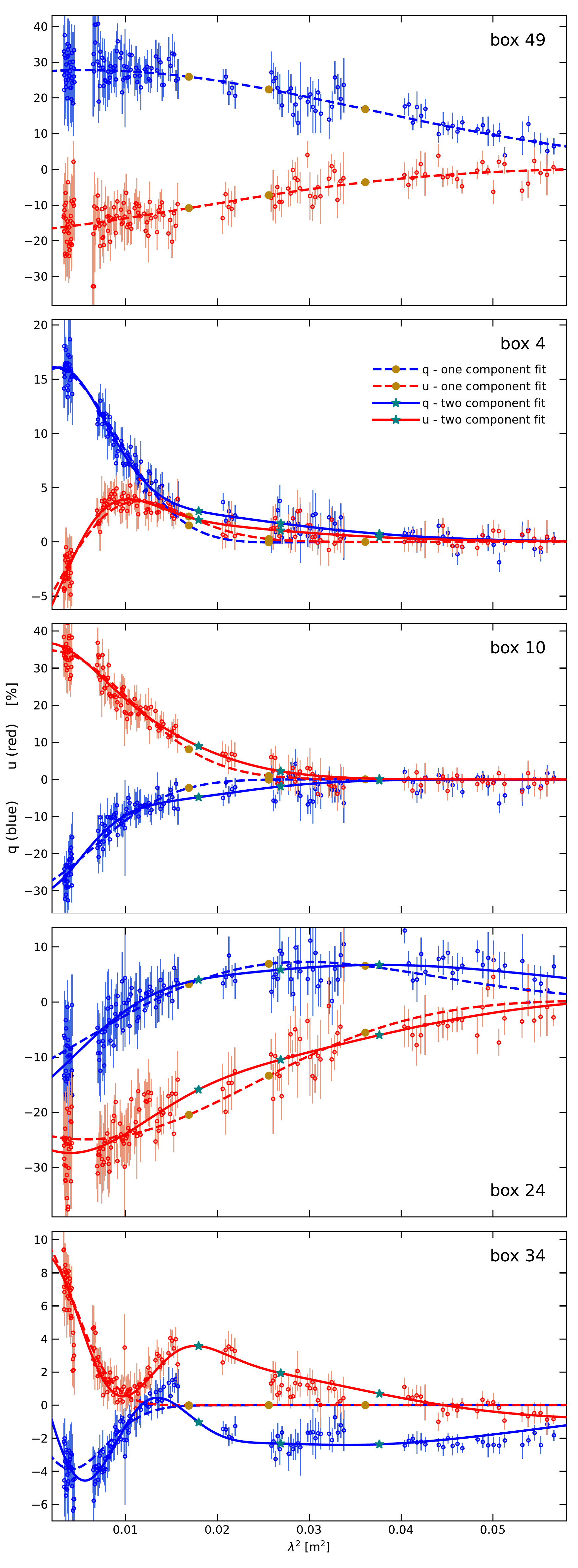}
   \vspace{-0.2cm}
   \caption{
     Comparison between RM-synthesis and QU-fitting for boxes 49, 4, 10, 24, and 34. {\it Left}: Cleaned (black) and dirty (green) Faraday dispersion functions (FDF) obtained using RM-synthesis. The red dash line is drawn at the $8\sigma_{\rm QU}$ level. The magenta lines indicate the peak positions ($\phi_{\rm max}$) of the FDFs. {\it Right}: Corresponding QU-fitting spectra, the fractional Stokes $q$ (blue) and $q$ (red) with the dots showing the QU-spectra measured in the boxes and the dashed and solid lines are the one and two component fits, respectively. We also mark the one and two component fits with brown circles and dark blue asterisks to clearly indicate the significance of these markers used in subsequent images.  For boxes 4, 10, 24, and 34 the QU-spectra are better fitted with two components. Correspondingly, RM-synthesis shows broader FDF for these boxes. For simple regions, exemplary box 49 is shown, both RM-synthesis and QU-fitting seems consistent with a single Faraday component.
     }
   \label{QURM}
 \end{figure*}

 %
 The polarized signal from the boxes introduced in the previous section is well suited for this approach: According to the generally accepted scenario for the origin of radio relics, electrons are accelerated at cluster-sized merger shocks and radiate in a comparably thin layer downstream of the shock. In the sky area covered by one box, we expect to see only a small part of the merger shock front. The width of each box corresponds to a physical size of 32\,kpc. If the line-of-sight to a box intersects the shock front only once and the front is inclined to the line-of-sight, the emission received from the box area originates from a volume which is rather small compared to the cluster dimensions. Only if the shock front is seen very close to edge-on, the volume will be significantly extended along the line-of-sight. Based on this scenario, we expect that the Faraday distribution in each box is reasonably well described by a single component in Faraday space. The ICM and the intergalactic medium (IGM), intersecting the line-of-sight to the emission volume, determine the position of the component in Faraday space and its width, the latter manifesting itself by the depolarization of the emission.

 \begin{figure*}[!thbp]
   \centering
   \includegraphics[width=1.0\textwidth]{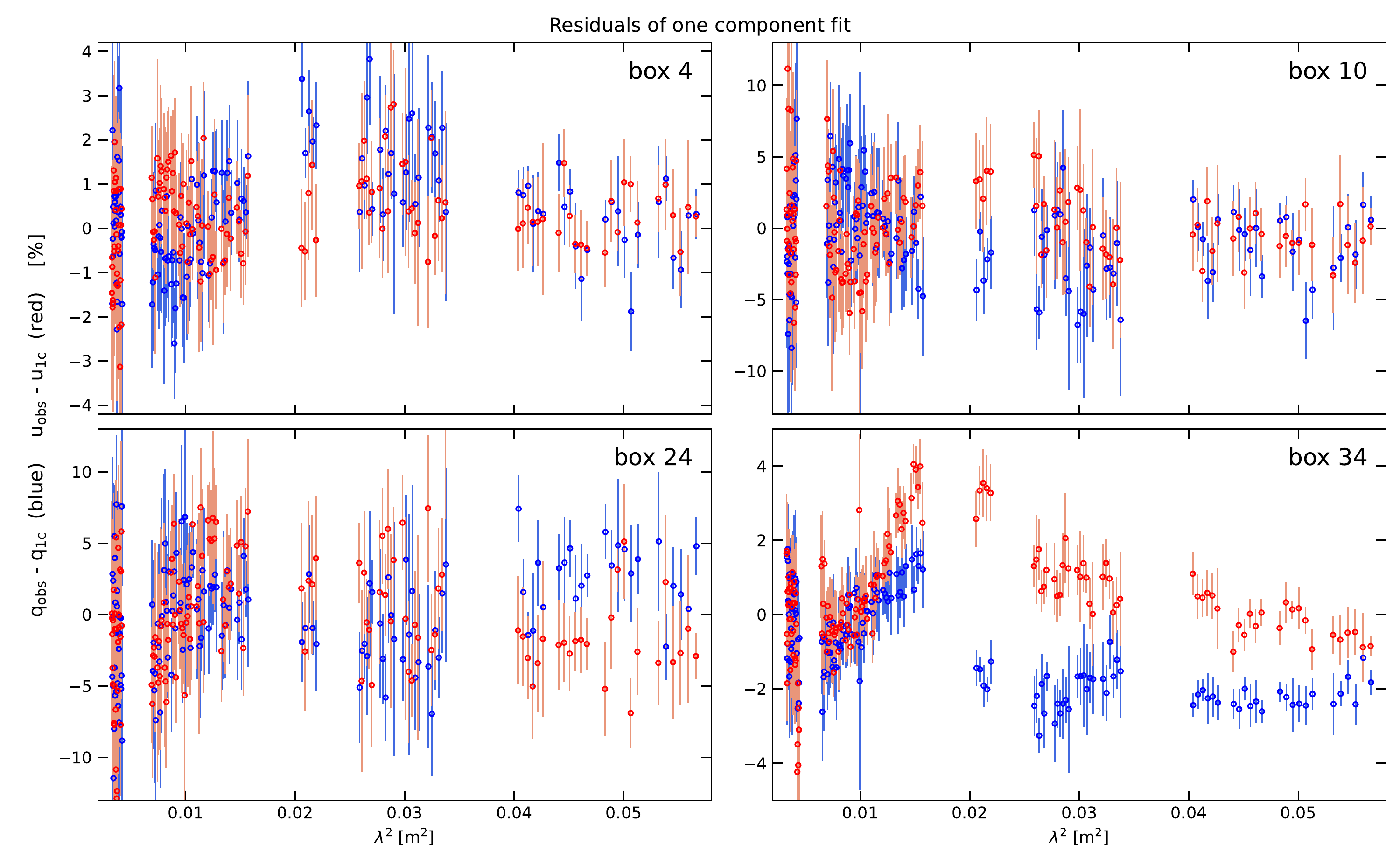}
   \caption{
   Residual QU-spectra for the boxes 4, 10, 24, and 34 obtained after subtracting the best one component fit. The blue and red circles show the observed fractional $q$ minus the best one component fit ($q_{\rm 1c}$) and the observed fractional $u$ minus the best one component fit ($u_{\rm 1c}$), respectively. The systematic differences between the QU-spectra measured in the boxes and the fits are evident.}
   \label{fig::residuals}
 \end{figure*} 

 %
 We assume, that the complex polarization $P$ of a single Gaussian component in Faraday space can be described by the expression 
 \begin{equation}
   P_{\rm 1c}(\lambda^2)
   = 
   I(\lambda^2) \, p \, 
   e^{-2\, \sigma_{\phi}^2 \, \lambda^{4}}
   e^{2i\, (\psi+\phi_{\rm c}\lambda^{2})},
   \label{eq::singleRM}
 \end{equation}
 where $I(\lambda^2)$ is the total intensity as a function of $\lambda^2$, $p$ the intrinsic polarization fraction, $\phi_{\rm c}$ the average Faraday depth of the emission (i.e., the position of the center of the Gaussian component) and $\sigma_\phi$ the Faraday dispersion (i.e., the width of the Gaussian component). In a more general model, also, the intrinsic polarization fraction and intrinsic polarization angle $\psi$ could be wavelength dependent. We emphasize, that the single Gaussian component ansatz is based on the scenario that we observe in each box only a small emission volume with a screen in front of it showing a Gaussian Faraday depth distribution. A more complex distribution of the emission or a non-Gaussian Faraday depth distribution of the screen would require a more complex description. Based on the motivation detailed above, we expect the model to provide a good approximation to the data. 
 
 %
 To eliminate spectral effects, we use the fractional properties $q = \text{Re}(P)/I$ and $u = \text{Im}(P)/I$. The `one Gaussian component' model functions become: 
 \begin{subequations}
 \begin{align}
   q_{\rm 1c}(\lambda^2) = 
   p \, 
   \cos({2\psi+ 2\,\phi_{\rm c}\,\lambda^{2})} \,
   e^{-2 \, \sigma_\phi^{2} \, \lambda^{4}}, 
   \\
   u_{\rm 1c}(\lambda^2) = 
   p \, 
   \sin({2\psi+ 2\,\phi_{\rm c}\,\lambda^{2})} \,
   e^{-2 \, \sigma_\phi^{2} \, \lambda^{4}}
   \:
   .
   \label{eq::qufract}
 \end{align}
 \end{subequations} 
 We approximate the QU-spectra in three steps: (i) First, we scan the four-dimensional parameter space ($p$, $\psi$, $\phi_{\rm c}$, $\sigma_\phi$) and compute the difference between model and data $\chi^2$ for each set of parameters.  The difference is computed according to
 \begin{equation}
     \chi^2
     =
     \sum_{i,x}
     \frac{(x_{\rm meas}(\lambda_i^2) - x_{\rm 1c}(\lambda_i^2))^2 }{ \sigma_{x,i}^2}
     \, ,
 \end{equation}
 where $\lambda_i$ denotes the central wavelength of the spectral channel $i$, $x$ the two fractional properties $q$ and $u$, and $\sigma_{x,i}$ the  uncertainty of the measurement for $q_{\rm meas}(\lambda_i^2)$ and $u_{\rm meas}(\lambda_i^2)$ in the boxes.
 (ii) Starting from the parameter set with the lowest $\chi^2$, we run a Levenberg-Marquardt parameter optimization. (iii) Starting from the optimized parameter set, we finally run a Markov chain Monte Carlo (MCMC).

 \begin{figure*}[!thbp]
   \centering
   \includegraphics[width=1.0\textwidth]{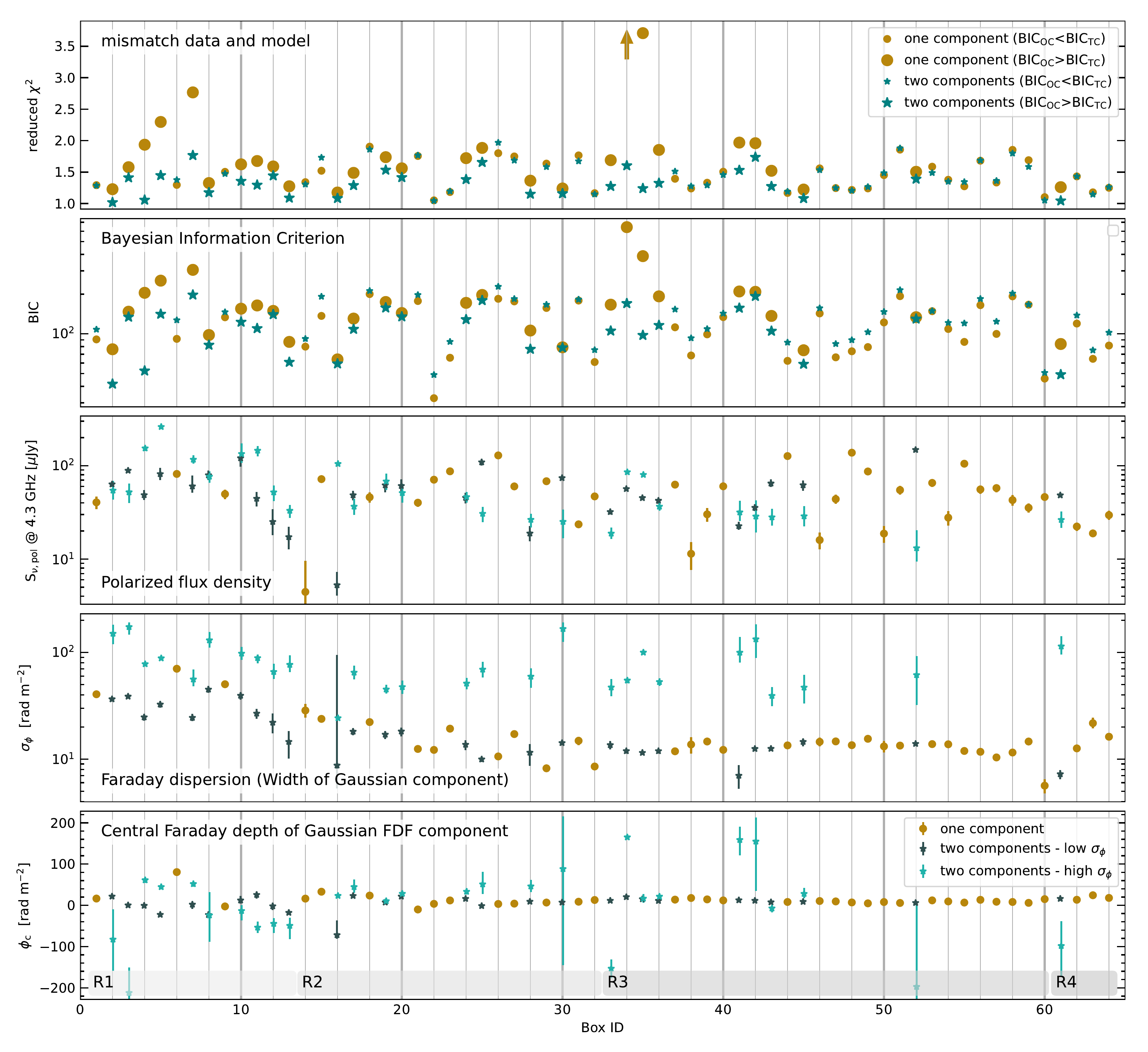}
   \vspace{-0.8cm}
   \caption{
   Results from QU-fitting. The best-fit parameters obtained by fitting a single and two independent one-component models are shown. The one component model fits are indicated with brown filled circles and the two component models with dark blue and cyan asterisks. The resulting reduced $\chi^{2}$ for each box is shown in the top panel and the BIC in the second one. If the BIC of the two component fit lower than the one component fit, the former is adopted for further analysis. For theses boxes the best fit parameters are shown with cyan and dark blue asterisks where the cyan color is assigned to the component with the higher Faraday dispersion. The third, forth and fifth panel shows the intrinsic polarized luminosity, $S_{\nu,\,\rm pol}$, the Faraday dispersion and the position for each of the Gaussian components. Finally, the Boxes used for extracting the Stokes $IQU$ values are shown Fig\,\ref{regions}. The plot shows that many regions in R1 and R2 are fitted better using two components.}
   \label{fig::QUFitParameter}
 \end{figure*} 

  \begin{figure}[!t]
   \centering
   \includegraphics[width=0.48\textwidth]{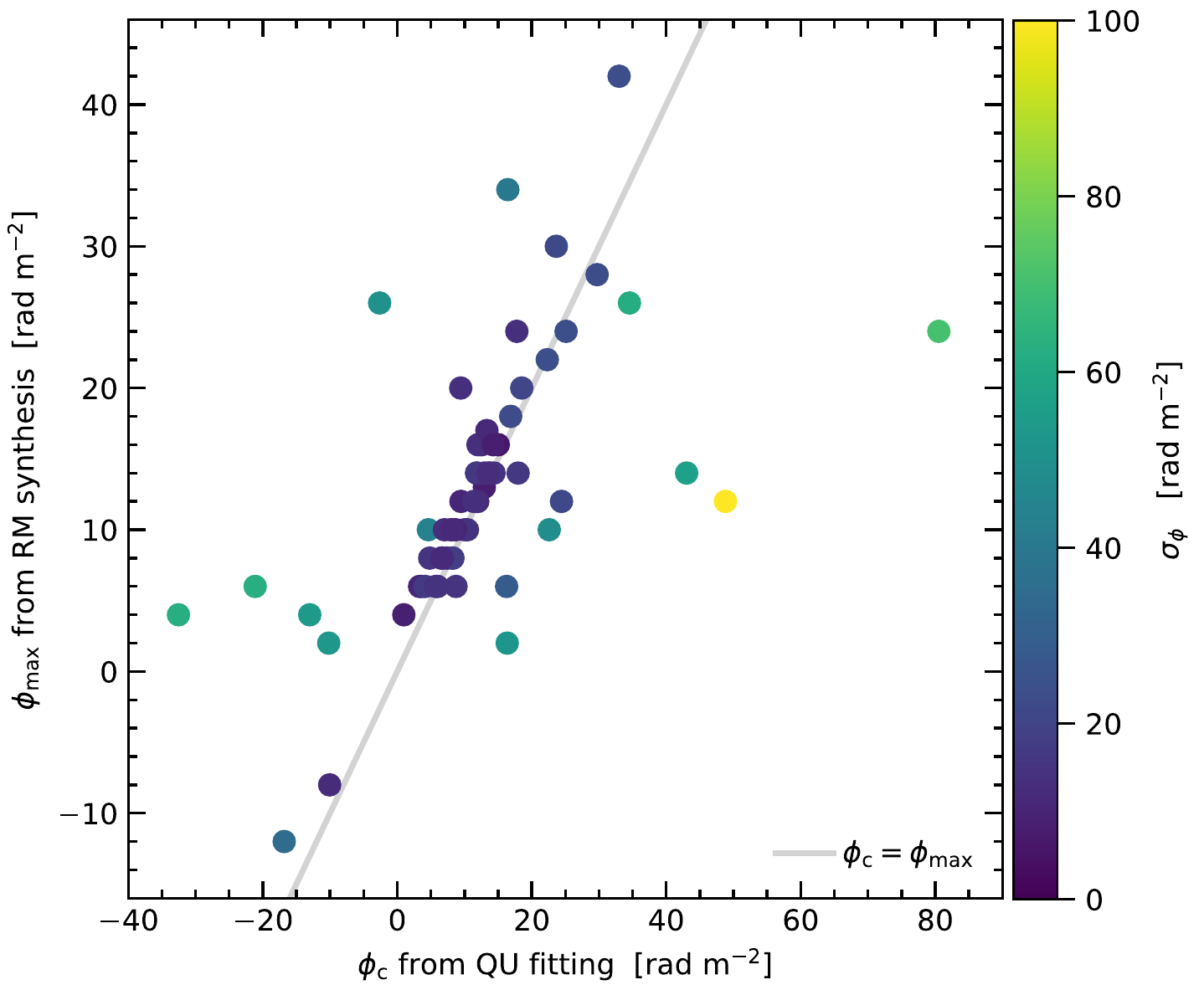}
      \vspace{-0.2cm}
   \caption{Central Faraday depth, $\phi_{c}$, position of the single component from QU-fitting (see Fig.~\ref{fig::QUFitParameter} brown markers, note we use here the one component fit for all boxes) versus peak position, $\phi_{\rm max}$ in the RM-synthesis spectra. The color indicates the Faraday depth width of the QU-fitting component. Evidently, QU-fitting and RM-synthesis gives similar results in case of a low Faraday dispersion. Larger differences between QU-fitting and RM-synthesis correspond to higher Faraday dispersions.}
   \label{fig::RMSynth_vs_QUfit}
 \end{figure}  
 
 %
 In the right panel of Fig.\,\ref{QURM}, we show the QU-fitting results for boxes 49, 4, 10, 24, and 34. The top panel, box 49, shows the QU-spectra of a region from the southern part of the relic, namely the R3 region. As evident, the one-component model provides a very good fit. For the other boxes, we find that there are systematic differences between the data and model, indicating that the actual Faraday distribution is more complex than the one Gaussian component ansatz. 
 
 In Fig.\,\ref{fig::residuals}, we show the residuals after subtracting the best one-component fit for the boxes 4, 10, 24, and 34. The systematic differences between the measured QU-spectra and the fits are evident. As an ad-hoc model, we assume that the actual Faraday distribution can be better fitted by two Gaussian components in Faraday space. We, therefore, approximate the QU-spectra with the sum of two independent one component models. To determine the model parameters, we follow the steps as given above but using eight independent parameters.
 
 Fig.\,\ref{fig::QUFitParameter} (upper panel) shows the resulting reduced $\chi^2$ for each box fitted with one and with two Gaussian components. For some regions, the one component model (brown circles) already results in a reduced $\chi^2$ close to one, suggesting that the data are reasonably well approximated. We note that the reduced $\chi^2 =1$ is only achieved for a perfect match between data and model if the uncertainties of the data reflect the uncertainty of independent measurements. It is beyond the scope of this work to study in detail if the Stokes $IQU$ data are indeed fully independent. Therefore, even a perfectly matched model may show a reduced $\chi^2$ slightly deviating from one. Some boxes show a reduced $\chi^2$ much larger than one, for example, boxes 4, 7, and 35, indicating a poor fit. As expected, the two Gaussian component model (dark blue asterisks in Fig.\,\ref{fig::QUFitParameter}) better matches the data, generally resulting in a lower reduced $\chi^2$. However, we adopt only the two component model with the significantly larger number of free parameters if the fit is substantially better. We therefore employ the Bayesian information criterion (BIC):  
 \begin{equation}
     {\rm BIC}
     =
     N \, \ln ( \chi^2 / N )
     +
     \ln(N) / N_{\rm var},
     \label{eq::BIC}
 \end{equation}
 where $N$ denotes the number of independent data points and $N_{\rm var}$ the number of free parameters in the fit. Fig.\,\ref{fig::QUFitParameter} second row shows the BIC for the one ($\rm BIC_{OC}$) and the two Gaussian component ($\rm BIC_{TC}$) model for all boxes. For about half of the boxes the BIC of the two component model is lower (large symbols in Fig.\,\ref{fig::QUFitParameter}) indicating a substantially better fit. We note that for some boxes, for example boxes 31 and 53, the reduced $\chi^2$ is lower for the two components, the BIC, in contrast is higher for the two components. Since the decision is based on the BIC, we adopt in these cases the one component model for further analysis, underlining that the BIC requires a significantly better fit for adopting two components for the further analysis.

 In Fig.\,\ref{fig::QUFitParameter}, we also show the model polarized flux density at 4\,GHz (third row) for all boxes, the Faraday dispersion obtained for each box (fourth row), and the central Faraday depth of the Gaussian components (fifth row). Based on the BIC, we adopt the two Gaussian component model only if it fits the data substantially better.

\subsection{RM-synthesis versus QU-fitting}
 
 The QU-fitting has revealed that for most of the boxes a single Gaussian component provides a reasonable approximation, even if for about half of the boxes provide an even better approximation. Since we used two methods, namely QU-fitting and RM-synthesis, to determine the Faraday structure of the emission in the boxes, it is interesting to compare the results of the two methods. Since the one component fit provides a reasonable fit, that is the reduced $\chi^2$ does not differ very much from one, see Fig.\,\ref{fig::QUFitParameter} first panel, we do expect from the RM-synthesis method also a single peak for most of the boxes.
 
 Since the RM-synthesis shows a single peak for most of the boxes, we do expect that the peak in the FDF obtained with RM-synthesis does correspond to the central Faraday depth of a single Gaussian component fit. We note that we use here for simplicity the single component fit for all boxes. The peak Faraday depth from RM-synthesis is read from the uncleaned (dirty) spectra. Fig.~\ref{fig::RMSynth_vs_QUfit} shows a reasonable agreement between the peak $\phi_{\rm max}$ and the central Faraday depth. Apparently, large differences occur only for boxes with a very broad single component, that is, where the Faraday dispersion $\sigma_\phi$ is large and the emission is depolarized at longer wavelengths. For instance, box 4 shows a $\phi_c$ of $43\,\rm rad \, m^{-2}$ and a $\phi_{\rm max}$ of $14\,\rm rad \, m^{-2}$, the Faraday dispersion of a single component fit in the box is $56\,\rm rad \, m^{-2}$. 
 
 Evidently, the Faraday dispersions $\sigma_\phi$ of many single component fits are very small. For instance,  $\sigma_\phi$ in the boxes 47, 48, and 49 is about $18\,\rm rad\,m^2$ in observers frame. RM-synthesis does not allow us to clearly recover such small Faraday dispersions, that is widths of FDFs, even after RM cleaning (see Fig.\,\ref{QURM} box 49). 
 For box 34, from the QU-fitting the two component fit is preferred, showing one component at $\rm 20\,rad\,m^{-2}$ and one at  $\rm 165\,rad\,m^{-2}$. It is interesting to note that the central Faraday depth position of the two components agrees well with the two peaks found in RM-synthesis, see Fig.\,\ref{QURM}.

 We conclude that (i) in situations with one single dominating component and weak depolarization, the peak Faraday depth from RM-synthesis and the central Faraday depth from QU-fitting agree well, (ii) in situations of strong depolarization, the results may differ significantly, and (iii) two clearly separated components in  Faraday space can be recovered by both methods.  However, we would like to emphasize that for RM-synthesis no assumptions are made for FDF, so much more general distributions might be found which allow us to reproduce the QU-spectra. We also note that RM-synthesis is sometimes reported to fail  to find the underlying Faraday distribution for even the simple case of two components \citep{Farnsworth2011,OSullivan2012}.


\section{The Faraday distribution of the relic}
\label{sec::FDFrelic}

 \begin{figure*}[!thbp]
   \centering
   \includegraphics[width=1.00\textwidth]{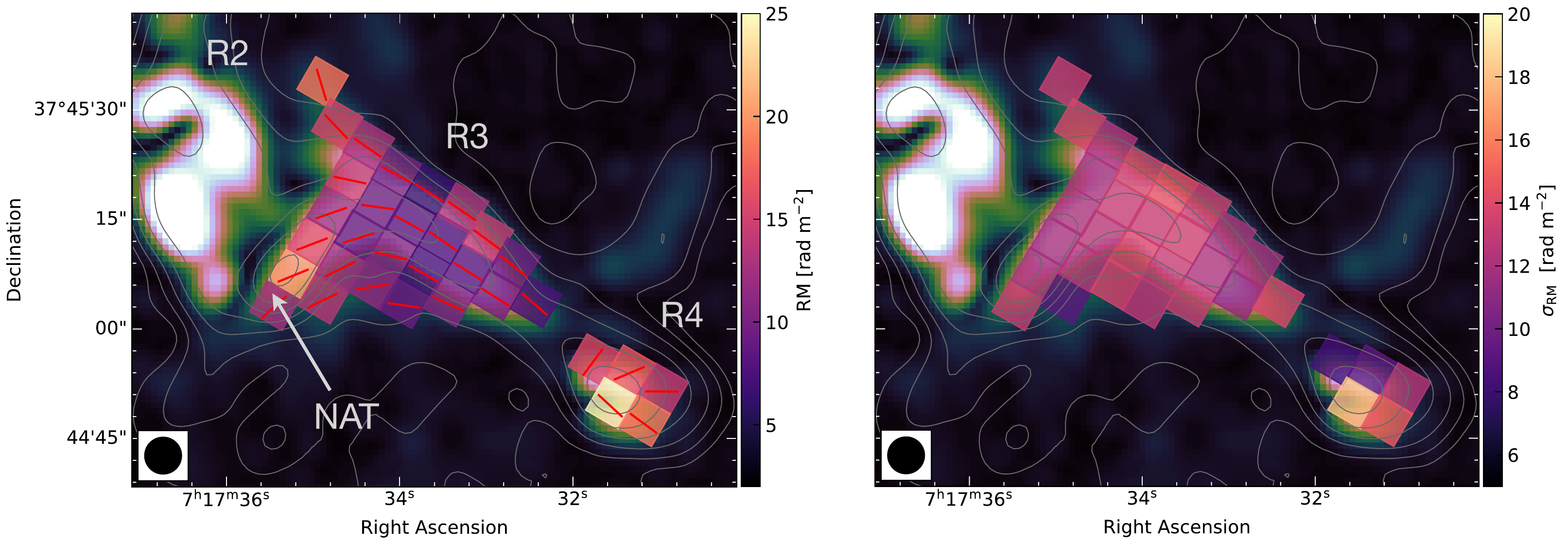}
      \vspace{-0.5cm}
   \caption{$\phi_{c}$ and $\sigma_{\phi}$ values across the R3 and R4 region of the relic obtained with QU-fitting. For almost all boxes the one Gaussian component model is preferred, except for the region where the NAT galaxy is located. We note that for the NAT region only the low dispersion component is shown in the figure, matching the $\phi_{c}$ of the relic. Moreover the left panel (red lines) shows the intrinsic polarization angle. The QU-fitting results are overlaid on the polarization intensity at 5\arcsec and contours of the VLA L-band Stokes $I$ image, see Table\,\ref{Tabel:imaging}, IM6, for image properties. Contour levels are drawn at $[1,2,4,8,\dots]\,\times\,5\,\sigma_{{\rm{ rms}}}$.  From these maps it is evident that $\phi_{c}$ and $\sigma_{\phi}$ are low in this region of the relic, in particular in the R3 region, and not changing significantly from box to box. The central Faraday depth of the Gaussian components are very close to the Galactic foreground RM, indicating very little Faraday rotation due material intervening the line-of-sight to cluster outside of the Milky Way. This suggests that this part of the relic is located at the periphery of the cluster towards the observer. 
     }
   \label{R3_RM}
 \end{figure*} 

 %
 In Fig.\,\ref{R3_RM}, we show the central Faraday depth and Faraday dispersion maps for the southern part of the relic obtained from QU-fitting. For the relic regions R3 and R4, QU-fitting and RM-synthesis (see Fig.\,\ref{rm_map}) consistently reveal a rather uniform distribution of the central Faraday depth, which apparently reflects the Galactic foreground of $+16 \rm\,rad \, m^{-2}$ \citep{Oppermann2012}. Except in the NAT regions, almost all boxes are well fitted with one component according to the BIC (see Fig\,\ref{fig::QUFitParameter}). In this part of the relic, the Faraday dispersion values is in the range $\sim 10 -20 \,~\rm rad\,m^{-2}$ (at the redshift of the observer). The depolarization in these boxes corresponds to a mean Faraday dispersion of about $\sigma_{\phi} \sim 12 \, \rm rad \, m^{-2}$ in the observer's frame. The Faraday dispersion likely reflect intrinsic depolarization in the relic (internal depolarization) or Faraday depth variations caused by the ICM intervening along the line-of-sight to the relic (external depolarization). In both scenarios, the source for depolarization is in the cluster, hence, we have to take into account the redshift effect discussed above (see Eq.\,\ref{eq::RMrest}). The mean Faraday dispersion at the location of the cluster amounts to $\sigma_{\phi, \rm R3+ R4} \sim 29 \, \rm rad \, m^{-2}$. Interestingly, the Sausage relic shows a similar Faraday dispersion \citep{DiGennaro2021}.

 %
 The central Faraday depth and Faraday dispersion maps for the northern part of the relic are shown in Fig.\,\ref{R1_RM}. RM-synthesis and QU-fitting reveal complex Faraday depth and dispersion distributions. The scatter of the central Faraday depth increases from south to north. This is consistent with the pixel-wise Faraday depth peak position analysis as shown in Fig.~\ref{rm_map} obtained from RM-synthesis. Moreover, in this region of the relic, the Faraday dispersion values vary up to $170\,\rm rad\,m^{-2}$. As can be seen from Fig.\,\ref{R1_RM}, lower panels, $\sigma_{\phi}$ systematically increases from south to north, that is from box 32 to box 1 (see also Fig.\,\ref{fig::QUFitParameter}).

 %
 From the parameters of R1, that is all boxes up to box 13, as shown in Fig.~\ref{fig::QUFitParameter}, we compute the average Faraday dispersion ($\sigma_\phi$) of all components. Weighting the parameters according to their uncertainty, we find an average of $47.4\,\rm rad \, m^{-2}$, that is $114\,\rm rad \, m^{-2}$ at the redshift of the cluster. Interestingly, the standard variation of the central Faraday depth is lower, namely $38.7\,\rm rad \, m^{-2}$, that is 93\,$\rm rad \, m^{-2}$ at the redshift of the cluster. The uncertainties of the parameters has again been used as weights. The fact that the average dispersion is a factor of 1.22 lower than the scatter of positions in Faraday space may indicate that the magnetic field along the line-of-sight shows significant fluctuations on scales of the size of the boxes (i.e. 32\, kpc) or smaller.  
 
  \begin{figure*}[!thbp] 
   \centering
   \includegraphics[width=0.99\textwidth]{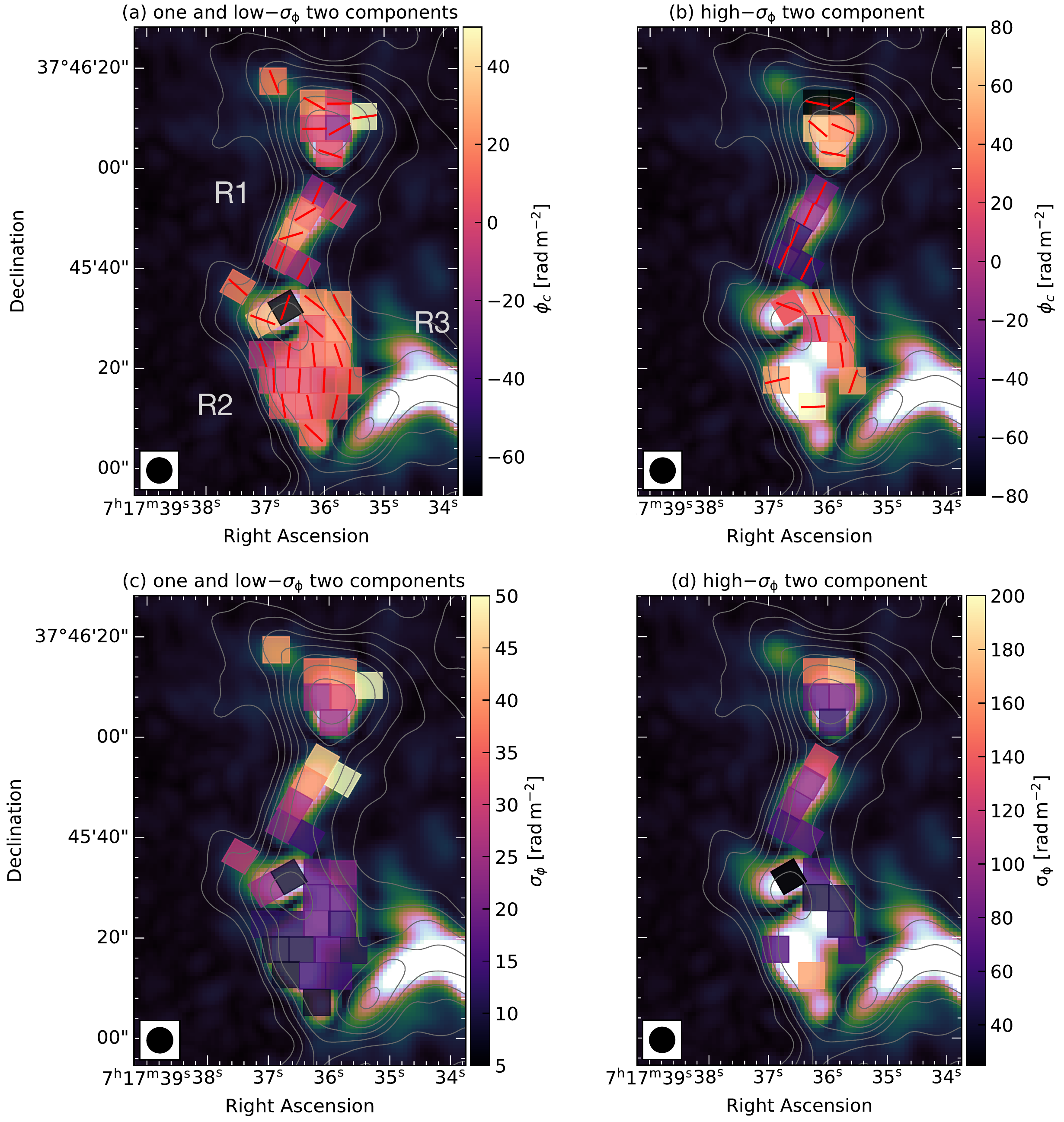}
      \vspace{-0.2cm}
   \caption{Central Faraday depth and Faraday dispersion values for the northern part of the relic obtained with QU-fitting. \textit{Top}: The $\phi_{c}$ values for one component and low Faraday dispersion components (panel a) and the high Faraday dispersion two components (panel b). For a large number of regions, we find the presence of two Faraday components; the one component or the low Faraday dispersion component in case of two components (left panels) and the high Faraday dispersion components (right panels). 
     \textit{Bottom}: The $\sigma_{\phi}$ values for one component and low Faraday dispersion components (panel c) and the high Faraday dispersion two components (panel d). The majority of the polarized emission is at rather high $\sigma_{\phi}$, suggesting that the northern part of the relic is located in or behind the ICM. Contours and image properties are as in Fig.\,\ref{R3_RM}.
     }
   \label{R1_RM}
 \end{figure*}  
 
 %
 In Fig.~\ref{fig::RMsigRM} we show the Faraday dispersion versus central Faraday depth for all boxes and all components. The plot shows again that the scatter of the central Faraday depth increases with increasing Faraday dispersion. The offset central Faraday depth due to the Galactic foreground of $+16 \rm\,rad \, m^{-2}$ is evident. The dashed lines indicate $\sigma_{\phi} = \pm 1.22 \cdot (\phi_{\rm c} - 16 \rm\,rad \, m^{-2})$, where the factor 1.22 is used according to the discussion above. The results in Fig.~\ref{fig::RMsigRM} corroborate that the standard deviation of the central Faraday depth of the components obtained for the boxes is lower than the Faraday dispersion obtained from the depolarization. 
 
 %
 The correlation between the scatter of the central Faraday depth and the Faraday dispersion is consistent with and gives evidence for the scenario that the emission with a higher Faraday depth is located deeper in the cluster or has a larger amount of ICM in front of it which causes the Faraday rotation. Interestingly, the parameters obtained for both components of the two component fit follow the correlation. With the scenario in mind that a slab of ICM in front of the emission causes both, the central Faraday depth and the Faraday dispersion, this could be interpreted as two different patches of ICM in front of a single emission found in one box, alternatively, there could be two different emission regions along the line-of-sight within one box.

 The high-resolution total power images (left and middle panels of Fig.\,\ref{filaments}) and spectral tomography analysis (right panel of Fig.\,\ref{filaments}) have revealed that the northern part of the relic is composed of multiple filaments \citep{vanWeeren2017b,Rajpurohit2021a}. These filaments are denoted with red arrows in Fig.\,\ref{filaments}. In these regions, there is a component that is almost always at or near the mean Galactic Faraday depth with a low Faraday dispersion while the second component shows much larger scatter in Faraday depth and a high value of Faraday dispersion. It could be that these features are part of the same shock front, located either in or behind the cluster, but there is also some emission closer to the observer. If true, the presence of two Faraday components may suggest that these filaments may be separated in Faraday space along the line-of-sight and we see them in projection. The second component, therefore, indicates two emission regions significantly separated along the line-of-sight. It is also worth noting that the intrinsic polarizations in the two component apparently show rather similar intrinsic polarization angles. This could be considered as an argument in favor of `two patches of ICM in front of single emission' scenario. 
 
 %
 In general, the Faraday distribution of the relic seems to be consistent with a tangled magnetic field in the ICM which is in front of the emission. It is beyond the scope of this work to draw conclusions from the correlation of the scatter of the central Faraday depth and the Faraday dispersion on the possible magnetic field distribution in the ICM. However, it is interesting to note that the ratio between dispersion and scatter is about 1.22. This may indicate that there is power on scales of the size of box or smaller in the power spectrum of the magnetic field distribution.

 The second component in boxes 33 and 34 shows a very low and very high central Faraday depth, respectively, clearly outside the correlation found for the other components.  The two boxes contain emission from the NAT, as discussed in more detail in Sect.\,\ref{NAT-relic}. It is therefore plausible to assume that the Faraday depth of these two components is not solely caused by the ICM.

 %
 Detailed Faraday rotation studies, over a sufficient frequency range, have been performed so far only for eight radio relics, namely, for Abell 2256 \citep{Owen2014,Ozawa2015}, the Coma relic \citep{Bonafede2013}, Abell 2255 \citep{Govoni2005,Pizzo2011}, RXC\,J1314.4-2515 \citep{Stuardi2019}, CIZA\,J2242.8+5301 \citep[aka the ``Sausage relic"; ][]{Kierdorf2016,Loi2017,DiGennaro2021}, Abell 2345 \citep{Stuardi2021}, Abell 2744 \citep{Rajpurohit2021c}, and 1RXS\,J0603.3+4214 \citep[aka ``Toothbrush relic" ;][]{vanWeeren2012a,Kierdorf2016}. In these relics, the Faraday rotation has been reported to be mainly caused by Galactic foreground, with no strong evidence for frequency-dependent depolarization, for example, the Sausage relic \citep{Kierdorf2016,DiGennaro2021}. In addition, the Faraday dispersion is mainly found to be below $40\,\rm rad\,m^{-2}$.

\begin{figure}[!thbp]
\centering
\includegraphics[width=0.49\textwidth]{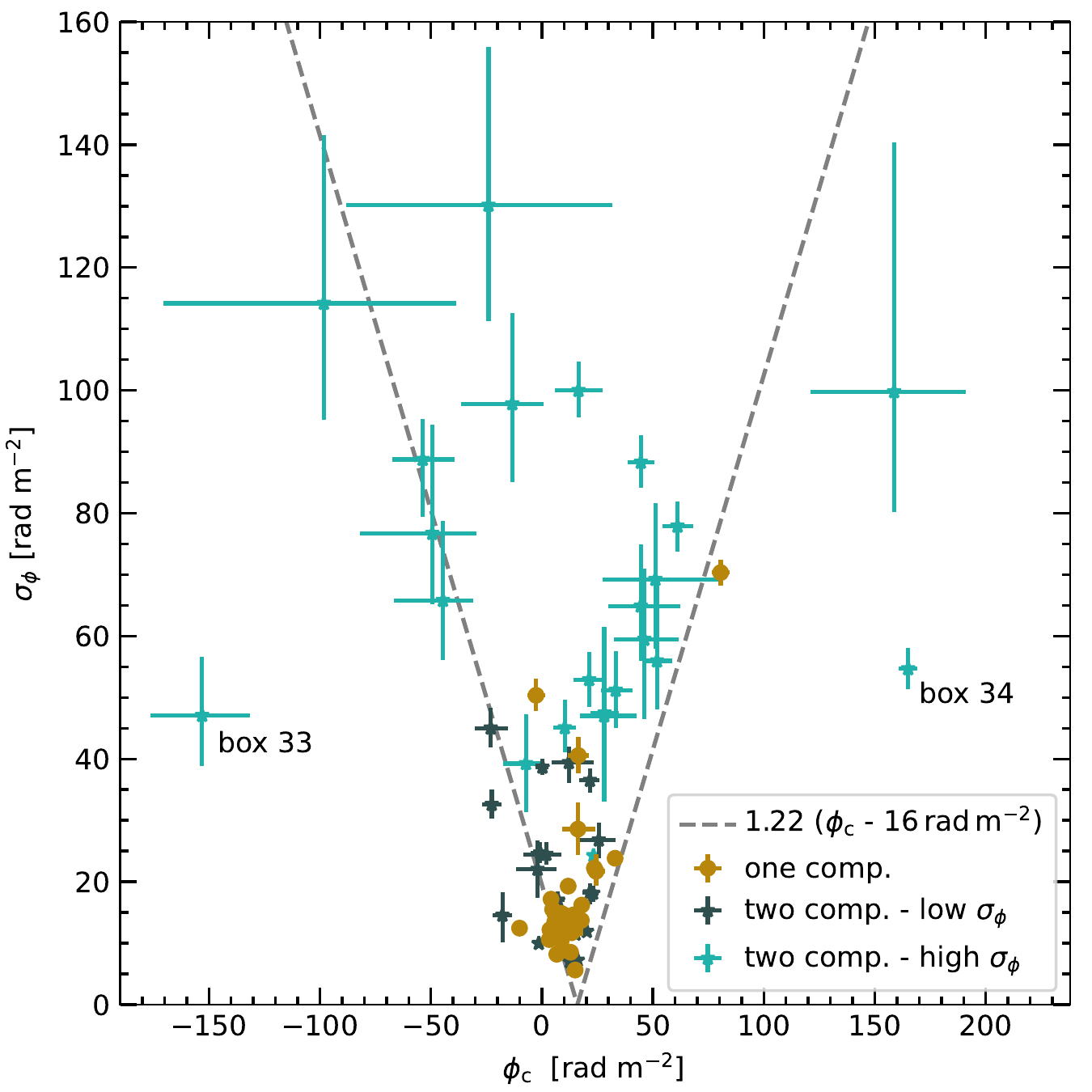}
\vspace{-0.5cm}
\caption{Central Faraday depth as a function of Faraday dispersion, measured in the observers frame. Dashed-lines indicate $\sigma_{\phi}=1.22\,(\phi_{c}-16\rm \,rad\,m^{-2})$ where $16\rm \,rad\,m^{-2}$ is the Faraday depth due to the Galactic foreground. One component models are indicated with brown filled circles and two component models with black and cyan stars, where black indicates the component with the lower Faraday dispersion and cyan the higher one. Components shown in Fig.\,\ref{fig::QUFitParameter} with a very large uncertainties are neglected here because they do not provide any constraint. The plot shows that the scatter of the central Faraday depth increases with increasing Faraday dispersion. }
\label{fig::RMsigRM}
\end{figure}

 To the best of our knowledge, only for parts of the Toothbrush and Abell\,2256 relics, do the observed RMs deviate from the Galactic foreground and show significantly high Faraday dispersion values \citep{vanWeeren2012a,Kierdorf2016,Ozawa2015}. Recently, we studied the Toothbrush relic at 18.6\,GHz and found strong depolarization between 4.9 and 18.5\,GHz, corresponding to a Faraday rotation measure dispersion of $212\pm23\,\rm rad\,m^{-2}$ \citep{Rajpurohit2020b}. The relic in MACS\,J0717.5$+$3745 is the first relic clearly showing a high $\sigma_{\phi}$ derived on the basis of a spatially well resolved analysis. For example, a median Faraday dispersion in R1 of 114\,$\rm rad \, m^{-2}$ in the rest-frame of the cluster. Such a high value of $\sigma_{\phi}$ implies that the ICM magnetic field is highly tangled even on scales as small as a few tens of kpc.



\section{Intrinsic polarization angle at the shock} 
\label{angle}

 The orientation of the observed polarization vectors provides valuable clues to the nature of diffuse emission. Merger-shock models predict the relic should be highly polarized only when viewed close to edge-on \citep{Skillman2013}. For the Sausage and the Toothbrush relics, which are very likely seen edge-on, high polarization fractions (55\,\%-70\,\%) have been reported \citep{vanWeeren2010,vanWeeren2012a,Loi2020,Rajpurohit2020b,DiGennaro2021}.  It is believed that the high polarization fraction in relics is due to shock wave, which compress and hence align isotropically distributed magnetic fields. For the relic in MACS\,J0717.5$+$3745, the polarization fraction reaches about 30\,\% or more in some regions. Such high degree of polarization rules out the possibility that the relic is seen face-on.

 In Fig.\,\ref{fig3b}, we show the magnetic field (B-field) orientation distribution at L-, S- and C-bands. The lines represent the plane of polarization (E-field) rotated by $90\degree$ to better visualize the magnetic field structure. We find that the B-field orientations are aligned across the relic, at all three frequencies. This implies that the magnetic field orientation is well correlated along the entire extent of the relic, and is mostly aligned with the shock front. The distribution of magnetic field orientations in the MACS\,J0717.5$+$3745 relic is very similar to what is found for the Sausage relic \citep{vanWeeren2010,DiGennaro2021}. 

 The magnetic field orientation distribution (intrinsic) obtained from QU-fitting is show in Figs.\,\ref{R3_RM} and \ref{R1_RM}; these are effectively corrected for the local and Galactic Faraday rotation. It is evident that the field orientations are aligned with the source extension for both single and two independent RM components. We note that for many boxes with two components the polarization angle is approximately the same for both components. Recently, using advanced simulations \cite{Paola2021a} found that intrinsic polarization angles (E-vectors) in relics strongly depend on the upstream properties of the medium and find that a turbulent medium can result in a highly aligned (anisotropic) magnetic field distribution at the shock front. The field orientation distribution in the MACS\,J0717.5$+$3745 is consistent with that simulation.  

 To resolve some fine structures, a high spatial resolution ($2\arcsec$) vector map is shown in Fig. \,\ref{vectors}. Even at full resolution, we still do not find any small-scale deviation of the magnetic field orientation from the source morphology.

 In the polarization images, R3 appears to be connected with R2 by a region of low surface brightness  emission; see the middle panel of Fig.\,\ref{fig3} and \ref{fig3b}. The change in the orientation of the B-field between the northern and southern part of the relic and this low surface brightness emission connecting them is evident. The observed gradual change in B-field orientation and the RM gradient from R1 to R4 hints that the northern and southern parts of the relic are part of the same physical structure rather than two independent sources seemingly aligned in projection. This point is further discussed in Sec.\,\ref{sim}.

  \begin{figure*}[!thbp]
   \centering
   \includegraphics[width=0.8\textwidth]{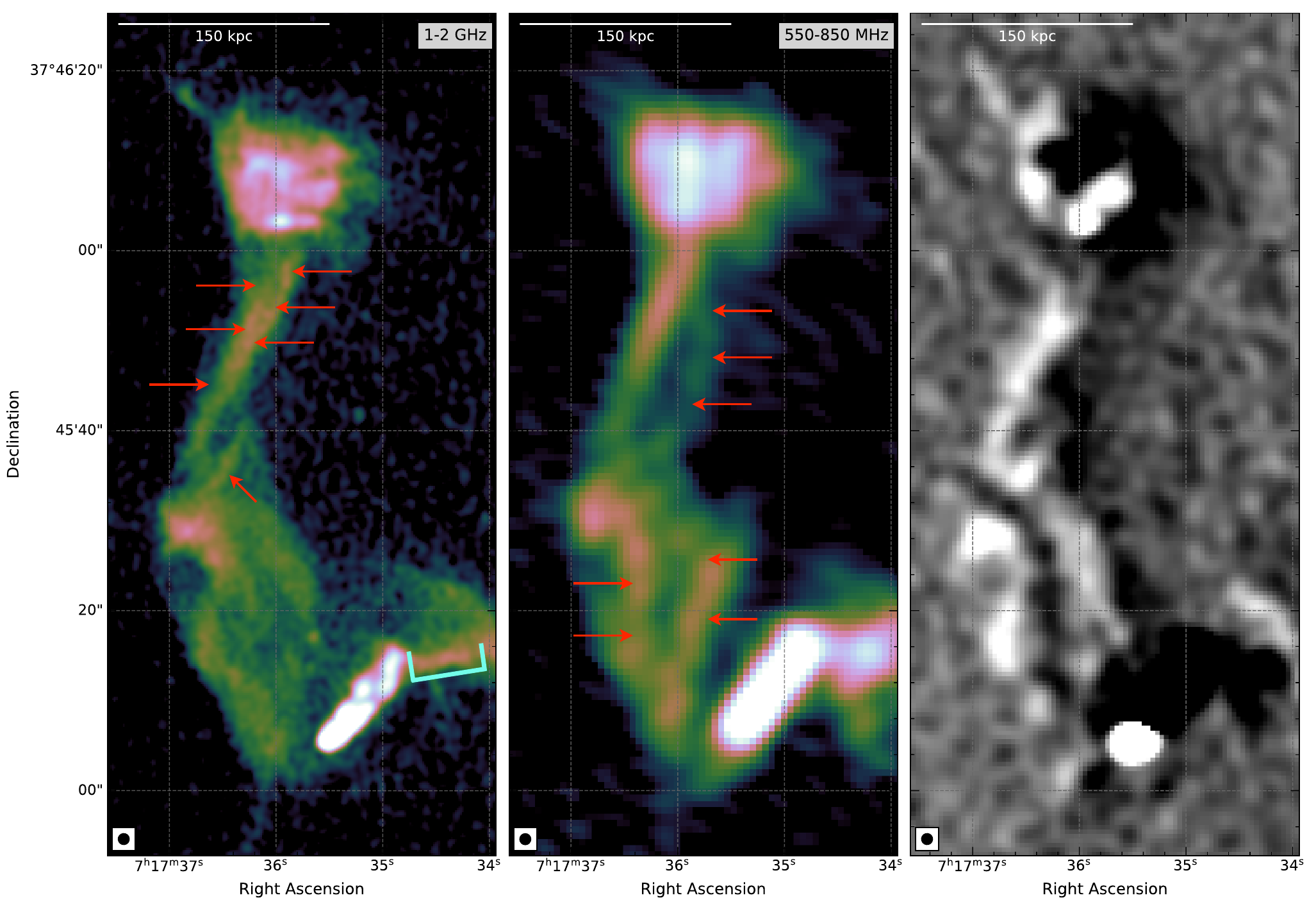}
   \vspace{-0.1cm}
   \caption{High resolution VLA \citep[left panel;][]{vanWeeren2017b} and uGMRT \citep[middle panel;][]{Rajpurohit2020a} images of the northern part of the relic. The red arrows show regions with fine filaments. The spectral tomography map of the same region is shown in the right panel, indicating that the relic is indeed composed of filaments with different spectral indices \citep{Rajpurohit2021a}. For the majority of these regions, the QU-fitting provides a better fit with two Faraday components.}
     \label{filaments}
 \end{figure*}

\begin{figure*}[!thbp]
    \centering
    \includegraphics[width=0.67\textwidth]{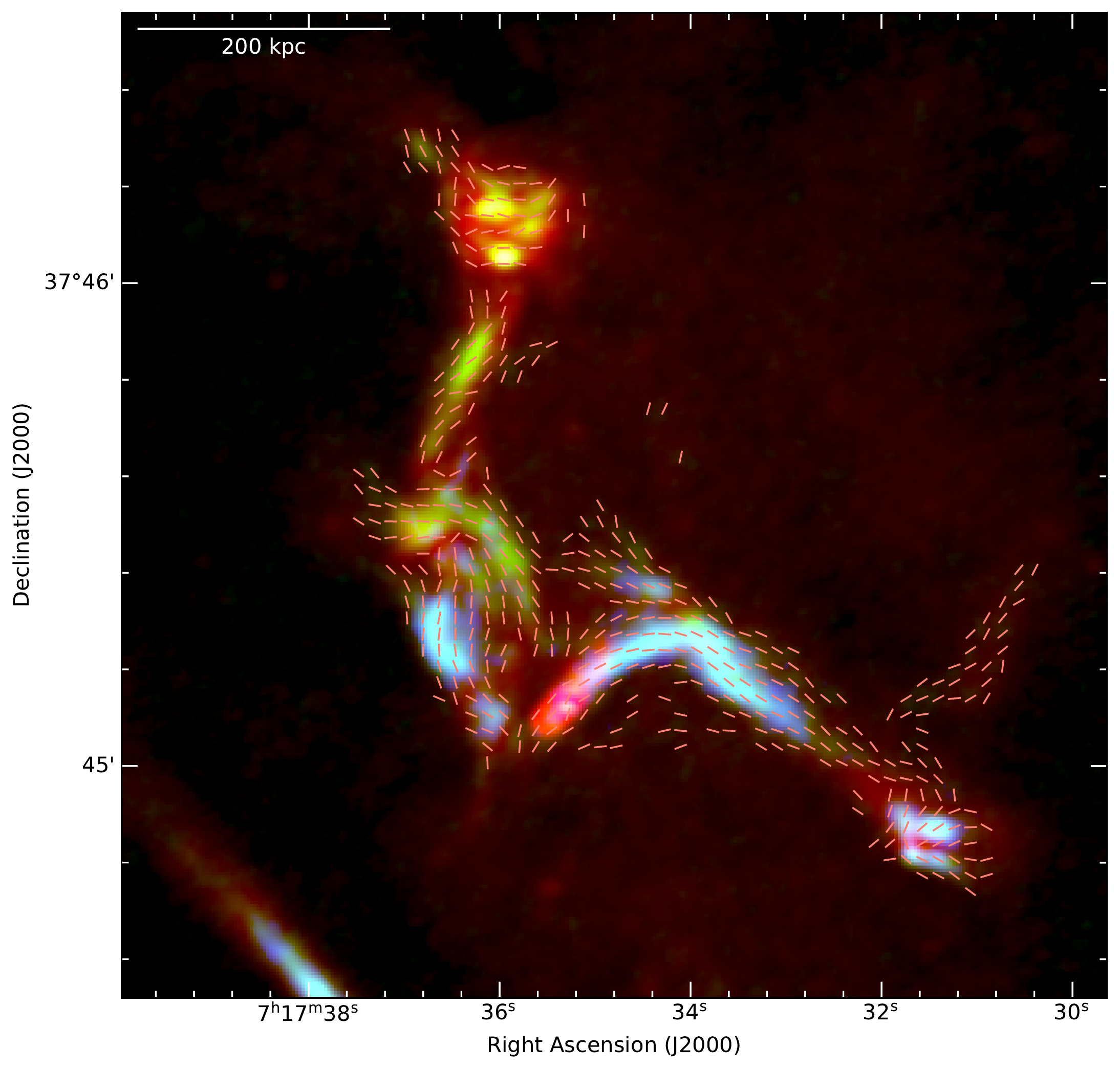}
        \vspace{-0.2cm}
    \caption{The Faraday corrected magnetic field  (B-field) orientation (pink lines) of the relic overlaid on the 2\arcsec resolution total intensity (red), S-band polarization (green), and L-band polarization (blue) images. The image demonstrates well aligned B-field vectors with the orientation of emission. This suggests a high degree of ordering of the B-field across the entire relic. The image properties are given in Table\,\ref{Tabel:imaging}, IM5, and IM9.}
      \label{vectors}
\end{figure*}

\begin{figure}[!thbp]
    \centering
    \includegraphics[width=0.45\textwidth]{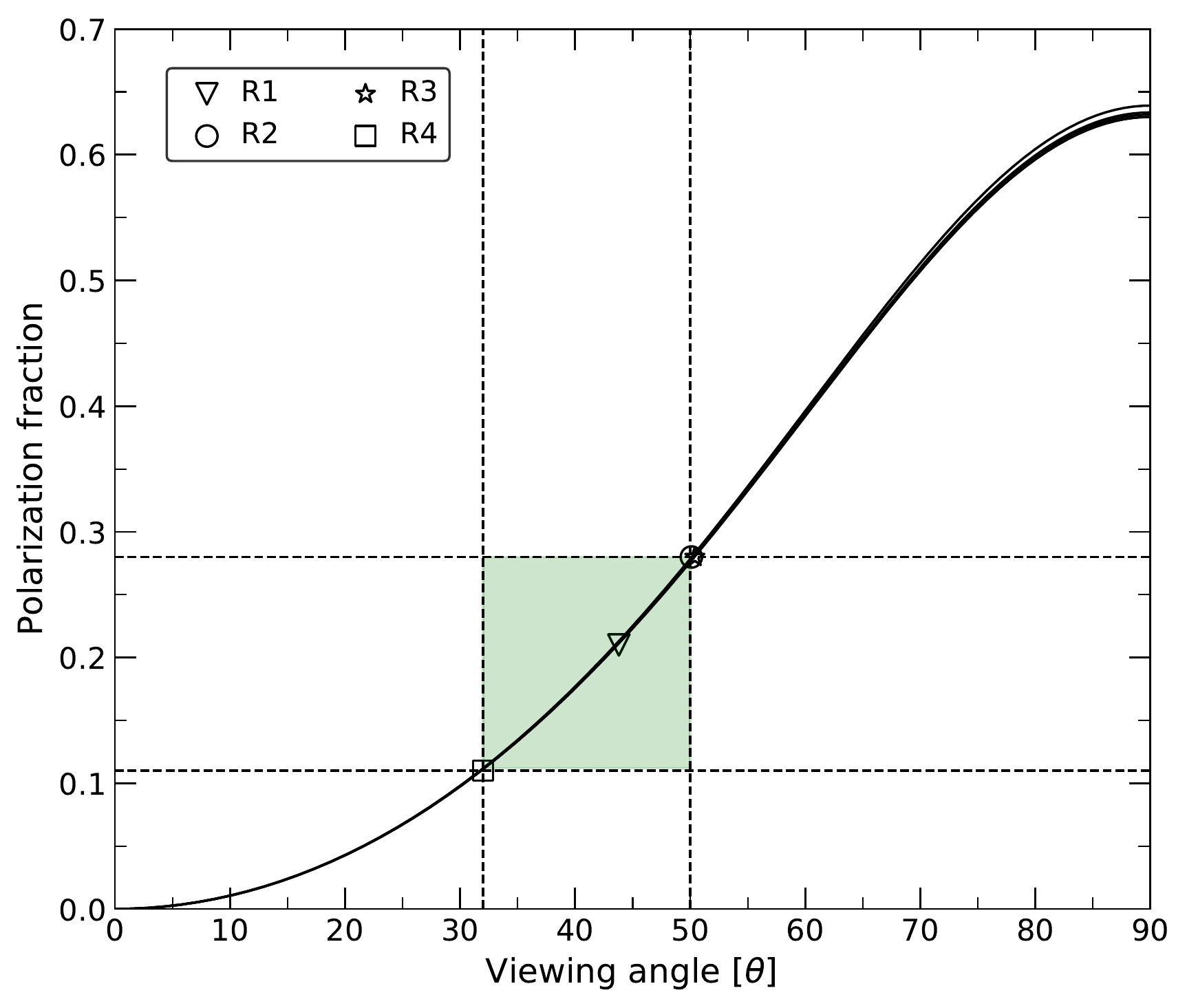}
    \caption{Viewing angle versus polarization fraction relation for the MACS\,J0717.5$+$3745 relic. A viewing angle of $90\degree$ implies that the merger is in the plane of the sky (i.e., perpendicular to the shock). The four solid curves, that are nearly identical, give the theoretical predictions for spectral indices of $\alpha = -1.18$, $-1.13$, $-1.17$, and $-1.16$, for the relic subregions. The triangle, circle, star, and square provide the corresponding estimates for the relic subregions R1, R2, R3, and R4. The dashed lines indicate the minimum and maximum value of the mean polarization fraction and the viewing angle.  The possible viewing angles of MACS\,J0717.5$+$3745 fall inside the green shaded region. The plot shows that the relic in MACS\,J0717.5$+$3745 is seen less edge-on.}
    \label{fig::viewingangle}
 \end{figure} 


\section{Connection between the NAT and the relic}
\label{NAT-relic}

 At the location of the NAT, (in particular for boxes 33-36), we find that a one-component Faraday screen provides a very poor fit to the QU-spectra while the two-component model instead allows us to fit $q$ and $u$ reasonably well. These regions cover the NAT core and the tails (about 94\,kpc distance from the core). We note that the tails of the NAT are even more extended at frequencies below 700\,MHz, and they bend to the south of R3. However, the bent tails are barely visible above 1.5\,GHz \citep{Rajpurohit2021a}. 
 
 \begin{figure*}
    \centering
    \includegraphics[width = 0.48\textwidth]{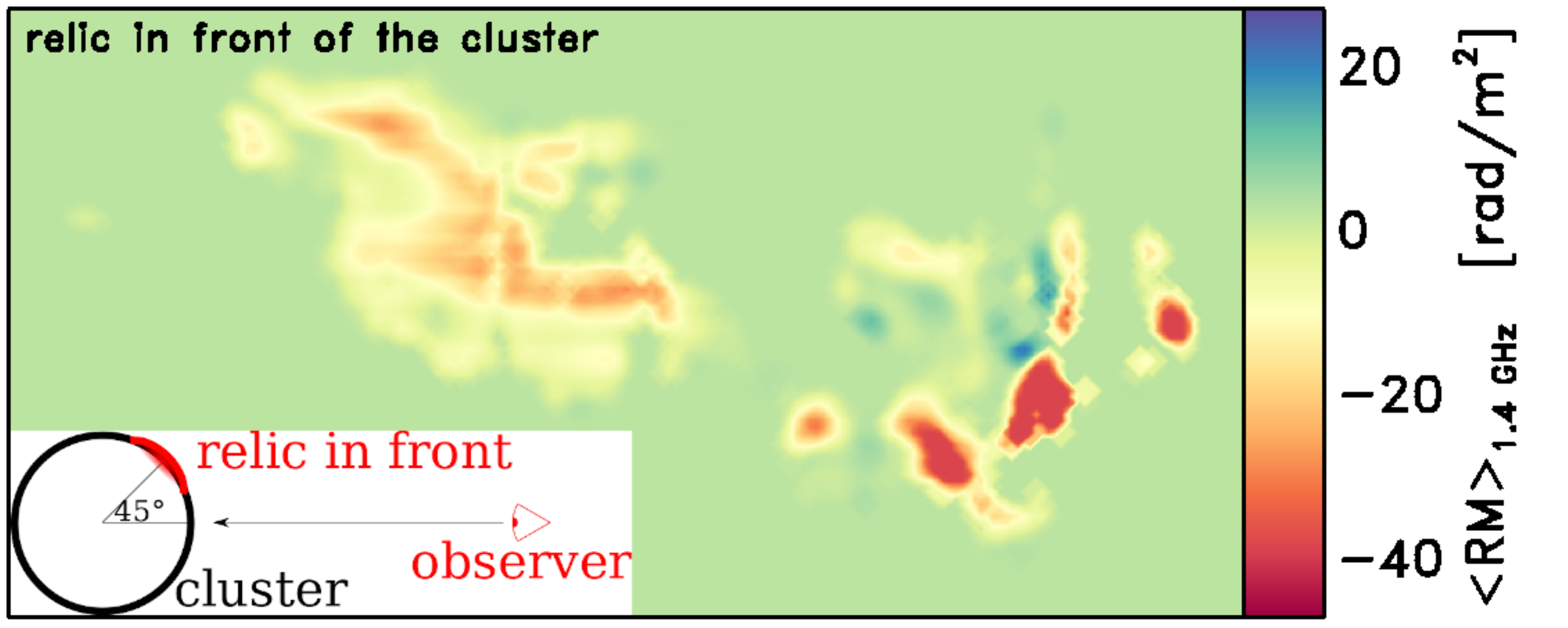}
    \includegraphics[width = 0.48\textwidth]{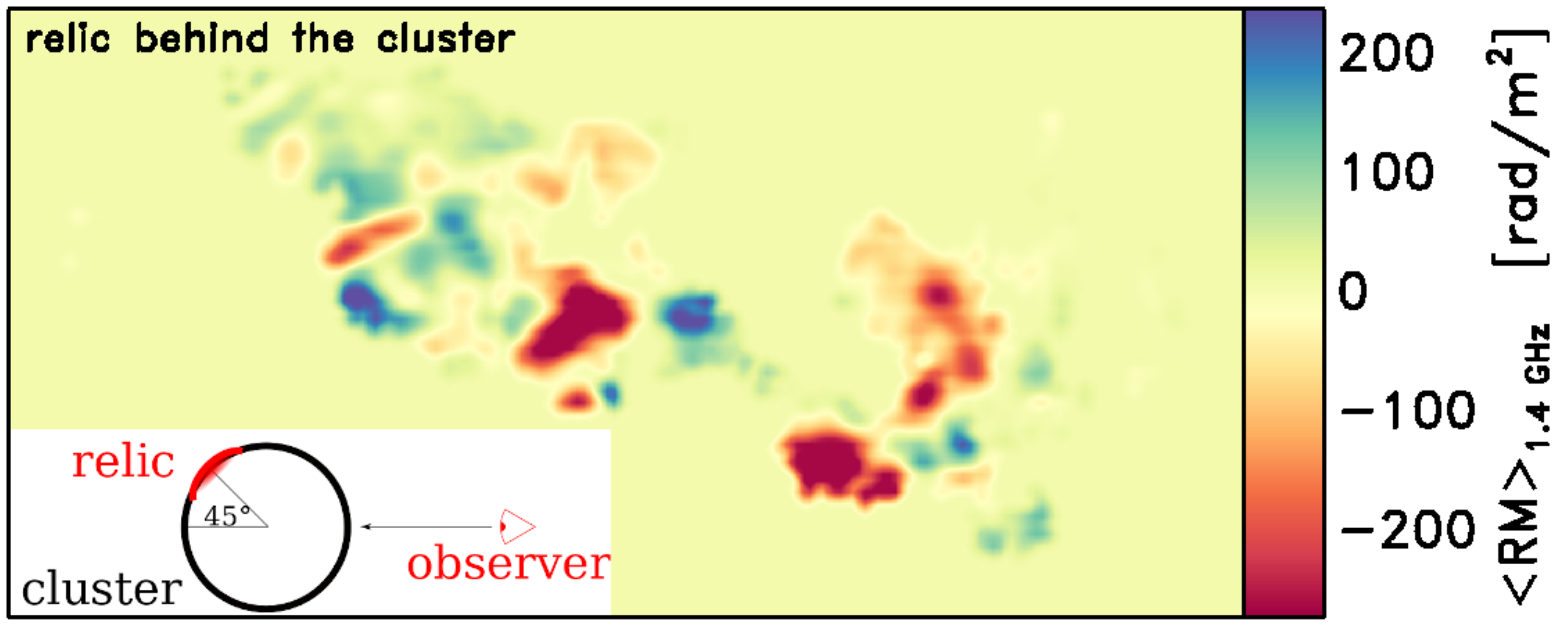} \\
    \caption{Simulations: maps of the average RM, radio weighted at $1.4 \ \mathrm{GHz}$. \textit{Left}: the RM distribution if the relic lies in front of the cluster, as depicted in the inset.  \textit{Right}: the RM distribution if the relic lies behind the cluster, as show in the inset. These maps show that strong fluctuations in the RM are expected if the relic is located within, or behind, the ICM screen.}
    \label{fig::simu_rm}
\end{figure*} 
 
 We find that boxes 33-36 and 41-43, and 45, coinciding with the NAT, show a clear second component with a high Faraday dispersion compared to the relic R3 region. The central Faraday depth of the high Faraday dispersion component varies significantly from box to box, see Fig.\,\ref{fig::QUFitParameter}, as expected from a Faraday screen with a tangled magnetic field. Therefore, it is plausible to assume that the Faraday dispersion of about $150\,\rm rad \, m^{-2}$ (at the redshift of the cluster), which indicates that these components have a significantly thicker ICM in front of them than the relic in that region. However, for boxes 33 and 34 we find that the second component falls outside the relation between Faraday dispersion and scatter of the central Faraday depth, found for almost all components. Possibly, part of the Faraday depth found for these components might be intrinsic to the emission and not caused by the ICM in front of it. Therefore, it is also conceivable that part of the Faraday dispersion has to be attributed to be intrinsic to the NAT and is not caused by the ICM in front of the emission. Therefore, any interpretation of the Faraday structure of these components has to be taken with some grains of salt.

 A plausible scenario for the structure and polarization properties of the NAT galaxy is that the galaxy is actually deep in the cluster or even at the rear side of the cluster. Evidently, the head-tail morphology indicates the interaction with the ICM. The Faraday dispersion of these components suggests a similar amount of magnetized ICM in front of the NAT than in front of many components of R1 and R2 and clearly a larger amount than in front of R3 and R4. 

 In the total power images, the emission located within box 44 appears to be connected to R3 through a bright, thin filamentary structure; see the region shown with cyan in Fig\,\ref{filaments} right panel. Box 44 is significantly polarized ($\sim30\%$)  with a typically low value of $\sigma_{\phi}$ ($\rm 13\, rad\,m^{-2}$). The low $\sigma_{\phi}$ suggests less path length through the ICM: that is the emission must be lying in the cluster periphery towards the observer. The high degree of polarization rather indicates that the contribution of the NAT is likely faint. We do not find any evidence that the NAT is moving diagonally through the cluster where the R3 region of the relic is located. If this were the case in boxes 33, 34, 35, 36, 37, and 44, we would expect to see a component from a region with high $\sigma_{\phi}$ propagating towards a region with a lower $\sigma_{\phi}$. We do not find any hint of such a component. 
 
Our analysis supports a scenario in which the NAT is moving through the cluster, coming from the observer, and is deep in the cluster causing the high $\sigma_{\phi}$. In contrast to the NAT, R3 shows a polarized emission component with low $\sigma_{\phi}$, indicating very little Faraday rotating intervening material, implying that the relic is located in front of the ICM or in the cluster periphery. This suggests that the NAT and R3 are well separated in Faraday space and, thus not connected physically. The spectral analysis of the same region also revealed two spectral components \citep{Rajpurohit2021a}.

 On the basis of the polarization and spectral analysis, we suggest that the NAT and R3 overlap only in the projection; in the cluster they are separated. This makes it unlikely that the NAT can be the source of seed electrons which are re-accelerated by the shock front at the relic.


\section{Viewing angle of the Merger}

 Based on the theoretical model by \citet{Ensslin1998}, it is possible to estimate the orientation of the merger axis using the average degree of polarization \citep[e.g.,][]{Hoang2018}. Here, we use the approximation for weak magnetic fields  \citep[Eq. 3.2. in][]{Ensslin1998} to estimate the viewing angle of the relic in MACS\,J0717.5$+$3745. We estimated the viewing angle for the four regions of the relic: R1, R2, R3, and R4. Our results are given in Fig.\,\ref{fig::viewingangle}. The three black curves give the theoretical estimate of the average polarization fraction depending on the viewing angle for spectral indices of $-1.18$, $-1.17$, $-1.16$, and $-1.13$ of the four subregions (see Table\,\ref{Tabel:Tabel2}). The theoretical predictions for the four indices do not differ significantly.

 We use the average intrinsic polarization fractions measured for the four subregions from QU-fitting (see the symbols in Fig. \ref{fig::viewingangle}). We found that the regions R2 and R4 are seen at an angle of $50.1\degree$ to $50.3\degree$, while R1 and R4 appear to be seen at angles of $43.8\degree$ and $32\degree$, respectively. Radio relics are not straight sheet-like structures, but have rather complex 3D-shapes \citep[e.g.,][]{Skillman2013,wi17,2019arXiv190911329W,2020arXiv200913514D}. Hence, it is very likely that different parts of the relic are seen under different viewing angles. The radio emission from relics are not spherically symmetric and thus their morphology depends on the viewing angle of the radio emission. The fact that the different regions of the MACS\,J0717.5$+$3745 relic are seen at such different viewing angles, could well explain its chair-like structure.

 The average intrinsic polarization fraction (obtained from QU-fitting) implies that the relic in MACS\,J0717.5$+$3745 is seen less edge-on compared to some other relics, for example the Sausage \citep{vanWeeren2010,Kierdorf2016,Loi2020,DiGennaro2021} and Toothbrush \citep{vanWeeren2012a,Kierdorf2016,Rajpurohit2020b} relics which show rather very high degree of polarization and are most likely seen edge-on. We note that the intrinsic polarization could be an underestimation because fractional polarization may still suffer from beam depolarization. The viewing angle of the MACS\,J0717.5$+$3745 relic is in the range $32\degree\leq\theta\leq51\degree$, this is also consistent with the one obtained from the radio color-color analysis \citep{Rajpurohit2021a}.

\section{Comparison with simulations}
\label{sim}

 We compare the polarization properties of the MACS\,J0717.5$+$3745 relic with the simulated relic from \citet{2019arXiv190911329W}. As discussed above, the viewing angle of the relic is at least $45\degree$ (i.e., the viewing angle of the merger is $45\degree$ with respect to the plane of the sky). Therefore in the following section, we compare the polarization properties of the simulated relic when seen at $45\degree$ to that of the MACS\,J0717.5$+$3745 relic.

 The simulation was carried out with the cosmological magneto-hydrodynamical ENZO code \citep[][]{ENZO_2014}. It belongs to a sample of high-resolution simulations of galaxy clusters that was used to study magnetic field properties in the ICM \citep[][]{vazza2018dynamo,2019MNRAS.486..623D}. These simulations self-consistently evolve complex magnetic field patterns during cluster mergers, starting from an assumed uniform ``primordial" seed field with of $0.1$ comoving nG at $z=40$. The model to produce synthetic radio observations is explained in detail in \citet{2019arXiv190911329W}; the model accounts for the acceleration of a tiny fraction of electrons from the thermal pool (based on the thermal leakage model) via DSA at the shock front, and for the aging of cosmic-ray electrons in the downstream region of the shock front. As the simulated relic was found to mimic several spectral properties of the MACS\,J0717.5$+$3745 relic, we investigate here RM fluctuations and gas perturbations in the shock front region. 

 Fig.\,\ref{fig::simu_rm} shows the RM distribution of the simulated relic. In the simulation, we study two different scenarios for the location of the relic with respect to the observer:  the relic either in front or behind the cluster (see insets in the RM maps in Fig. \ref{fig::simu_rm}). If the relic lies behind the cluster, the X-ray emission provides a good proxy for the amount of ICM the radio emission has to transverse. In Fig.\,\ref{fig::simu_rm}, we plot the maps of the average RM along the line-of-sight of the simulated relic. The RM maps are weighted with the radio power at 1.4\,GHz and, hence, they are meant to reflect the RM measured at the brightest region of the relic along the line-of-sight.

 If the relic lies in front of the cluster, the resulting RM values are similar to the southern part of the relic in MACS\,J0717.5$+$3745, except for a larger value close to the dense sub-clump in the region \citep[see Fig.\,1 of][]{2019arXiv190911329W}.  On the other hand, if the relic is located behind the cluster, the RM values in the simulations are similar to the ones measured for the relic in MACS\,J0717.5$+$3745. These larger RM values are attributed to the fact that the simulated relic lies deeper inside the ICM. Furthermore, the simulated relic shows an RM trend when moving from north to south (see Fig. \ref{fig::simu_rm}). The relative  decrease is also similar to the observed decrease from region R1 to R4 in the MACS\,J0717.5$+$3745 relic. Such a gradient is missing if the relic is in front of the cluster. 

 If the relic is instead located behind the cluster, the RM increases from the north of the relic to the south. The reason for this behavior is twofold. First, the top part is located in a less dense ICM. Second, the relic is tilted with respect to the line-of-sight and, hence, the top part lies closer to the observer. Therefore, such an RM trend is expected if a part of the relic is located in or behind a denser ICM, or at a larger distance from the observer as  a denser ICM can boost the strength of the RM gradient.

 Similar trends in RM are observed for the relic in MACS\,J0717.5$+$3745:  the northern part of the relic in MACS\,J0717.5$+$3745 is located in or behind the dense ICM, whereas the southern part extends into the low density ICM; see discussion in Sect\,\ref{sec::FDFrelic}. We emphasize that the simulated relic is a single structure caused by the same shock front. This raises an important question: are R1+R2 and R3+R4 a connected structure simply inclined (or tilted) towards the line-of-sight, or do they belong to two different structures that appear to be connected in projection? If part of the relic is at a larger distance to the observer, the emitting structure is either disconnected or tilted with respect to the line-of-sight. The latter seems to be a more likely explanation for the relic in MACS\,J0717.5$+$3745 (see discussion in Sect.\,\ref{angle}); however the current data do not allow us to rule out that the northern and southern parts are disconnected.

   \begin{figure*}[!thbp]
    \centering
    \includegraphics[width=1.0\textwidth]{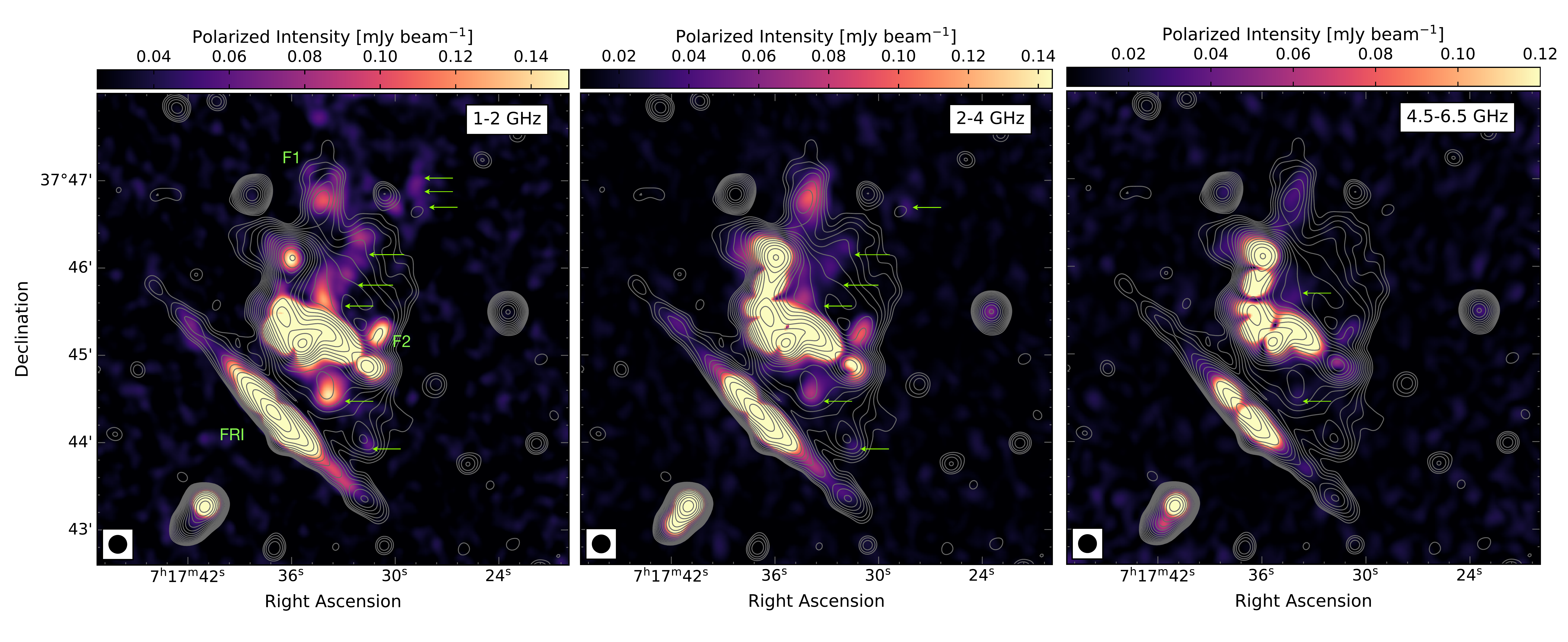}
    \vspace{-0.7cm}
    \caption{Low resolution ($12.5\arcsec$ resolution) polarization intensity overlaid with the total intensity contours,  showing that a part of the halo is polarized (shown with green arrows). Contour levels are drawn at $\sqrt{[1,2,4,8,\dots]}\,\times\,4\,\sigma_{{\rm{ rms}}}$ and are from the S-band Stokes $I$ image. The beam sizes are indicated in the bottom left corner of the image.The image properties are given in Table\,\ref{Tabel:imaging}, IM4, IM8, and IM12. }
      \label{fig4}
  \end{figure*}  
   

\section{Magnetic field estimates}

 The combination of observed Faraday dispersion and central Faraday depth trends across the relic can be used to constrain the magnetic field properties of the ICM. Both the strength and the morphology of magnetic fields affect the Faraday depth of radio sources. Under a few simplifying assumptions, we can use the observed Faraday dispersion ($\sigma_{\phi}$) to estimate the magnetic field values following, for example, \cite{Sokoloff1998,Kierdorf2016}:
 \begin{equation}
  \sigma_{\phi}
  =
  \sqrt{(1/3)}\,0.81\,\langle n_{\rm e}\rangle\,B_{\rm turb}\,(L\,t/f)^{0.5},
\label{sigmaRM}
\end{equation}
 where $\langle n_{\rm e}\rangle$ is the average thermal electron density of the ionized gas along the line-of-sight in $\rm cm^{-3}$, $B_{\rm turb}$ is the magnetic field strength in $\upmu$G, and $f$ the volume filling factor of the Faraday-rotating plasma. $L$ and $t$ are the path length through the thermal gas and turbulence scale, respectively, in pc. 
 
 For the relic in MACS\,J0717.5$+$3745, we find fluctuations in the polarization intensity on a scale as small as 10\,kpc (see Sect.\,\ref{results}), particularly for the northern part of the relic. This may hint that the turbulence coherence length is of the same scale. Therefore, we adopt t=10\,kpc. We assume that the path which is dominating the Faraday depth scatter is where the density is highest along the path (i.e., $L$) and has a length of 1\,Mpc. Depending on the line of sight and density, this could be close to the cluster center with high density or rather in the periphery. We consider the thermal electron density of $\langle n_{e} \rangle=10^{-3}\rm\,cm^{-3}$. Finally, the filling factor is assumed to be 0.5 following \cite{Murgia2004,Kierdorf2016}

A significant number of boxes in R1 and R2 regions provide a better fit with two Faraday components, however, for the magnetic field estimate we only considered the low Faraday dispersion component. For the northern part of the relic, the mean $\sigma_{\phi}$ of the low Faraday dispersion component is about $\rm 120\,rad\,m^{-2}$ (in the rest-frame of the cluster). By inserting all values in Eq.\,\ref{sigmaRM}, we obtained $B_{\rm turb,\,R1+R2}\sim 1.8\,\upmu \rm G$. When using $\langle n_{e} \rangle=10^{-4}\rm\,cm^{-3}$, we get a very high value of magnetic field ($\sim 18\,\upmu \rm G$) which is very unlikely because in this case we would expect much higher value of $\sigma_{\phi}$.

For the southern part of the relic, the mean Faraday dispersion is about $29\,\rm rad\,m^{-2}$ (in the rest-frame of the cluster). We note that the region contaminated by the NAT is excluded. Since the $\sigma_{\phi}$ is low compared to the northern part of the relic and also the fluctuations in the polarization intensity, we assume $\langle n_{e}\rangle=10^{-4}\rm\,cm^{-3}$, $L=1.5$\,Mpc and t=100\,kpc. For this part of the relic, we obtained $B_{\rm turb,\,R3+R4}\sim 1.2\,\upmu \rm G$.

If the estimated strength of the turbulent magnetic field and the turbulent scale reflects the properties of the ICM, this suggests that the northern component of the relic is embedded in a moderately dense ICM. Since the Faraday dispersion and the magnetic field are not very high (at least for the low Faraday dispersion component), it is very unlikely that this part of the relic is located behind the ICM. On the other hand, the southern part favors a geometry in which the line-of-sight passes through a slightly less denser region of the ICM or in front of the ICM.


\section{Is the halo emission polarized?}
\label{halopol}

 The halo in MACS\,J0717.5$+$3745 exhibits significant substructure \citep[e.g.,][]{vanWeeren2017b}. It remains uncertain whether or not those filamentary substructures can be entirely attributed to the halo (tracing regions of increased turbulence) or whether they are similar to relics (tracing shock waves). We note that the estimated viewing angle suggests that some structures might indeed be only seen in projections with the halo.   

 Polarized emission from radio halos has been very difficult to detect. Cluster-wide polarized emission from halos has not yet been detected from any cluster. So far, polarized filamentary structures have been detected only in three halos, namely MACS\,J0717.5$+$3745, Abell\,2255, and Abell\,523,  \citep{Govoni2005,Bonafede2009a,Girardi2016}. However, in these three cases it is not clear that the polarized emission is truly associated with the halos. The absence of polarization in halos has been interpreted as the result of internal Faraday rotation and beam depolarization. The low surface brightness also limits the possibility of detecting the polarized signal from a radio halo. The halo in MACS\,J0717.5$+$3745 is one of the most powerful known halos; since the measured surface brightness is typically higher in powerful halos, this cluster offers one of the best opportunities to detect any polarized emission that may be present \citep{Govoni2013}. 

 \cite{Bonafede2009a} reported polarized emission from the radio halo associated with MACS\,J0717.5$+$3745, with a mean fractional polarization of 5\% at 1.4\,GHz from a $20\arcsec$ resolution image. However, it is not yet fully clear whether this polarized emission truly comes from the halo emission, or rather from the relic emission projected along the line-of-sight. 
 
 The high-resolution polarization intensity images reveal only a few patches of polarized emission in the halo region; see Fig.\,\ref{fig3a} and \ref{fig3b}. Since radio halos typically have a low surface brightness and are often not detected in high resolution images, we also created moderate-resolution ($12.5\arcsec$) polarized intensity images. In Figure \,\ref{fig4}, we show these moderate-resolution polarized intensity maps at L-, S- and C-bands. At this resolution, we are much more sensitive to low surface brightness emission. 

 As shown in Fig.\,\ref{fig4}, we detect more polarized emission in the halo region in our low resolution images. This indicates that the magnetic field is very likely not tangled on scales smaller than the beam size in these regions. We detect patches of polarized emission from the halo (indicated with green arrows). We note that the regions of the halo with filamentary substructures, visible in Stokes $I$, are polarized with a fractional polarization of 8-36\%. Excluding these polarized patches, the polarized halo emission is below the $1\sigma$ level. 

 To calculate an upper limit on the fractional polarization of the halo, we used our moderate-resolution polarization image at 3\,GHz. We exclude polarized regions that are shock related and projected on the halo, as mentioned above. For the $12.5\arcsec$ resolution polarized intensity image at 3\,GHz, the $\sigma_{Q,U}$ is $9\,\upmu$Jy\,beam$^{-1}$\,RMSF$^{-1}$. We found no polarized emission at greater than $5\sigma_{Q,U}$ within the region of the radio halo. From the Stokes $I$ image, we find a peak halo flux density of $1.9\,\upmu$Jy\,beam$^{-1}$ at 3\,GHz. Therefore, our $5\sigma_{Q,U}$ upper limit on the fractional polarization for the halo at 3\,GHz is $3\%$. It is important to emphasize here that these polarization images were obtained using deep observations (45 hours on-source VLA observations combining A,B,C, and D configurations) with high sensitivity to low surface brightness emission.    

 The polarized intensity map (see Fig.\,\ref{fig4}) clearly shows a discontinuity between between the polarized structures in the halo region and the relic. In addition, there also exists a clear separation between the polarized emission detected in the halo emission and filaments F1 and F2. The low-resolution Faraday map is shown in the right panel of Fig\,\ref{rm_map}. The Faraday map shows a uniform distribution across the halo region with a mean Faraday depth of $+18\,\rm rad\,m^{-2}$. The $\sigma_{\phi}$ in these regions is of the order of $11-16\,\rm rad\,m^{-2}$. If this polarized emission were from a halo, we would expect high values of $\sigma_{\phi}$ with significant fluctuations arising from the dense ICM; we do not find any evidence of such behavior. The observed Faraday depth values are consistent with being Galactic in origin, which suggests that the emission is not experiencing significant Faraday rotation from the ICM. We also find similar Faraday depth values for filaments F1 and F2. Moreover, at least F1 is apparently located in the cluster outskirt. Towards the cluster center, the radio emission experiences greater Faraday rotation, thus we expect a high value of Faraday dispersion value. The low $\sigma_{\phi}$ values found for polarized patches in the halo region suggest that these regions are in fact located toward the cluster outskirts. 

 The filamentary substructures, including F1 and F2, found in the halo region are significantly polarized. We find that the magnetic field vectors are highly ordered in these polarized regions, typical for shock-related structures; see Fig.\,\ref{fig3b}. In addition, F1 shows a distinct behavior in the point-to-point radio versus X-ray surface brightness relation \citep{Rajpurohit2021b}, suggesting the emission in F1 is not associated with the halo and has a different origin. Considering all the evidences presented in this section, we suggest that these polarized filaments filaments are shock-related structures projected onto the cluster center and the halo region.


\section{Summary and Conclusions}
\label{summary}

In this work, we have presented VLA L-, S-, and C-band polarimetric radio observations of the galaxy cluster  MACS\,J0717.5$+$3745. Thanks to the very wide bandwidth of the combined observations and the high angular resolution of the images, it has been possible to reveal the complexity of diffuse radio emission in polarized intensity. Polarization and Faraday depth maps, with resolutions ranging from 2\arcsec to 12.5\arcsec, and QU-spectra in 64 regions were constructed and analyzed to study the origin of the diffuse emission in the cluster. We summarize the overall results as follows:

\begin{enumerate} 

\item{} The relic is highly polarized with a polarization fraction ${>}30\%$ in some regions. Between 2-6.5\,GHz, polarized emission is detected along the whole extent of the relic. The polarized emission is clumpier than the total power emission. The fractional polarization changes systematically increases from R1 to R3.\\

\item{} By comparing Rotation Measure (RM) Synthesis and QU-fitting results, we find a reasonable agreement when the Faraday dispersion functions are simple and the depolarization is low or modest.\\

\item{} A strong wavelength-dependent Faraday depolarization is detected between 1\,GHz and 6.5\,GHz for the northern part of the relic (R1 and R2). The underlying Faraday dispersion may originate from an intervening magnetized screen that arises from dense ICM containing a tangled magnetic fields. The high depolarization of the northern part of the relic, corresponding to a Faraday dispersion ($\sigma_{\phi}$) of about $30{-}170\,\rm rad\,m^{-2}$ (in the observer-frame) suggests that the northern part of the relic is located in or behind the ICM. \\

\item{} For the southern part of the relic (R3 and R4), we find a single Faraday component and a low value of $\sigma_{\phi}$ (below $20\,\rm rad\,m^{-2}$, in the observer-frame). The Faraday depth ($\phi_{\rm c}$) values are very close to the Galactic foreground, indicating very little Faraday-rotating material along the line-of-sight toward this part of the relic. This suggests that the southern part of the relic is likely to reside in a lower-density region of the ICM, or is located in the front of the cluster.\\

\item{} For a number of subregions (`boxes') in the relic regions R1 and R2, we find that the QU-spectra are significantly better modeled when two Gaussian Faraday components are adopted instead of one. The second component exhibits rather high $\sigma_{\phi}$  (as high as $\sim 170\,\rm rad\,m^{-2}$, in the observer-frame). We find strong spatial variation in both $\phi_{\rm c}$ and $\sigma_{\phi}$. These fluctuations are consistent with an magnetic field in the ICM which is tangled on scales of the few tens of kpc.\\

\item{} From spatially resolved analysis, we find that the scatter of Faraday depth correlates with the depolarization, corroborating that a tangled field in the ICM causes the depolarization.\\

\item{} The magnetic field orientations derived from the polarization angle are well-aligned along the relic structure. This indicates that the magnetic field distribution in the plane of the sky within the relic is highly anisotropic, very likely due to compression by the passage of the merger shock wave. \\

\item{} We find evidence of two clear Faraday components along line-of-sight passing through the NAT galaxy. The presence of two Faraday components suggests that there is also a relic component at that location. We suggest that the NAT and the R3 region of the relic are separated in Faraday space, such that the NAT is either located in or behind the ICM while the relic component (associated with R3) lies in front of the ICM. If true, this implies that the relic is not seeded by the shock re-acceleration of fossil electrons from the NAT galaxy.\\

\item{} The degree of polarization across filaments F1 and F2 is ${>}15\%$. We find generally low Faraday depth values across both of these structures, confirming that they are located in the cluster outskirts. The magnetic field in these structures is well ordered, as typical for relics. Therefore, these filaments are very likely shock-related structures. \\

\item{} We detect polarized emission from filamentary structures found within the halo region in both high- and low-resolution VLA maps. The absence of significant Faraday rotation, the aligned magnetic field orientations within the emitting region and the generally low $\sigma_{\phi}$ suggests that these polarized emission features, previously considered to be part of the halo, are related to shocks projected onto the cluster center and the halo region.   \\

\item{} We do not detect any significant polarized emission truly associated with the halo in our deep (45 hours on-source time) and highly sensitive VLA observations. The upper limit on the fractional polarization for the halo at 3\,GHz is $3\%$. \\

\end{enumerate}

The spatially resolved polarization and Faraday analysis  of the complex merging galaxy cluster MACS\,J0717.5$+$3745 suggests that the ICM magnetic field is highly tangled. The observed depolarization and high value of Faraday dispersion are consistent with an intervening magnetized screen that arises from the dense ICM. Based on the spectral and polarization properties, we conclude that several of the observed properties of this system are dominated by a superposition of plasma mediums containing tangled fields along the line-of-sight. 

\begin{acknowledgements}
MH thanks A. Basu for enlightening discussions. 
KR, FV, M. Brienza, and NL acknowledge financial support from the ERC Starting Grant ``MAGCOW", no. 714196. MH and AD acknowledge financial support by the BMBF with the grant number 05A20STA. DW acknowledge financial support by the Deutsche Forschungsgemeinschaft (DFG, German Research Foundation) - 441694982. RJvW and A. Botteon acknowledges support from the VIDI research programme with project number 639.042.729, which is financed by the Netherlands Organization for Scientific Research (NWO). LR acknowledge support, in part from US National Science Foundation grant AST 17-14205 to the University of Minnesota. W. Forman acknowledges support from the Smithsonian Institution,  the High Resolution Camera Project through NASA contract NAS8-03060, and NASA Grants 80NSSC19K0116, GO1-22132X, and GO9-20109X. CJR, EB, M. Brienza, and A. Bonafede acknowledge support from the ERC through the grant ERC-Stg DRANOEL n. 714245. PDF acknowledge supported by the National Research Foundation (NRF) of Korea through grants 2016R1A5A1013277 and 2020R1A2C2102800. FL acknowledges financial support from the Italian Minister for Research and Education (MIUR), project FARE, project code R16PR59747, project name FORNAX-B. A part of the data reduction was performed using computer facilities at Th\"uringer Landessternwarte Tautenburg, Germany. This research made use of computer facility  on the HPC resources at the Physical Research Laboratory (PRL), India. The cosmological simulations in this work were performed using the ENZO code (http://enzo-project.org). The authors gratefully acknowledge the Gauss Centre for Supercomputing e.V. (www.gauss-centre.eu) for supporting this project by providing computing time through the John von Neumann Institute for Computing (NIC) on the GCS Supercomputer JUWELS at J\"ulich Supercomputing Centre (JSC), under projects no. HHH42 and {\it stressicm} (PI F.Vazza) as well as HHH44 (PI D. Wittor). We also acknowledge the usage of online storage tools kindly provided by the INAF Astronomical Archive (IA2) initiave (http://www.ia2.inaf.it).

The National Radio Astronomy Observatory is a facility of the National Science Foundation operated under cooperative agreement by Associated Universities. 

Finally, we wish to acknowledge the developers of the following python packages, which were used extensively during this project: \texttt{aplpy} \citep{Robitaille2012}, \texttt{astropy} \citep{Astropy2013}, \texttt{matplotlib} \citep{Hunter2007}, \texttt{numpy} \citep{Numpy2011} and \texttt{scipy} \citep{Jones2001}.

\end{acknowledgements}

\bibliographystyle{aa}

\bibliography{ref.bib}

\begin{thebibliography}{88}
\expandafter\ifx\csname natexlab\endcsname\relax\def\natexlab#1{#1}\fi

\bibitem[{{Anderson} {et~al.}(2016){Anderson}, {Gaensler}, \&
  {Feain}}]{Anderson2016}
{Anderson}, C.~S., {Gaensler}, B.~M., \& {Feain}, I.~J. 2016, \apj, 825, 59

\bibitem[{{Anderson} {et~al.}(2015){Anderson}, {Gaensler}, {Feain}, \&
  {Franzen}}]{Anderson2015}
{Anderson}, C.~S., {Gaensler}, B.~M., {Feain}, I.~J., \& {Franzen}, T.~M.~O.
  2015, \apj, 815, 49

\bibitem[{{Anderson} {et~al.}(2018){Anderson}, {Gaensler}, {Heald},
  {O'Sullivan}, {Kaczmarek}, \& {Feain}}]{Anderson2018}
{Anderson}, C.~S., {Gaensler}, B.~M., {Heald}, G.~H., {et~al.} 2018, \apj, 855,
  41

\bibitem[{{Astropy Collaboration} {et~al.}(2013){Astropy Collaboration},
  {Robitaille}, {Tollerud}, {Greenfield}, {Droettboom}, {Bray}, {Aldcroft},
  {Davis}, {Ginsburg}, {Price-Whelan}, {Kerzendorf}, {Conley}, {Crighton},
  {Barbary}, {Muna}, {Ferguson}, {Grollier}, {Parikh}, {Nair}, {Unther},
  {Deil}, {Woillez}, {Conseil}, {Kramer}, {Turner}, {Singer}, {Fox}, {Weaver},
  {Zabalza}, {Edwards}, {Azalee Bostroem}, {Burke}, {Casey}, {Crawford},
  {Dencheva}, {Ely}, {Jenness}, {Labrie}, {Lim}, {Pierfederici}, {Pontzen},
  {Ptak}, {Refsdal}, {Servillat}, \& {Streicher}}]{Astropy2013}
{Astropy Collaboration}, {Robitaille}, T.~P., {Tollerud}, E.~J., {et~al.} 2013,
  \aap, 558, A33

\bibitem[{{Basu} {et~al.}(2018){Basu}, {Mao}, {Fletcher}, {Kanekar},
  {Shukurov}, {Schnitzeler}, {Vacca}, \& {Junklewitz}}]{2018MNRAS.477.2528B}
{Basu}, A., {Mao}, S.~A., {Fletcher}, A., {et~al.} 2018, \mnras, 477, 2528

\bibitem[{{Beck}(2015)}]{Beck2015}
{Beck}, R. 2015, \aapr, 24, 4

\bibitem[{{Blandford} \& {Eichler}(1987)}]{Blandford1987}
{Blandford}, R. \& {Eichler}, D. 1987, \physrep, 154, 1

\bibitem[{{Bonafede} {et~al.}(2018){Bonafede}, {Br{\"u}ggen}, {Rafferty},
  {Zhuravleva}, {Riseley}, {van Weeren}, {Farnes}, {Vazza}, {Savini}, {Wilber},
  {Botteon}, {Brunetti}, {Cassano}, {Ferrari}, {de Gasperin}, {Orr{\'u}},
  {Pizzo}, {R{\"o}ttgering}, \& {Shimwell}}]{Bonafede2018}
{Bonafede}, A., {Br{\"u}ggen}, M., {Rafferty}, D., {et~al.} 2018, \mnras, 478,
  2927

\bibitem[{{Bonafede} {et~al.}(2009){Bonafede}, {Feretti}, {Giovannini},
  {Govoni}, {Murgia}, {Taylor}, {Ebeling}, {Allen}, {Gentile}, \&
  {Pihlstr{\"o}m}}]{Bonafede2009a}
{Bonafede}, A., {Feretti}, L., {Giovannini}, G., {et~al.} 2009, \aap, 503, 707

\bibitem[{{Bonafede} {et~al.}(2014){Bonafede}, {Intema}, {Br{\"u}ggen},
  {Girardi}, {Nonino}, {Kantharia}, {van Weeren}, \&
  {R{\"o}ttgering}}]{Bonafede2014}
{Bonafede}, A., {Intema}, H.~T., {Br{\"u}ggen}, M., {et~al.} 2014, \apj, 785, 1

\bibitem[{{Bonafede} {et~al.}(2013){Bonafede}, {Vazza}, {Br{\"u}ggen},
  {Murgia}, {Govoni}, {Feretti}, {Giovannini}, \& {Ogrean}}]{Bonafede2013}
{Bonafede}, A., {Vazza}, F., {Br{\"u}ggen}, M., {et~al.} 2013, \mnras, 433,
  3208

\bibitem[{{Botteon} {et~al.}(2020){Botteon}, {Brunetti}, {Ryu}, \&
  {Roh}}]{Botteon2020a}
{Botteon}, A., {Brunetti}, G., {Ryu}, D., \& {Roh}, S. 2020, \aap, 634, A64

\bibitem[{Botteon {et~al.}(2018)Botteon, Gastaldello, \&
  Brunetti}]{Botteon2018}
Botteon, A., Gastaldello, F., \& Brunetti, G. 2018, Monthly Notices of the
  Royal Astronomical Society, 476, 5591

\bibitem[{{Botteon} {et~al.}(2016){Botteon}, {Gastaldello}, {Brunetti}, \&
  {Kale}}]{Botteon2016b}
{Botteon}, A., {Gastaldello}, F., {Brunetti}, G., \& {Kale}, R. 2016, \mnras,
  463, 1534

\bibitem[{{Brentjens} \& {de Bruyn}(2005)}]{Brentjens2005}
{Brentjens}, M.~A. \& {de Bruyn}, A.~G. 2005, \aap, 441, 1217

\bibitem[{{Briggs}(1995)}]{Briggs1995}
{Briggs}, D.~S. 1995, in American Astronomical Society Meeting Abstracts, Vol.
  187, American Astronomical Society Meeting Abstracts, 112.02

\bibitem[{{Brunetti} \& {Jones}(2014)}]{Brunetti2014}
{Brunetti}, G. \& {Jones}, T.~W. 2014, International Journal of Modern Physics
  D, 23, 1430007

\bibitem[{{Brunetti} {et~al.}(2001){Brunetti}, {Setti}, {Feretti}, \&
  {Giovannini}}]{Brunetti2001}
{Brunetti}, G., {Setti}, G., {Feretti}, L., \& {Giovannini}, G. 2001, \mnras,
  320, 365

\bibitem[{{Bryan} {et~al.}(2014){Bryan}, {Norman}, {O'Shea}, {Abel}, {Wise},
  {Turk}, {Reynolds}, {Collins}, {Wang}, {Skillman}, {Smith}, {Harkness},
  {Bordner}, {Kim}, {Kuhlen}, {Xu}, {Goldbaum}, {Hummels}, {Kritsuk}, {Tasker},
  {Skory}, {Simpson}, {Hahn}, {Oishi}, {So}, {Zhao}, {Cen}, {Li}, \& {Enzo
  Collaboration}}]{ENZO_2014}
{Bryan}, G.~L., {Norman}, M.~L., {O'Shea}, B.~W., {et~al.} 2014, \apjs, 211, 19

\bibitem[{{Burn}(1966)}]{Burn1966}
{Burn}, B.~J. 1966, \mnras, 133, 67

\bibitem[{{Carilli} \& {Taylor}(2002)}]{Carilli2002}
{Carilli}, C.~L. \& {Taylor}, G.~B. 2002, \araa, 40, 319

\bibitem[{{Di Gennaro} {et~al.}(2018){Di Gennaro}, {van Weeren}, {Hoeft},
  {Kang}, {Ryu}, {Rudnick}, {Forman}, {R{\"o}ttgering}, {Br{\"u}ggen},
  {Dawson}, {Golovich}, {Hoang}, {Intema}, {Jones}, {Kraft}, {Shimwell}, \&
  {Stroe}}]{Gennaro2018}
{Di Gennaro}, G., {van Weeren}, R.~J., {Hoeft}, M., {et~al.} 2018, \apj, 865,
  24

\bibitem[{{Di Gennaro} {et~al.}(2021){Di Gennaro}, {van Weeren}, {Rudnick},
  {Hoeft}, {Br{\"u}ggen}, {Ryu}, {R{\"o}ttgering}, {Forman}, {Stroe},
  {Shimwell}, {Kraft}, {Jones}, \& {Hoang}}]{DiGennaro2021}
{Di Gennaro}, G., {van Weeren}, R.~J., {Rudnick}, L., {et~al.} 2021, arXiv
  e-prints, arXiv:2102.06631

\bibitem[{{Dom{\'\i}nguez-Fern{\'a}ndez}
  {et~al.}(2020){Dom{\'\i}nguez-Fern{\'a}ndez}, {Br{\"u}ggen}, {Vazza},
  {Banda-Barrag{\'a}n}, {Rajpurohit}, {Mignone}, {Mukherjee}, \&
  {Vaidya}}]{2020arXiv200913514D}
{Dom{\'\i}nguez-Fern{\'a}ndez}, P., {Br{\"u}ggen}, M., {Vazza}, F., {et~al.}
  2020, arXiv e-prints, arXiv:2009.13514

\bibitem[{{Dom{\'\i}nguez-Fern{\'a}ndez}
  {et~al.}(2021){Dom{\'\i}nguez-Fern{\'a}ndez}, {Br{\"u}ggen}, {Vazza},
  {Hoeft}, {Banda-Barrag{\'a}n}, {Rajpurohit}, {Wittor}, {Mignone},
  {Mukherjee}, \& {Vaidya}}]{Paola2021a}
{Dom{\'\i}nguez-Fern{\'a}ndez}, P., {Br{\"u}ggen}, M., {Vazza}, F., {et~al.}
  2021, arXiv e-prints, arXiv:2108.06343

\bibitem[{{Dom{\'{\i}}nguez-Fern{\'a}ndez}
  {et~al.}(2019){Dom{\'{\i}}nguez-Fern{\'a}ndez}, {Vazza}, {Br{\"u}ggen}, \&
  {Brunetti}}]{2019MNRAS.486..623D}
{Dom{\'{\i}}nguez-Fern{\'a}ndez}, P., {Vazza}, F., {Br{\"u}ggen}, M., \&
  {Brunetti}, G. 2019, \mnras, 486, 623

\bibitem[{{Donnert} {et~al.}(2018){Donnert}, {Vazza}, {Br{\"u}ggen}, \&
  {ZuHone}}]{Donnert2018}
{Donnert}, J., {Vazza}, F., {Br{\"u}ggen}, M., \& {ZuHone}, J. 2018, \ssr, 214,
  122

\bibitem[{{Drury}(1983)}]{Drury1983}
{Drury}, L.~O. 1983, Reports on Progress in Physics, 46, 973

\bibitem[{{Ensslin} {et~al.}(1998){Ensslin}, {Biermann}, {Klein}, \&
  {Kohle}}]{Ensslin1998}
{Ensslin}, T.~A., {Biermann}, P.~L., {Klein}, U., \& {Kohle}, S. 1998, \aap,
  332, 395

\bibitem[{{Farnsworth} {et~al.}(2011){Farnsworth}, {Rudnick}, \&
  {Brown}}]{Farnsworth2011}
{Farnsworth}, D., {Rudnick}, L., \& {Brown}, S. 2011, \aj, 141, 191

\bibitem[{{George} {et~al.}(2012){George}, {Stil}, \& {Keller}}]{George2012}
{George}, S.~J., {Stil}, J.~M., \& {Keller}, B.~W. 2012, \pasa, 29, 214

\bibitem[{{Girardi} {et~al.}(2016){Girardi}, {Boschin}, {Gastaldello},
  {Giovannini}, {Govoni}, {Murgia}, {Barrena}, {Ettori}, {Trasatti}, \&
  {Vacca}}]{Girardi2016}
{Girardi}, M., {Boschin}, W., {Gastaldello}, F., {et~al.} 2016, \mnras, 456,
  2829

\bibitem[{{Govoni} \& {Feretti}(2004)}]{Govoni2004}
{Govoni}, F. \& {Feretti}, L. 2004, International Journal of Modern Physics D,
  13, 1549

\bibitem[{{Govoni} {et~al.}(2005){Govoni}, {Murgia}, {Feretti}, {Giovannini},
  {Dallacasa}, \& {Taylor}}]{Govoni2005}
{Govoni}, F., {Murgia}, M., {Feretti}, L., {et~al.} 2005, \aap, 430, L5

\bibitem[{{Govoni} {et~al.}(2013){Govoni}, {Murgia}, {Xu}, {Li}, {Norman},
  {Feretti}, {Giovannini}, \& {Vacca}}]{Govoni2013}
{Govoni}, F., {Murgia}, M., {Xu}, H., {et~al.} 2013, \aap, 554, A102

\bibitem[{{Heald}(2009)}]{Heald2009}
{Heald}, G. 2009, in IAU Symposium, Vol. 259, IAU Symposium, ed. K.~G.
  {Strassmeier}, A.~G. {Kosovichev}, \& J.~E. {Beckman}, 591--602

\bibitem[{{Hoang} {et~al.}(2018){Hoang}, {Shimwell}, {van Weeren}, {Intema},
  {R{\"o}ttgering}, {Andrade-Santos}, {Akamatsu}, {Bonafede}, {Brunetti},
  {Dawson}, {Golovich}, {Best}, {Botteon}, {Br{\"u}ggen}, {Cassano}, {de
  Gasperin}, {Hoeft}, {Stroe}, \& {White}}]{Hoang2018}
{Hoang}, D.~N., {Shimwell}, T.~W., {van Weeren}, R.~J., {et~al.} 2018, \mnras,
  478, 2218

\bibitem[{{Hoeft} \& {Br{\"u}ggen}(2007)}]{Hoeft2007}
{Hoeft}, M. \& {Br{\"u}ggen}, M. 2007, \mnras, 375, 77

\bibitem[{{Hunter}(2007)}]{Hunter2007}
{Hunter}, J.~D. 2007, Computing in Science and Engineering, 9, 90

\bibitem[{{Johnson} {et~al.}(2020){Johnson}, {Rudnick}, {Jones}, {Mendygral},
  \& {Dolag}}]{Johnson2020}
{Johnson}, A.~R., {Rudnick}, L., {Jones}, T.~W., {Mendygral}, P.~J., \&
  {Dolag}, K. 2020, \apj, 888, 101

\bibitem[{Jones {et~al.}(2001)Jones, Oliphant, Peterson, {et~al.}}]{Jones2001}
Jones, E., Oliphant, T., Peterson, P., {et~al.} 2001, {SciPy}: Open source
  scientific tools for {Python}, [Online; accessed <today>]

\bibitem[{{Kang} \& {Ryu}(2011)}]{Kang2011}
{Kang}, H. \& {Ryu}, D. 2011, \apj, 734, 18

\bibitem[{{Kierdorf} {et~al.}(2017){Kierdorf}, {Beck, R.}, {Hoeft, M.}, {Klein,
  U.}, {van Weeren, R. J.}, {Forman, W. R.}, \& {Jones, C.}}]{Kierdorf2016}
{Kierdorf}, M., {Beck, R.}, {Hoeft, M.}, {et~al.} 2017, A\&A, 600, A18

\bibitem[{{Kim} {et~al.}(2016){Kim}, {Lilly}, {Miniati}, {Bernet}, {Beck},
  {O'Sullivan}, \& {Gaensler}}]{Kim2016}
{Kim}, K.~S., {Lilly}, S.~J., {Miniati}, F., {et~al.} 2016, \apj, 829, 133

\bibitem[{{Klein} \& {Fletcher}(2015)}]{Klein2015}
{Klein}, U. \& {Fletcher}, A. 2015, {Galactic and Intergalactic Magnetic
  Fields}

\bibitem[{{Laing}(1980)}]{Laing1980}
{Laing}, R.~A. 1980, \mnras, 193, 439

\bibitem[{{Loi} {et~al.}(2017){Loi}, {Murgia}, {Govoni}, {Vacca}, {Feretti},
  {Giovannini}, {Carretti}, {Gastaldello}, {Girardi}, {Vazza}, {Concu},
  {Melis}, {Paladino}, {Poppi}, {Valente}, {Boschin}, {Clarke},
  {Colafrancesco}, {En{\ss}lin}, {Ferrari}, {de Gasperin}, {Gregorini},
  {Johnston-Hollitt}, {Junklewitz}, {Orr{\`u}}, {Parma}, {Perley}, \&
  {Taylor}}]{Loi2017}
{Loi}, F., {Murgia}, M., {Govoni}, F., {et~al.} 2017, \mnras, 472, 3605

\bibitem[{{Loi} {et~al.}(2020){Loi}, {Murgia}, {Vacca}, {Govoni}, {Melis},
  {Wittor}, {Beck}, {Kierdorf}, {Bonafede}, {Boschin}, {Brienza}, {Carretti},
  {Concu}, {Feretti}, {Gastaldello}, {Paladino}, {Rajpurohit}, {Serra}, \&
  {Vazza}}]{Loi2020}
{Loi}, F., {Murgia}, M., {Vacca}, V., {et~al.} 2020, arXiv e-prints,
  arXiv:2008.03314

\bibitem[{{Mohan} \& {Rafferty}(2015)}]{Mohan2015}
{Mohan}, N. \& {Rafferty}, D. 2015, {PyBDSM: Python Blob Detection and Source
  Measurement}, Astrophysics Source Code Library

\bibitem[{{Murgia} {et~al.}(2004){Murgia}, {Govoni}, {Feretti}, {Giovannini},
  {Dallacasa}, {Fanti}, {Taylor}, \& {Dolag}}]{Murgia2004}
{Murgia}, M., {Govoni}, F., {Feretti}, L., {et~al.} 2004, \aap, 424, 429

\bibitem[{{Ogrean} {et~al.}(2013){Ogrean}, {Br{\"u}ggen}, {van Weeren},
  {R{\"o}ttgering}, {Croston}, \& {Hoeft}}]{Ogrean2013}
{Ogrean}, G.~A., {Br{\"u}ggen}, M., {van Weeren}, R.~J., {et~al.} 2013, \mnras,
  433, 812

\bibitem[{{Oppermann} {et~al.}(2012){Oppermann}, {Junklewitz}, {Robbers},
  {Bell}, {En{\ss}lin}, {Bonafede}, {Braun}, {Brown}, {Clarke}, {Feain},
  {Gaensler}, {Hammond}, {Harvey-Smith}, {Heald}, {Johnston-Hollitt}, {Klein},
  {Kronberg}, {Mao}, {McClure-Griffiths}, {O'Sullivan}, {Pratley}, {Robishaw},
  {Roy}, {Schnitzeler}, {Sotomayor-Beltran}, {Stevens}, {Stil}, {Sunstrum},
  {Tanna}, {Taylor}, \& {Van Eck}}]{Oppermann2012}
{Oppermann}, N., {Junklewitz}, H., {Robbers}, G., {et~al.} 2012, \aap, 542, A93

\bibitem[{{O'Sullivan} {et~al.}(2012){O'Sullivan}, {Brown}, {Robishaw},
  {Schnitzeler}, {McClure-Griffiths}, {Feain}, {Taylor}, {Gaensler}, {Land
  ecker}, {Harvey-Smith}, \& {Carretti}}]{OSullivan2012}
{O'Sullivan}, S.~P., {Brown}, S., {Robishaw}, T., {et~al.} 2012, \mnras, 421,
  3300

\bibitem[{{Owen} {et~al.}(2014){Owen}, {Rudnick}, {Eilek}, {Rau}, {Bhatnagar},
  \& {Kogan}}]{Owen2014}
{Owen}, F.~N., {Rudnick}, L., {Eilek}, J., {et~al.} 2014, \apj, 794, 24

\bibitem[{{Ozawa} {et~al.}(2015){Ozawa}, {Nakanishi}, {Akahori}, {Anraku},
  {Takizawa}, {Takahashi}, {Onodera}, {Tsuda}, \& {Sofue}}]{Ozawa2015}
{Ozawa}, T., {Nakanishi}, H., {Akahori}, T., {et~al.} 2015, \pasj, 67, 110

\bibitem[{{Pandey-Pommier} {et~al.}(2013){Pandey-Pommier}, {Richard}, {Combes},
  {Dwarakanath}, {Guiderdoni}, {Ferrari}, {Sirothia}, \&
  {Narasimha}}]{PandeyPommier2013}
{Pandey-Pommier}, M., {Richard}, J., {Combes}, F., {et~al.} 2013, \aap, 557,
  A117

\bibitem[{{Pasetto} {et~al.}(2018){Pasetto}, {Carrasco-Gonz{\'a}lez},
  {O'Sullivan}, {Basu}, {Bruni}, {Kraus}, {Curiel}, \& {Mack}}]{Pasetto2018}
{Pasetto}, A., {Carrasco-Gonz{\'a}lez}, C., {O'Sullivan}, S., {et~al.} 2018,
  \aap, 613, A74

\bibitem[{{Pearce} {et~al.}(2017){Pearce}, {van Weeren}, {Andrade-Santos},
  {Jones}, {Forman}, {Br{\"u}ggen}, {Bulbul}, {Clarke}, {Kraft}, {Medezinski},
  {Mroczkowski}, {Nonino}, {Nulsen}, {Randall}, \& {Umetsu}}]{Pearce2017}
{Pearce}, C.~J.~J., {van Weeren}, R.~J., {Andrade-Santos}, F., {et~al.} 2017,
  \apj, 845, 81

\bibitem[{{Perley} \& {Butler}(2013)}]{Perley2013}
{Perley}, R.~A. \& {Butler}, B.~J. 2013, \apjs, 204, 19

\bibitem[{{Petrosian}(2001)}]{Petrosian2001}
{Petrosian}, V. 2001, \apj, 557, 560

\bibitem[{{Pizzo} {et~al.}(2011){Pizzo}, {de Bruyn}, {Bernardi}, \&
  {Brentjens}}]{Pizzo2011}
{Pizzo}, R.~F., {de Bruyn}, A.~G., {Bernardi}, G., \& {Brentjens}, M.~A. 2011,
  \aap, 525, A104

\bibitem[{{Rajpurohit} {et~al.}(2021{\natexlab{a}}){Rajpurohit}, {Brunetti},
  {Bonafede}, {van Weeren}, {Botteon}, {Vazza}, {Hoeft}, {Riseley},
  {Bonnassieux}, {Brienza}, {Forman}, {R{\"o}ttgering}, {Rajpurohit},
  {Locatelli}, {Shimwell}, {Cassano}, {Di Gennaro}, {Br{\"u}ggen}, {Wittor},
  {Drabent}, \& {Ignesti}}]{Rajpurohit2021b}
{Rajpurohit}, K., {Brunetti}, G., {Bonafede}, A., {et~al.} 2021{\natexlab{a}},
  \aap, 646, A135

\bibitem[{{Rajpurohit} {et~al.}(2018){Rajpurohit}, {Hoeft}, {van Weeren},
  {Rudnick}, {R{\"o}ttgering}, {Forman}, {Br{\"u}ggen}, {Croston},
  {Andrade-Santos}, {Dawson}, {Intema}, {Kraft}, {Jones}, \&
  {Jee}}]{Rajpurohit2018}
{Rajpurohit}, K., {Hoeft}, M., {van Weeren}, R.~J., {et~al.} 2018, \apj, 852,
  65

\bibitem[{{Rajpurohit} {et~al.}(2020{\natexlab{a}}){Rajpurohit}, {Hoeft},
  {Vazza}, {Rudnick}, {van Weeren}, {Wittor}, {Drabent}, {Brienza},
  {Bonnassieux}, {Locatelli}, {Kale}, \& {Dumba}}]{Rajpurohit2020a}
{Rajpurohit}, K., {Hoeft}, M., {Vazza}, F., {et~al.} 2020{\natexlab{a}}, \aap,
  636, A30

\bibitem[{{Rajpurohit} {et~al.}(2020{\natexlab{b}}){Rajpurohit}, {Vazza},
  {Hoeft}, {Loi}, {Beck}, {Vacca}, {Kierdorf}, {van Weeren}, {Wittor},
  {Govoni}, {Murgia}, {Riseley}, {Locatelli}, {Drabent}, \&
  {Bonnassieux}}]{Rajpurohit2020b}
{Rajpurohit}, K., {Vazza}, F., {Hoeft}, M., {et~al.} 2020{\natexlab{b}}, \aap,
  642, L13

\bibitem[{{Rajpurohit} {et~al.}(2021{\natexlab{b}}){Rajpurohit}, {Vazza}, {van
  Weeren}, {Hoeft}, {Brienza}, {Bonnassieux}, {Riseley}, {Brunetti},
  {Bonafede}, {Br{\"u}ggen}, {Formann}, {R{\"o}ttgering}, {Drabent},
  {Rajpurohit}, {Dom{\'\i}nguez-Fern{\'a}ndez}, {Wittor}, \&
  {Andrade-Santos}}]{Rajpurohit2021c}
{Rajpurohit}, K., {Vazza}, F., {van Weeren}, R.~J., {et~al.}
  2021{\natexlab{b}}, arXiv e-prints, arXiv:2104.05690

\bibitem[{{Rajpurohit} {et~al.}(2021{\natexlab{c}}){Rajpurohit}, {Wittor}, {van
  Weeren}, {Vazza}, {Hoeft}, {Rudnick}, {Locatelli}, {Eilek}, {Forman},
  {Bonafede}, {Bonnassieux}, {Riseley}, {Brienza}, {Brunetti}, {Br{\"u}ggen},
  {Loi}, {Rajpurohit}, {R{\"o}ttgering}, {Botteon}, {Clarke}, {Drabent},
  {Dom{\'\i}nguez-Fern{\'a}ndez}, {Di Gennaro}, \&
  {Gastaldello}}]{Rajpurohit2021a}
{Rajpurohit}, K., {Wittor}, D., {van Weeren}, R.~J., {et~al.}
  2021{\natexlab{c}}, \aap, 646, A56

\bibitem[{{Robitaille} \& {Bressert}(2012)}]{Robitaille2012}
{Robitaille}, T. \& {Bressert}, E. 2012, {APLpy: Astronomical Plotting Library
  in Python}, Astrophysics Source Code Library

\bibitem[{{Sarazin} {et~al.}(2013){Sarazin}, {Finoguenov}, \&
  {Wik}}]{Sarazin2013}
{Sarazin}, C.~L., {Finoguenov}, A., \& {Wik}, D.~R. 2013, Astronomische
  Nachrichten, 334, 346

\bibitem[{{Skillman} {et~al.}(2013){Skillman}, {Xu}, {Hallman}, {O'Shea},
  {Burns}, {Li}, {Collins}, \& {Norman}}]{Skillman2013}
{Skillman}, S.~W., {Xu}, H., {Hallman}, E.~J., {et~al.} 2013, \apj, 765, 21

\bibitem[{{Sokoloff} {et~al.}(1998){Sokoloff}, {Bykov}, {Shukurov},
  {Berkhuijsen}, {Beck}, \& {Poezd}}]{Sokoloff1998}
{Sokoloff}, D.~D., {Bykov}, A.~A., {Shukurov}, A., {et~al.} 1998, \mnras, 299,
  189

\bibitem[{{Stuardi} {et~al.}(2021){Stuardi}, {Bonafede}, {Lovisari},
  {Dom{\'\i}nguez-Fern{\'a}ndez}, {Vazza}, {Br{\"u}ggen}, {van Weeren}, \& {de
  Gasperin}}]{Stuardi2021}
{Stuardi}, C., {Bonafede}, A., {Lovisari}, L., {et~al.} 2021, \mnras, 502, 2518

\bibitem[{{Stuardi} {et~al.}(2019){Stuardi}, {Bonafede}, {Wittor}, {Vazza},
  {Botteon}, {Locatelli}, {Dallacasa}, {Golovich}, {Hoeft}, {van Weeren},
  {Br{\"u}ggen}, \& {de Gasperin}}]{Stuardi2019}
{Stuardi}, C., {Bonafede}, A., {Wittor}, D., {et~al.} 2019, \mnras, 489, 3905

\bibitem[{{Tribble}(1991)}]{1991MNRAS.250..726T}
{Tribble}, P.~C. 1991, \mnras, 250, 726

\bibitem[{van~der Walt {et~al.}(2011)van~der Walt, Colbert, \&
  Varoquaux}]{Numpy2011}
van~der Walt, S., Colbert, S.~C., \& Varoquaux, G. 2011, Computing in Science
  Engineering, 13, 22

\bibitem[{{van Weeren} {et~al.}(2017{\natexlab{a}}){van Weeren},
  {Andrade-Santos}, {Dawson}, {Golovich}, {Lal}, {Kang}, {Ryu}, {Br{\`i}ggen},
  {Ogrean}, {Forman}, {Jones}, {Placco}, {Santucci}, {Wittman}, {Jee}, {Kraft},
  {Sobral}, {Stroe}, \& {Fogarty}}]{vanWeeren2017a}
{van Weeren}, R.~J., {Andrade-Santos}, F., {Dawson}, W.~A., {et~al.}
  2017{\natexlab{a}}, Nature Astronomy, 1, 0005

\bibitem[{{van Weeren} {et~al.}(2016{\natexlab{a}}){van Weeren}, {Brunetti},
  {Br{\"u}ggen}, {Andrade-Santos}, {Ogrean}, {Williams}, {R{\"o}ttgering},
  {Dawson}, {Forman}, {de Gasperin}, {Hardcastle}, {Jones}, {Miley},
  {Rafferty}, {Rudnick}, {Sabater}, {Sarazin}, {Shimwell}, {Bonafede}, {Best},
  {B{\^i}rzan}, {Cassano}, {Chy{\.z}y}, {Croston}, {Dijkema}, {En{\ss}lin},
  {Ferrari}, {Heald}, {Hoeft}, {Horellou}, {Jarvis}, {Kraft}, {Mevius},
  {Intema}, {Murray}, {Orr{\'u}}, {Pizzo}, {Sridhar}, {Simionescu}, {Stroe},
  {van der Tol}, \& {White}}]{vanWeeren2016a}
{van Weeren}, R.~J., {Brunetti}, G., {Br{\"u}ggen}, M., {et~al.}
  2016{\natexlab{a}}, \apj, 818, 204

\bibitem[{{van Weeren} {et~al.}(2019){van Weeren}, {de Gasperin}, {Akamatsu},
  {Br{\"u}ggen}, {Feretti}, {Kang}, {Stroe}, \& {Zandanel}}]{vanWeeren2019}
{van Weeren}, R.~J., {de Gasperin}, F., {Akamatsu}, H., {et~al.} 2019, \ssr,
  215, 16

\bibitem[{{van Weeren} {et~al.}(2016{\natexlab{b}}){van Weeren}, {Ogrean},
  {Jones}, {Forman}, {Andrade-Santos}, {Bonafede}, {Br{\"u}ggen}, {Bulbul},
  {Clarke}, {Churazov}, {David}, {Dawson}, {Donahue}, {Goulding}, {Kraft},
  {Mason}, {Merten}, {Mroczkowski}, {Murray}, {Nulsen}, {Rosati}, {Roediger},
  {Randall}, {Sayers}, {Umetsu}, {Vikhlinin}, \& {Zitrin}}]{vanWeeren2016b}
{van Weeren}, R.~J., {Ogrean}, G.~A., {Jones}, C., {et~al.} 2016{\natexlab{b}},
  \apj, 817, 98

\bibitem[{{van Weeren} {et~al.}(2017{\natexlab{b}}){van Weeren}, {Ogrean},
  {Jones}, {Forman}, {Andrade-Santos}, {Pearce}, {Bonafede}, {Br{\"u}ggen},
  {Bulbul}, {Clarke}, {Churazov}, {David}, {Dawson}, {Donahue}, {Goulding},
  {Kraft}, {Mason}, {Merten}, {Mroczkowski}, {Nulsen}, {Rosati}, {Roediger},
  {Randall}, {Sayers}, {Umetsu}, {Vikhlinin}, \& {Zitrin}}]{vanWeeren2017b}
{van Weeren}, R.~J., {Ogrean}, G.~A., {Jones}, C., {et~al.} 2017{\natexlab{b}},
  \apj, 835, 197

\bibitem[{{van Weeren} {et~al.}(2009){van Weeren}, {R{\"o}ttgering},
  {Br{\"u}ggen}, \& {Cohen}}]{vanWeeren2009}
{van Weeren}, R.~J., {R{\"o}ttgering}, H.~J.~A., {Br{\"u}ggen}, M., \& {Cohen},
  A. 2009, \aap, 505, 991

\bibitem[{{van Weeren} {et~al.}(2010){van Weeren}, {R{\"o}ttgering},
  {Br{\"u}ggen}, \& {Hoeft}}]{vanWeeren2010}
{van Weeren}, R.~J., {R{\"o}ttgering}, H.~J.~A., {Br{\"u}ggen}, M., \& {Hoeft},
  M. 2010, Science, 330, 347

\bibitem[{{van Weeren} {et~al.}(2012){van Weeren}, {R{\"o}ttgering}, {Intema},
  {Rudnick}, {Br{\"u}ggen}, {Hoeft}, \& {Oonk}}]{vanWeeren2012a}
{van Weeren}, R.~J., {R{\"o}ttgering}, H.~J.~A., {Intema}, H.~T., {et~al.}
  2012, \aap, 546, A124

\bibitem[{{Vazza} {et~al.}(2018){Vazza}, {Brunetti}, {Br{\"u}ggen}, \&
  {Bonafede}}]{vazza2018dynamo}
{Vazza}, F., {Brunetti}, G., {Br{\"u}ggen}, M., \& {Bonafede}, A. 2018, \mnras,
  474, 1672

\bibitem[{{Wardle} \& {Kronberg}(1974)}]{Wardle1974}
{Wardle}, J.~F.~C. \& {Kronberg}, P.~P. 1974, \apj, 194, 249

\bibitem[{{Wittor} {et~al.}(2019){Wittor}, {Hoeft}, {Vazza}, {Br{\"u}ggen}, \&
  {Dom{\'\i}nguez-Fern{\'a}ndez}}]{2019arXiv190911329W}
{Wittor}, D., {Hoeft}, M., {Vazza}, F., {Br{\"u}ggen}, M., \&
  {Dom{\'\i}nguez-Fern{\'a}ndez}, P. 2019, \mnras, 490, 3987

\bibitem[{{Wittor} {et~al.}(2017){Wittor}, {Vazza}, \& {Br{\"u}ggen}}]{wi17}
{Wittor}, D., {Vazza}, F., \& {Br{\"u}ggen}, M. 2017, \mnras, 464, 4448

\bibitem[{{Wittor} {et~al.}(2020){Wittor}, {Vazza}, {Ryu}, \&
  {Kang}}]{2020MNRAS.495L.112W}
{Wittor}, D., {Vazza}, F., {Ryu}, D., \& {Kang}, H. 2020, \mnras, 495, L112

\end{thebibliography}

\end{document}